\documentclass[twoside,12pt]{article}
\usepackage{epsfig}

\usepackage{color}

\usepackage{authblk}

\usepackage{appendix}

\usepackage{comment}

\usepackage{cite}

\usepackage{amssymb}
\usepackage{graphics}
\usepackage{graphicx}
\usepackage{amsmath}
\usepackage{amsfonts}
\usepackage{bm}
\usepackage{slashed}

\def\ss{\mbox{\boldmath $\sigma$}}
\newcommand{\be}{\begin{equation}}
\newcommand{\ee}{\end{equation}}
\newcommand{\bea}{\begin{eqnarray}}
\newcommand{\eea}{\end{eqnarray}}

\begin{document}

\title{\vspace{1cm} Precision Gravity Tests and the Einstein Equivalence Principle}
\author{G. M.\ Tino,$^{1}$ L.\ Cacciapuoti,$^{2}$ S.\ Capozziello,$^{3,3a}$ G.\ Lambiase,$^{4,4a}$ F.\ Sorrentino $^{5}$\\ 
$^1$ Dipartimento di Fisica e Astronomia and LENS Laboratory\\
Universit\`a di Firenze and INFN-Sezione di Firenze\\ via  Sansone 1,  Sesto Fiorentino, Italy\\
$^2$ European Space Agency, Keplerlaan 1 - P.O. Box 299, 2200 AG Noordwijk ZH, The Netherlands\\
$^3$ Dipartimento di Fisica "E. Pancini" , Universit\`a di Napoli “Federico II”, via Cinthia 9, I-80126, Napoli, Italy. \\
$^{3a}$ INFN, Sezione di Napoli, via Cinthia 9, I-80126 Napoli, Italy \\
$^4$ Dipartimento di Fisica E.R: Caianiello, Universit\`a di Salerno,  Via Giovanni Paolo II 132, I-84084, Fisciano (SA), Italy. \\
$^{4a}$ INFN, Sezione di Napoli, Gruppo collegato di Salerno, Via Giovanni Paolo II 132, I-84084 Fisciano (SA), Italy\\
$^5$  INFN Sezione di Genova, Via Dodecaneso 33, I-16146 Genova, Italy}
\maketitle
\begin{abstract} 
General Relativity is today the best theory of gravity 
addressing a wide range of phenomena. Our
understanding of  physical  laws, from cosmology  to local scales, cannot be properly formulated without taking into account its concepts, procedures and formalim. It is based on one of the most fundamental principles of Nature, the Equivalence Principle, which represents the core of the Einstein theory of gravity describing, under the same standard, the metric and geodesic structure of the spacetime. The confirmation of its validity at different  scales and in different contexts  represents one of the main challenges of  modern physics both from  the theoretical and the experimental points of view. 

A major issue related to this principle is the fact that we actually do not know if it is valid or not at quantum level. We are simply assuming its validity at fundamental scales. This question is crucial in any self-consistent theory of gravity. 

Furthermore, recent  progress on relativistic theories of gravity, including deviations from General Relativity at various scales, such as  extensions and alternatives to the Einstein theory,  have to take into  account, besides the Equivalence Principle,  new issues like  Dark Matter and Dark Energy, as well as  the validity of fundamental principles like local Lorentz and position invariance. The general trend is that high precision experiments  are conceived and realized  to test both Einstein's  theory and its alternatives at fundamental level using established and novel methods. For example, experiments based on quantum sensors (atomic clocks, accelerometers, gyroscopes, gravimeters, etc.) allow to  set stringent constraints on well established symmetry laws (e.g. CPT and Lorentz invariance), on the physics beyond the Standard Model of particles and interactions, and on General Relativity and its possible extensions.\\
In this review, we discuss precision tests of gravity in General Relativity and alternative theories and their relation with the Equivalence Principle.
In the first part, we discuss  the Einstein Equivalence Principle according to its  weak and strong formulation. We recall some basic topics of General Relativity and the necessity of its extension. Some models of modified gravity are presented in some details. This provides  us the ground for discussing the Equivalence Principle also in the framework of extended and alternative theories of gravity. In particular,  we focus on the possibility to violate the Equivalence Principle at finite temperature, both in the frameworks of General Relativity and of modified gravity. Equivalence Principle violations in the Standard Model Extension are also discussed.
The second part of the paper is devoted to the experimental tests of the Equivalence Principle in its weak formulation. We present the  results and methods used in high-precision  experiments, and discuss the potential and prospects for future experimental tests.
\end{abstract}

\eject
\tableofcontents
\eject

\section{Introduction}

General Relativity (GR) relies  on the  assumption that space and
time are entangled into a unique structure, i.e.    the spacetime. It is  assigned on  a pseudo-Riemannian manifold endowed with a Lorenzian signature.
Dynamics has to reproduce, in  the absence of a gravitational field, the Minkowski spacetime.

 GR, as an extension of classical mechanics,  has to match some minimal
requirements to be considered  a self-consistent  theory of gravitation: it has to reproduce results of   Newton's  physics  in the
weak-energy limit, hence it must be able to explain  dynamics related to the planetary motion  and the gravitating
structures such as stars, galaxies, clusters of galaxies. Moreover, it has to pass   observational tests in  the  Solar System. These facts  constitute the  experimental foundation on which GR is based. They are usually called the "classical tests of GR" \cite{Will1993,Will2014}.

Beside the above "mechanical issues", GR has  to explain  Galactic dynamics,
considering  baryonic constituents, like stars,  planets, dust and gas,
radiation. These components are tied together by the  Newtonian potential, which is  supposed to work at  any Galactic scales. Also, GR has to  address the  large scale structure formation and dynamics. 
  At  cosmological scales, GR is required to  address  dynamics of the whole universe and  correctly reproduce
cosmological parameters like  the Hubble expansion rate, the density parameters and the accelerated (decelerated) behavior of cosmic fluid. 
Cosmological and astrophysical observations actually probe only the standard baryonic matter, the
radiation and the attractive overall interaction of gravity  acting at all
scales.  

Furthermore, starting from  Galileo, 
the free-fall  of different bodies  is assumed to be independent of  the nature of massive bodies  on the Earth. The  free-fall acceleration is unique and implies that  gravitational and  inertial mass ratio is identical for different bodies. This experimental result  is one of the foundations  of Einstein's GR  as well as of any  metric  theory  of gravity.
After Galileo's experiment from the leaning tower of Pisa,  the free-fall acceleration uniqueness has been  verified in many experiments,
%
as widely discussed in the second part of this review. Summarizing, we can say that  GR is based on four main  assumptions:
\begin{quote}
The "{\it  Relativity Principle}" -  there is no  preferred
inertial frames, i.e. all frames (accelerated or not)  are good frames for Physics.
\end{quote}
\begin{quote}
The "{\it Equivalence Principle}" (EP) -
inertial effects are locally indistinguishable from
gravitational effects (which means the equivalence between the
inertial and the gravitational masses). In other words, any gravitational field can be locally cancelled.
\end{quote}
\begin{quote}
The "{\it  General Covariance Principle}" - field equations must be "covariant" in form, i.e. they must be  invariant  under the action  of any spacetime diffeomorphisms.
\end{quote}
\begin{quote}
The "{\it Causality Principle}" - each point of space-time admits a universal notion of past, present and future.
\end{quote}

\noindent On these bases, Einstein postulated that, in  a
four-dimensional spacetime manifold, the
gravitational field is described by the metric tensor  $ds^2 = g_{\mu\nu}dx^{\mu}dx^{\nu}$, with the same signature of
Minkowski metric.  The  metric coefficients are the  physical  gravitational potentials. Moreover, 
spacetime is curved by the distribution of  energy-matter sources, e.g.,  the distribution of celestial bodies.

An important historical remark is necessary at this point.  E. Kretschmann, in 1917 \cite{k1917}, criticized the General Covariance Principle. In demanding General Covariance, he asserted that Einstein  placed no constraint on the physical content of his theory.   Kretschmann stressed that  any spacetime theory could be formulated in  a generally covariant way without any physical principle. In formulating GR, Einstein used  tensor calculus. Kretschmann pointed out that  this calculus  allowed for  general covariant formulations of theories while  Einstein discussed  general covariance as the form invariance of   theory's equations as soon as  the spacetime coordinates are transformed. This  can be considered as a  "passive" point of view  of General Covariance: if we have some system of fields, we can change our spacetime coordinate system as we like and the new descriptions of the fields in the new coordinate systems will still solve the theory's equations. The answer by  Einstein was that the  form invariance of the theory's equations also allows  a second version, the so-called "active" General Covariance. It involves no transformation of the spacetime coordinate system. In fact, active General Covariance gives rise to  the generation of  new solutions of the equations of the theory in the same coordinate system once one solution is  given. According to this approach, General Covariance Principle can be considered a physical principle.

The above  principles  require
that the spacetime structure has to be determined by either one or
both of the  two following fields:  a Lorentzian metric $g$ and a linear
connection $\Gamma$, assumed by Einstein  to be torsionless because, at that time, the spin of particles was not considered a possible source for the gravitational field. The physical meaning of these two fields is the following: The metric  $g$  fixes the spacetime causal structure, that is the
light cones. According to this statement,  metric relations, i.e. clocks and rods, are possible. On the other hand, 
the connection  $\Gamma$  fixes the free-fall of objects, that is  the local
inertial observers according to the Equivalence Principle. Both fields, of course, have to satisfy some
compatibility relations like the  requirement that photons
follow null geodesics of $\Gamma$. This means that  $\Gamma$ and $g$ can be
independent, {\it a priori}, but they are constrained, {\it a posteriori},
according to  some physical restrictions which  impose that
$\Gamma$ has  to be the Levi-Civita connection of  $g$. However, in more general approaches, $\Gamma$ and $g$ can be 
independent  \cite{Ferraris1982}.

Despite  the self-consistency and the solid foundation,  there are several issues for GR, both from the
 theoretical and the experimental (observational)  points of view.
The latter   clearly points out  that  GR is not  capable of addressing
Galactic, extra-galactic and cosmic dynamics unless a huge quantity of  some
exotic forms of matter-energy is considered. These  ingredients are  usually called   {\it dark matter} and {\it dark
energy}  and constitute up to  $95\%$ of the total  amount of cosmic matter-energy \cite{Capozziello2013,Capozziello2011}.

On the other hand, instead of changing the source side of the  Einstein field equations,  a "geometrical view" can be taken into account to fit  the missing matter-energy  of the  observed Universe.  In such a case, the dark side could be addressed  by extending GR including further  geometric invariants into the  Hilbert - Einstein Action besides the Ricci curvature scalar $R$. These effective Lagrangians can be  justified at the fundamental level considering the quantization of fields  on curved spacetimes \cite{Capozziello2011}.
However, at the present stage of the research,    there is  no final probe discriminating between dark matter-energy picture  and extended (alternative) gravity\footnote{An important remark is useful at this point. With the term {\it Extended Gravity}, we mean any class of theories by which it is possible to recover Einstein GR as a particular case or in some post-Einstenian limit as in the case of $f(R)$ gravity. With {\it Alternative Gravity}, we mean a class of theories which considers different approach with respect to GR, for example the Teleparallel Equivalent Gravity considering the torsion scalar instead of curvature scalar to describe dynamics.}.
Furthermore, the bulk of observations to be considered is very large and then   an effective Lagrangian or a single new particle, addressing  the whole phenomenology at all astrophysical and cosmic scales, would be very difficult to find.

An important discussion is related to the choice of the dynamical variables. 
In formulating GR, Einstein assumed that the metric $g$  is the fundamental object to describe gravity.
The connection $\Gamma^\alpha_{\mu\nu}=\Big\{ \begin{array}{c} \alpha \\ \mu\nu \end{array} \Big\}_g$ is assumed, by construction, 
with no dynamics. Only $g$ has
dynamics. This means that the metric $g$ determines,  at the same time,  the causal structure (light
cones), the measurements (rods and clocks) and the free fall of test particles (geodesic structure).
Spacetime is given by  a couple of mathematical objects $\{{\cal M},g\}$ constituted by a Riemann  manifold and a metric.
 Einstein realized that gravity induces  free fall  and that 
the EP  selects an object that cannot be a tensor because the connection $\Gamma$  can be  switched off and set to zero at least in a point. According to this consideration, Einstein  was obliged to
choose the Levi - Civita connection determined by the metric structure.

Alternatively, in the Palatini formalism a (symmetric) connection $\Gamma$ and a metric $g$ are assumed  and these two fields can  varied independently. According to this picture, spacetime is a triple
$\{{\cal M}, g, \Gamma\}$ where the metric determines causal structure  while $\Gamma$ determines the free fall.
This means that, in the Palatini formalism,  connections  are differential  equations determining dynamics.
From this point of view,  $\Gamma$ is  the Levi-Civita connection of $g$  as  an outcome of the field equations.

The connection is the fundamental field in the Lagrangian while  the metric $g$ enters the Lagrangian as  the need  to define lengths and distances to make experiments. It defines  the causal structure but  has no dynamical role. As a consequence, there is no reason  to assume $g$ as the  potential of $\Gamma$. 

With this consideration in mind, we discuss here the role of the EP in the debate of theories of gravity both from a theoretical and experimental point of view.

This review is organized as follows. In Section 2 we discuss the different formulations of the EP. After summarizing the main topics of GR  and Quantum Field Theory (QFT) in curved spacetimes, we  discuss  metric theories of gravity considering possible extensions and modifications of GR. Motivations, both theoretical and experimental, suggesting  generalizations of  GR, are considered. Specifically, these theories have been introduced to account for shortcomings of GR, both at early and late phases of the Universe evolution: Cosmological Inflation, Dark Matter and Dark Energy represent  the main issues of this debate.  From the other side, GR is not a fundamental theory of physics because it should require the inclusion of quantum effects. It is then natural to ask whether the Equivalence Principle still holds in the framework of any modified gravity approach aimed to enclose quantum physics under the standard of gravitational interaction. According to this requirement, we  discuss the possibility to violate the EP by considering  QFT at finite temperature. Besides,  violations of the EP also occur in the framework of the extensions of Standard Model of particles. Section 3 is essentially devoted to experimental tests. We present a wide class of experiments aiming to test the EP, in particular its weak formulation, with a  high accuracy. These include free falling tests, measurements based on  Earth-to-Moon and Earth-to-satellite distances, cold atoms and particles interferometry tests, spin-gravity coupling tests, matter-antimatter tests. Conclusions are drawn in Section 4.

\section{The Foundation of the Equivalence Principle }

The EP is  related to the above considerations and  plays a  relevant role  to discriminate among concurring theories of gravity. In particular,  the role of $g$ and $\Gamma$ are related to the validity of EP.  Specifically, precise measurements of EP could say if $\Gamma$ is only Levi - Civita or if a more general connection, disentangled from $g$, is necessary to describe gravitational dynamics.
Furthermore, possible violation of EP   can put in evidence other dynamical fields like torsion discriminating among the fundamental structure of spacetime that can be Riemannian or not.

Summarizing, the relevance of EP comes from  the following points:

\begin{itemize}
  \item Competing theories of gravity can be discriminated according to the validity of EP;
  \item EP holds at classical level but  it could be violated at  quantum level;
  \item EP allows to investigate  independently  geodesic and causal structure  of spacetime.If it is violated at fundamental level, such structures could be independent.
\end{itemize}

\noindent From a theoretical point of view, EP constitutes   the  foundation of metric theories. The first formulation of EP comes out from the  formulation of gravity  by Galileo and Newton, i.e. the Weak Equivalence Principle (WEP) which states that the inertial mass $m_i$ and the gravitational mass  $m_g$ of  physical objects are equivalent.
The WEP implies that it is impossible to distinguish, locally, between the effects of a gravitational field from those experienced in uniformly accelerated frames using the straightforward  observation of the free fall of physical objects.

The first generalization of WEP states that  Special Relativity is  locally valid. Einstein obtained, in the framework of  Special Relativity, that  mass can be reduced to a manifestation of energy and momentum. As a consequence, it is impossible to distinguish between an uniform acceleration and an external gravitational field, not only for free-falling objects,  but whatever is the experiment. According to this observation,
Einstein EP states:
\begin{itemize}
  \item The WEP is valid.
  \item The outcome of any local non-gravitational test experiment is independent of the velocity of free-falling apparatus.
  \item The outcome of any local non-gravitational test experiment is independent of where and when  it is performed in the universe.
\end{itemize}
One can define a "local non-gravitational experiment" as that  performed in a small-size of a free-falling laboratory.
Immediately, it is possible to realize  that  gravitational interaction depends on the curvature of spacetime. It means that the postulates of  metric gravity theories  have to be satisfied.
Hence  the following statements hold:
\begin{itemize}
  \item Spacetime is endowed with a metric $g_{\mu\nu}$ that constitutes the dynamic variables.
  \item The world lines of test bodies are geodesics of the metric.
  \item In local freely falling frames, i.e. the local Lorentz frames, the non-gravitational laws of physics are those of Special Relativity.
\end{itemize}
One of the predictions of this principle is the gravitational red-shift, experimentally probed  by Pound and Rebka \cite{Will1993}.  It is worth noticing that gravitational interactions are excluded from WEP and Einstein EP.

To classify extended  and alternative theories of gravity, the gravitational WEP and the Strong Equivalence Principle (SEP) is introduced.
The SEP extends the Einstein EP by including all the laws of physics. It  states:

\begin{itemize}
  \item WEP is valid for self-gravitating bodies as well as for test bodies (gravitational WEP).
  \item The outcome of any local test  is independent of the velocity of the free-falling apparatus.
  \item The outcome of any local test  is independent of where and when  it is performed in the universe.
\end{itemize}
 The Einstein EP is recovered from SEP as soon as the gravitational forces are neglected.
Several  authors claim that the only theory coherent with SEP is GR and then WEP has to be deeply investigated.

A very important issue is the consistency of EP with respect to  Quantum Mechanics.
GR is not the only gravity theory  and several alternatives  have been investigated starting from the 60's \cite{Capozziello2011}. Some of them are effective descriptions coming from quantum field theories on curved spacetime. 
Considering the spacetime as  special relativistic at a background level, gravitation can be treated  as a Lorentz-invariant perturbation field on the background.  Assuming the possibility of GR extensions and alternatives, two different classes of experiments can be conceived:

\begin{itemize}
  \item Tests for  the foundations of gravity  according to  the various formulations of  EP.
  \item Tests of  metric theories where spacetime is  endowed with a metric tensor and where the Einstein EP is assumed  valid.
\end{itemize}
The  difference between the two classes of experiments consists in  the fact that EP can be postulated "a priori" or "recovered" from the self-consistency of the theory.
What   is  clear is that, for several fundamental reasons, extra fields are necessary to describe  gravity with respect to  other interactions. Such fields can be   scalar fields or higher-order corrections of curvature and torsion invariants \cite{Capozziello2011}.
According to  these reasons,   two sets of field equations can be considered: The first  couples the gravitational field to  non-gravitational fields, i.e. the matter distribution, the electromagnetic fields, etc. The second set of equations considers  dynamics of non-gravitational fields. In  the framework of metric theories, these laws depend only on the metric and this is a consequence of the Einstein EP.
In the case where gravitational field equations are modified with respect to the Einstein ones,  and matter field are minimally coupled with gravity, we are dealing with the  {\it Jordan frame}. In the case where Einstein field equations are preserved and matter field are non-minimally coupled, we are dealing with the  {\it Einstein frame}. Both frames are conformally related but the very final issue is to  understand if passing from one frame to the other (and vice versa) is physically significant. Clearly, EP plays a fundamental role in this discussion.  In particular, the main question is  if EP is  valid in any case or it  is violated at quantum level.

\subsection{The debate on gravitational theories}


As  discussed  before,  GR provides a
comprehensive description of space, time, gravity,
and matter  under the same standard at macroscopic level.  Einstein formulated it in such a
way that space and time are dynamical and entangled quantities determined by the 
distribution and motion of matter and energy. As a consequence, GR is related to a new conception of the universe which can be considered as a dynamical system where precise  physical measurements are possible.

In this perspective, cosmology is not only a philosophical branch of knowledge  but can be  legitimately incorporated  into 
science. Investigating  scientifically the universe has led, along the last century, 
to the formulation of the Standard Big Bang Model
\cite{Weinberg1972}
which, in principle,   matched most of the available
cosmological observations until more or less twenty years ago.

Deapite these successes, several shortcomings of Einstein's  theory  emerged recently at ultraviolet and infrared scales  and scientists considered the hypothesis  whether GR is the only fundamental theory
 of  gravitational interaction. This new point of view comes from 
cosmology (infrared scales)  and quantum field theory (ultraviolet scales). In  the first case, the
Big Bang singularity, the flatness, horizon, and
monopole  problems \cite{Guth1980} 
led to the conclusion that
a cosmological model based on GR and 
the Standard Model of particles is inadequate to describe
the universe in  extreme energy-curvature regimes. Furthermore, GR  
cannot work as a fundamental theory of gravity
if a quantum description of spacetime is required. The Einstein theory is essentially a classical description. Due to these reasons, and, in particular
due to the lack of a self-consistent
quantum theory of gravity, various alternative and extensions of GR were  proposed. The general approach is   to formulate,  at least,
a  semiclassical effective theory  where GR and its positive results can
be recovered in some limit (e.g. the weak field limit or the Solar System scales). A fruitful approach is 
the
so-called {\it Extended Theories of Gravity} (ETGs)  which have recently  become a paradigm to study the
gravitational  interaction. Essentially they  are based on corrections
and extensions  of  Einstein's GR. The paradigm consistsin adding higher order curvature invariants and
minimally or non-minimally coupled scalar fields into the
dynamics.  In this sense, we can deal with  effective 
gravity actions emerging from quantum field theory \cite{Capozziello2010,Capozziello2011}.

Other reasons to modify GR are related   to the issue of
incorporating  Mach's principle into the theory. GR is  only
partially  Machian  and allows  solutions that are explicitly
anti-Machian, e.g. the G\"{o}del solution \cite{Godel1949} 
or exact $pp$-waves \cite{Ozsvath1962}. 

 Mach's  principle states that  local inertial frames are determined by the
average motions of distant astronomical objects
\cite{Bondi1952}. This  implies that the gravitational
coupling can be
determined by the surrounding matter distribution and, therefore,  becomes a
spacetime function which can assume the form of a scalar  field. As a
consequence,   inertia and 
Equivalence Principle are concepts that have to be revised. Brans-Dicke theory
\cite{Brans1961} is the first alternative to
GR and the first attempt to fully incorporate the Mach principle. It is considered the prototype of alternative
theories of gravity and a straightforward GR extension. The 
gravitational ``constant'' is assumed "variable" and 
corresponds to a scalar field   non-minimally coupled to
 geometry. This approach constitute  a more satisfactory
implementation of
Mach's principle than GR \cite{Brans1961,Capozziello1996,Sciama1953}.

Furthermore, any scheme  unifying  fundamental interactions with gravity, such as
superstrings,  supergravity, or Grand-Unified Theories (GUTs)
produces effective
actions where non-minimal couplings to the  geometry are present. Also
higher order  curvature invariants are
present in general. They emerge  as
loop corrections in  high-curvature
regimes. This scheme has been
adopted in quantizing matter fields
on  curved spacetimes and the result is that  interactions
between quantum fields and  background geometry, or 
gravitational self-interactions  give rise to corrections in the
Hilbert-Einstein Lagrangian \cite{Birrell1982}. Furthermore,
these corrections are unavoidable
in the effective quantum gravity actions \cite{Vilkovisky1992} and then GR extensions are necessary. All
these  models do not constitute a self-consistent  quantum gravity theory, but are useful  working schemes towards it.

To summarize, higher order curvature  invariants
like
$R^{2}$, $R^{\mu\nu} R_{\mu\nu}$,
$R^{\mu\nu\alpha\beta}R_{\mu\nu\alpha\beta}$, $R \,\Box R$, $R
\,\Box^{k}R$, or non-minimally couplings  between matter
fields and geometry such as $\phi^{2}R$ have
to be added to the  gravitational Lagrangian as soon as
quantum corrections are introduced. For example, these terms
occur in the low-energy limit of string   Lagrangian  or in  Kaluza-Klein theories where extra spatial
dimensions are  taken into account  \cite{Gasperini1991}.

Moreover, from a conceptual  viewpoint, there
is no  {\it a priori} reason to restrict the gravitational
Lagrangian
to a  function, linear in  the Ricci scalar $R$, minimally coupled
to  matter \cite{Magnano1987}. This concept is in agreement with the idea that there
are no ``exact'' laws of physics. It this case, 
 the effective Lagrangians of physical interactions
would be given by the average quantities arising from the stochastic behaviour of fields  at a  microscopic level.
This approach  means that the local gauge invariances and
the conservation laws are  approximated and emerge 
only in the low-energy limit. In this perspective,  fundamental constants
of physics can be assumed variable.

Furthermore, besides  fundamental physics motivations,  ETGs are  interesting in cosmology because  they
exhibit  inflationary behaviours able to overcome 
shortcomings of   Standard Big Bang model. The
related inflationary scenarios are  realistic and 
match  current observations coming from 
the cosmic microwave background
(CMB) \cite{Starobinsky1980,Duruisseau1986}.
It can be been shown that, by  conformal
transformations, the higher order and non-minimally coupled terms
 correspond to Einstein gravity plus one or more than one scalar
field(s)  minimally coupled to the curvature
\cite{Teyssandier1983,Maeda1988,Wands1993}. Specifically,after conformal transformations, 
higher order and non-minimally coupled terms   appear as  equivalent
scalar fields in the Einstein field equations. For example, in the Jordan frame,  a  term like
$R^{2}$ in the Lagrangian  gives fourth order field equations, $ R \ \Box R$ gives sixth order
equations \cite{Amendola1993}, $R \,\Box^{2}R$ yields
eighth order equations \cite{BattagliaMayer1993}, and so  on.
After  a conformal transformation,  second order
derivative  terms corresponds to a scalar field:\footnote{Dynamics of any of    these  scalars fields   is determined by a second
order  Klein-Gordon  equation.} specifically, fourth order
gravity is conformally equivalent to Einstein gravity plus a scalar
field; sixth order gravity is conformally equivalent to GR plus two scalar fields;
and so on \cite{Schmidt1990}.

Furthermore, it is also possible to show that $f(R)$ gravity  to the 
Einstein theory minimally coupled to an ideal fluid \cite{Capozziello2006}. This
feature  is useful if multiple
inflationary events are necessary for structure formation. In fact,  an early stage could
produce large-scale structure with very long
wavelengths which after  give rise  to the observed clusters of galaxies.
A later stage could select smaller scales observed as galaxies today \cite{Amendola1993}.
The underlying philosophy is that any inflationary era  is related to the
dynamics of a related scalar field. Finally, these
extended schemes  could solve the graceful
exit problem, avoiding the shortcomings of other  inflationary
models \cite{Amendola1992}.

The  revision of early cosmological scenarios, leading to
inflation, can imply that a new approach is necessary also at
late epochs:   ETGs play a fundamental role also in today observed universe. In fact,  the increasing amount of observational data, accumulated
over the past decades,  has given rise to  a new
cosmological model,  the so called {\it Concordance
Model} or $\Lambda$-Cold Dark Matter ($\Lambda$CDM) model.

The Hubble diagram of type Ia Supernovae (hereafter SNeIa)
was the first  evidence that the universe is
today  undergoing an accelerated expansion phase. Furthermore,
balloon-born experiments
\cite{Stompor2001} determined the location of the first two
Doppler peaks in the spectrum of CMB anisotropies. These features strongly suggest a spatially  flat universe also if some recent data could question this result. If  combined with  constraints
on   matter density parameter $\Omega_M$, these data point out that the universe is
dominated by an un-clustered fluid, with negative pressure,
usually  referred to as {\it dark energy}. Such a  fluid drives
the accelerated expansion. This picture has been 
strengthened by other precise measurements on  CMB
spectrum and by the extension of the SNeIa Hubble diagram to redshifts higher
than one.

A huge amount   of papers appeared following these
observational results. They  present several
models attempting to explain  the cosmic acceleration. The
simplest explanation is  the well-known
cosmological constant $\Lambda$. Although this ingredient
provides the best-fit to most of the available astrophysical data
\cite{Spergel2003}, the $\Lambda$CDM model fails  in
explaining why the  value of $\Lambda$ is so tiny
(120 orders of magnitude lower) if compared with the typical
 vacuum energy density predicted by
particle physics,  and why its present value  is comparable to
the matter density. This constitutes   the so-called {\it coincidence problem}.

A possible  solution is  replacing the
cosmological constant with a scalar field $\varphi$ rolling
slowly down a
flat section of a potential $V(\varphi)$ and giving rise to the
models known  as {\it quintessence} \cite{Padmanabhan2003,Copeland2006}. Also if it is 
successfully   fitting  data with many models, the quintessence
approach to dark energy is still  plagued by the coincidence problem since the
dark energy and dark matter densities evolve in a different way  and reach
comparable values only during a very short time  of the
history of the universe. In particular, they coincide right  at  present era.
In other words, the quintessence 
is tracking matter and evolves in the same way for a long time;
at late times,  it  changes its behaviour
and no longer tracks the dark matter but  dominates dynamics 
as a  cosmological constant. This is, specifically, 
the quintessence coincidence problem.

The origin of this quintessence  scalar field is one of the big  mystery of modern cosmology.  Although several
models have been proposed,  a great deal of uncertainty is related to
the choice of the scalar field potential $V(\varphi)$
necessary to achieve the late-time acceleration of the universe. The 
elusive nature of  dark energy has led many authors to  look for a  different explanation of
the cosmic acceleration  without introducing
exotic components. It is worth stressing that the
present-day cosmic acceleration   requires a negative
pressure  that has  to dominate  dynamics after 
the matter era. However, we do not anything about the  nature and the number of the cosmic
fluids  filling the universe. This consideration
suggests us that  the accelerated
expansion could be explained with  a single cosmic fluid characterized by an
equation of state  acting like dark matter at
high densities, and like  dark energy at low
densities. The relevant  feature of these models, 
referred  as {\it Unified Dark
Energy} (UDE) or {\it Unified Dark Matter} (UDM) models, is that
the coincidence problem is  naturally solved. Examples are
the  Chaplying gas \cite{Kamenshchik2001},  tachyon fields
\cite{Padmanabhan2002}, and condensate cosmology \cite{Bassett2002}.
These models are extremely
interesting  because they can be interpreted both in the framework of
UDE models or as two-fluid models  representing the dark matter and
the  dark energy regime. A main feature of this
approach is  that a generalized equation of state can  
be easily  obtained and the fit of  observational data can be
achieved.

There is another approach  to face the problem of
the cosmic  acceleration. As reported in \cite{Lue2003}, it is
possible that the observed acceleration is not related to
another cosmic ingredient, but rather the 
signal of a breakdown, at  infra-red scales, of the law of
gravitation as a scale invariant interaction. From this view point,  
modifying the Einstein-Friedmann equations,  fitting the astrophysical data with models
containing only standard matter and without exotic fluids is another approach.
Examples in this direction 
are the Cardassian model \cite{Freese2002} and
Dvali-Gabadadze-Porrati (DGP) cosmology \cite{Dvali2000}. In  the
same conceptual framework, it is possible to  find
alternative approaches  where   a quintessential behavior is
obtained by incorporating effective models coming from
fundamental physics and giving rise to extended gravity actions. For example, a cosmological constant can be recovered as a
consequence of  non-vanishing torsion fields. Also in this case, it is possible to build up  
models consistent with  the SNeIa Hubble diagram and
the  Sunyaev\,-\,Zel'dovich effect in galaxy
clusters
\cite{Capozziello2002}. SNeIa data can  also be  fitted
by including   higher-order
curvature invariants.
These  models provide   a cosmological
component with negative pressure which is originated by  the
geometry of the universe. According to this picture, we do not need new particle counterparts to  address the phenomenology.

The amount  of cosmological models which are 
viable  candidates to explain the observed accelerated expansion
is too wide to be reported here.  This
overabundance points out   that only a  few  number
of cosmological tests is available to discriminate
between competing approaches, so it is clear that there is a high degeneracy of models.
It s important to stress  that both  SNeIa Hubble diagram and  angular
size-redshift relation of compact radio sources
are distance-based probes of the cosmological
model and, therefore, systematic errors and biases could be
iterated. According to this consideration, it is interesting to search
for tests based on time-dependent observables.
 For example, we  can take into account the {\it lookback time}
to distant objects. This quantity  discriminates among
different cosmological models. The lookback time is
 estimated as the difference between the
  age of the universe and the age of a given object at
redshift $z$. This estimate becomes  realistic  when the object is a
galaxy observed in more than one photometric band because its color
is determined by the age as a consequence of stellar evolution.
Hence, it is possible to obtain the  galaxy age
by  measuring its magnitude in different bands and then using
stellar evolutionary codes to best reproduce the observed colors.

In general, in the case of weak-field limit, which essentially
coincides with  Solar System scales,  ETGs are expected to
reproduce GR which is precisely  tested  at these scales \cite{Will1993}. Even this limit  is a matter of debate because several  theories  do not
reproduce exactly the Einstein theory in its Newtonian
limit but, in some sense,
generalize it giving  rise to Yukawa-like
corrections to the Newtonian potential which could be 
 physically relevant already at Galactic scales
\cite{Bertolami2007,Capozziello2015,Capozziello2014,Lambiase2016,Blasone2018a,
Buoninfante2018}.

As a general remark,  relativistic gravity theories  give rise to
corrections to the weak-field gravitational potentials which, at the  post-Newtonian level and in the
Parametrized Post-Newtonian (PPN) formalism,
constitute a test bed for these theories \cite{Will1993}.
Furthermore, the  {\it gravitational lensing astronomy}
\cite{Ehlers1992} provide  additional tests  over
small, large, and very large scales which can provide
 measurements on the variation of the Newton constant
\cite{Krauss2003}, the potential of galaxies, clusters of galaxies,
and  other features of gravitating systems.
In principle, such data can be  capable of confirming or ruling
out  any alternative to GR.

In ETGs, the Einstein field equations can be modified in two ways: {\em i)} the geometry part can be
non-minimally coupled to some scalar field, and/or~{\em ii)}
higher than second order derivatives of the metric can appear.
In the former case, we  deal with scalar-tensor
theories of gravity; in the latter, with higher order
theories of gravity. Combinations of non-minimally coupled and
higher order components can also emerge.

From the  mathematical viewpoint, the problem of
 reducing  more general theories to the Einstein theory has
been widely discussed.  Through a Legendre transformation
on the metric, higher order theories with Lagrangians satisfying
some  regularity conditions assume the form
of GR with (possibly multiple)  scalar field(s) as sources
the  gravitational field ({\em e.g.}, 
\cite{Magnano1987,Sokolowski1989,Ferraris1988}). The formal equivalence between
models with variable gravitational coupling
and Einstein gravity via conformal
transformations is also well known 
\cite{Dicke1961}. This  gave rise to the debate on whether
the
mathematical equivalence between different conformal
representations is also a physical equivalence \cite{Faraoni2006a,Faraoni2006b}. 

Another important issue is the Palatini approach: this problem was
first proposed by Einstein himself, but it was
called the Palatini approach because the Italian mathematician Attilio Palatini formalized it \cite{Ferraris1982}. The main  idea  of
this  formalism is considering the
connection $\Gamma^{\mu}_{\alpha\beta}$   as
independent of the  metric $g_{\mu\nu}$. It is well
known that the Palatini  formulation of GR  is  equivalent
to the  metric formulation
\cite{Weinberg1972}. This result follows from the fact that the
field equations for  connection  $\Gamma^{\mu}_{\alpha\beta}$, also if 
assumed   independent of the metric, yield the
Levi-Civita  connection of  $g_{\mu\nu}$ in GR. Therefore,  the Palatini  variational principle in the Einstein theory gives exactly the same  field equations of  the metric variational principle. However, the situation  changes  if we consider  ETGs formulated as  functions of
curvature invariants, such as $f(R)$, or as  scalar-tensor theories.
There, the  Palatini and the metric variational principles   give rise to different field equations that could describe different physics. The relevance of the Palatini formulation 
has been recently  highlighted  according to cosmological applications \cite{Vollick2003}.

Another  crucial problem is related to the Newtonian potential in alternative
gravity and its relations with the conformal factor
\cite{Meng2003}. From a physical point of view, considering
the metric and the connection 
as independent fields corresponds  to decoupling the metric structure
of spacetime from the geodesic structure (with the
connection  being, in general,
different from the
Levi-Civita connection   of the metric. The causal  structure of
spacetime is governed by $g_{\mu\nu}$, while the
spacetime trajectories of  particles are governed
by $\Gamma^{\mu}_{\alpha\beta}$.

The decoupling of causal and geodesic structures enlarges the
spacetime geometry  and generalizes the 
metric formalism. This metric-affine structure  can be immediately translated, by means of the
Palatini field equations, into a bi-metric structure.
In addition to  the physical metric
$g_{\mu\nu}$,  a second metric $h_{\mu\nu}$ is present which  is
related, in  the case of $f(R)$ gravity, to the connection. As a
matter of  fact, the connection $\Gamma^{\mu}_{\alpha\beta}$
turns out  to be the Levi-Civita connection  of this second metric $h_{\mu\nu}$ and provides the geodesic structure of spacetime.

If we consider non-minimal couplings  in  gravitational Lagrangian, the further metric $h_{\mu\nu}$ is  related to the coupling.   According to the
Palatini formalism, non-minimal couplings and  scalar
fields entering the evolution of the gravitational field are
related by the metric structure of spacetime\footnote{Due to these features, the Palatini approach could play a main role in clarifying the physical aspects  of
conformal transformations \cite{Allemandi2004}.}

\subsection{Einstein's General Relativity}

The  Newton theory of gravity was the issue 
that Einstein needed to recover in the  weak field limit 
 and slow motion. In Newton formulation,   space and time are
absolute entities and require  particles to move, in a
preferred inertial frame, along  curved trajectories, the
curvature of which  ({\em i.e.}, the acceleration) is a function
of the intensity  of the sources through
the ``forces''. According to this requirements, Einstein 
postulated that the gravitational
forces have to  be described by the  curvature of the
metric tensor   $g_{\mu\nu}$ related to the line element
$ ds^2 = g_{\mu\nu} dx^{\mu} dx^{\nu} $ of a four-dimensional
spacetime manifold. This metric has the same signature of
the Minkowski metric (the {\em Lorentzian
signature} here assumed to be $\left( -,+,+,+ \right)$). Einstein
postulated also that spacetime curves onto itself and that
 curvature is locally determined by the distribution of the
sources, that is  by the
four-dimensional  generalization of the  matter stress-energy
tensor (another  rank-two symmetric tensor)  $T^{(m)}_{\mu\nu}$ of
continuum mechanics.

Once a metric $g_{\mu\nu}$ is assigned,  curvature is given by
the Riemann (or curvature) tensor
\begin{equation}
R_{\alpha \beta\mu}\hspace{0.01pt}^{\nu}=
\Gamma_{\alpha\beta,\mu}^{\nu}-
\Gamma_{\beta\mu,\alpha}^{\nu}
+\Gamma_{\alpha\mu}^{\sigma}
\Gamma_{\sigma\beta}^{\nu}
-\Gamma_{\beta\mu}^{\sigma}\Gamma_{\sigma\alpha}^{\nu}\,
\end{equation}
where the commas denote partial derivatives. Its contraction
\begin{equation}
R_{\alpha\mu}\equiv  {R_{\alpha\beta\mu}}^{\beta} \,,
\end{equation}
is the {\em Ricci tensor}, while the contraction
\begin{equation}
R \equiv {R^{\mu}}_{\mu}=g^{\mu\nu}R_{\mu\nu}
\end{equation}
is the  {\em Ricci curvature scalar} of
$g_{\mu\nu}$.
Einstein initially derived the
field equations $R_{\mu\nu} = \frac{\kappa}{2}
T^{(m)}_{\mu\nu}\label{Introwrong}$,
where $\kappa=8\pi G$ (in units in which $c=1$)
is the gravitational  coupling constant. These equations turned
out to be  inconsistent as pointed out by Levi-Civita. Furthermore
  Hilbert stressed that they do not derive
from an action principle \cite{Schrodinger2011}. In fact,  there is no action
reproducing them exactly through a variation.
Einstein's answer was  that he realized
that the equations were physically inconsistent, since they
were incompatible with the continuity equation deemed to
be satisfied by  reasonable forms of matter.

Assuming  matter consisting of  perfect fluids with stress-energy tensor
\begin{equation}
T^{(m)}_{\mu\nu} = \left( P + \rho \right) u_{\mu}u_{\nu} +  P\,
g_{\mu\nu} \;,
\end{equation}
where $u^{\mu} $  is the four-velocity of the particles,
$P$ and $\rho$  the pressure and  energy density of the
fluid, respectively,  the continuity equation requires
$T^{(m)}_{\mu\nu}$ to be
covariantly constant, {\em i.e.}, to
satisfy the conservation law
\begin{equation}\label{Introconservation}
\nabla^{\mu} T^{(m)}_{\mu\nu} = 0\,,
\end{equation}
where  $\nabla_{\alpha}$ denotes the covariant derivative
operator
of the metric $g_{\mu\nu}$.
 In fact,  $\nabla^{\mu} R_{\mu\nu}$
does not vanish, except in the trivial case  $R \equiv
0$. Einstein  concluded
that the field equations are
\begin{equation}\label{Introfield}
G_{\mu\nu} = \kappa \, T^{(m)}_{\mu\nu}
\end{equation}
where
\begin{equation}
G_{\mu\nu} \equiv R_{\mu\nu} - \frac{1}{2} \,  g_{\mu\nu}R
\end{equation}
is  the {\em Einstein tensor} of $g_{\mu\nu}$.
These equations can be derived also by minimizing an action containing $R$ and
satisfy the conservation law (\ref{Introconservation}) since 
\begin{equation}
\nabla^{\mu} G_{\mu\nu} = 0\,,
\end{equation}
holds as a contraction  of the  Bianchi identities that the
curvature tensor of $g_{\mu\nu}$ satisfies \cite{Weinberg1972}.

Specifically, the  Lagrangian that, if varied, produces the field equations
(\ref{Introfield}) is the sum of a ``matter'' Lagrangian
density ${\cal L}^{(m)}$,  whose  variational derivative  is
\begin{equation}
T^{(m)}_{\mu\nu} = \frac{\delta {\cal L}^{(m)} }{\delta
g^{\mu\nu}} ,
\end{equation}
and of the gravitational  ({\em Hilbert-Einstein}) {\em
Lagrangian} density
\begin{equation} \label{IntroHE}
{\cal L}_{HE} = \sqrt{-g} \,
g^{\mu\nu} R_{\mu\nu} =\sqrt{-g} \, R \,,
\end{equation}
where $g$ is the determinant of the metric $g_{\mu\nu}$.

 Einstein's choice  was  arbitrary but it was certainly the
simplest.  As clarified by Levi-Civita in 1919,
curvature is  not a purely metric notion but it is also
related to the linear connection  of  parallel
transport and covariant derivative.
In some sense, this idea is the precursor   of  ``gauge-theoretical framework'' \cite{Klein1938},
following the pioneering work by Cartan of 1925.

After, it was  clarified  that the  principles of relativity,
equivalence, and covariance, together with causality,  require
only that the spacetime structure can be determined by a Lorentzian metric  $g_{\mu\nu}$  and
a linear connection  $\Gamma^{\alpha}_{\mu\nu}$, assumed to be
torsionless for the sake of simplicity. The metric    fixes the
causal structure of spacetime (the light cones) as well as its metric
relations measured by clocks and rods and the lenghts of
four-vectors.  The connection  
determines the laws of   free fall, that is the
four-dimensional spacetime trajectories followed by locally
inertial observers. These  observers must satisfy some
compatibility relations including  the requirement that photons
follow null geodesics,  so that
$\Gamma^{\alpha}_{\mu\nu}$ and $g_{\mu\nu}$ can
{\it a priori} be independent, but constrained {\it a posteriori}
by the physics. These physical constraints, however, do not
necessarily impose that $\Gamma^{\alpha}_{\mu\nu}$ is
the Levi-Civita connection of  $g_{\mu\nu}$
\cite{Capozziello2011}.

\subsection{Quantum Gravity motivations}

A challenge of modern  physics is 
constructing a theory capable of describing the fundamental interactions
of nature under the same standard.
This goal has led to  formulate several 
unification schemes which  attempt  to describe gravity
together with  the other interactions. All
these schemes  describe the  fields under the conceptual apparatus of Quantum Mechanics. This is based
on the assumption  that the states of  physical systems are described by
vectors in a Hilbert space ${\cal H}$ and the physical fields are
linear operators defined on domains of ${\cal H}$.
Till now, any attempt to incorporate gravity into this scheme is
 failed or revealed unsatisfactory.  The main
  problem is that  gravitational field describes,
at the same time, the gravitational degrees of freedom
and the spacetime background where these degrees of freedom are defined.

Owing to the difficulties of building up a self-consistent theory
unifying interactions and particles,  the two fundamental theories of modern physics, GR and
Quantum Mechanics, have been critically
re-analyzed. On the one hand, we assume that  matter fields
(bosons and fermions) come out from superstructures
({\em e.g.}, Higgs bosons or superstrings) that, undergoing
certain phase transitions, generate the known particles. On
the other hand, it is assumed that the geometry
 interacts
directly with quantum matter  fields which  back-react on it. This
interaction necessarily  modifies the standard
gravitational theory, that is the Hilbert-Einstein Lagrangian. 
This fact leads to the ETGs.

From the  point of view of cosmology, the modifications of
GR provide inflationary scenarios of remarkable
interest.  In any case, a condition that
 such theories have to respect  in order to be physically acceptable is that
GR is recovered in the low-energy limit.

Although  conceptual progresses have been made assuming generalized gravitational theories, at
the same time   mathematical difficulties have increased.
The corrections   into the Lagrangian enlarge the
(intrinsic) non-linearity of the Einstein equations, making them
more difficult to study because  differential
equations of higher order than second are often obtained and
because it is extremely difficult to separate  geometry from  
matter degrees of freedom. To overcome these
difficulties and try to simplify the field equations, one often looks for symmetries of  dynamics  and identifies conserved quantities which, often,  allow to find out 
exact solutions. 

The necessity of  quantum gravity was recognized at
the end of  1950s, when physicists tried 
to deal with all interactions at a fundamental level and describe
them under the standard of quantum field theory.  The first
attempts to  quantize gravity adopted the canonical
approach and the covariant approache, which had been already applied
with success to electromagnetism. In the first
approach applied to electromagnetism,  one takes into account  electric
and magnetic fields
satisfying the Heisenberg  uncertainty principle and the quantum
states are gauge-invariant functionals, generated by the
vector potential, defined on 3-surfaces labeled with  constant time. In the
second approach, one  quantizes the
two degrees
of freedom of the Maxwell field without  (3+1) decomposition
of the metric, while the quantum states are elements of a Fock
space \cite{Itzykson1980}. These procedures are fully equivalent. The former
allows  a well-defined measure, whereas  the latter is more
convenient for perturbative calculations such as
the computation of  the  $S$-matrix in Quanrum Electrodynamics (QED).

These methods have been adopted also in GR, but
several difficulties arise in this case due to the fact that GR 
cannot be formulated as a quantum field
theory on a fixed Minkowski  background. To be 
specific, in GR the geometry of    background spacetime
cannot be given a priori: spacetime is itself the dynamical variable. 
To introduce the notions of causality, time, and
evolution, one has to  solve  equations of
motion and  build up the related spacetime. 
For example, to know 
if a particular initial condition will give
rise to a black hole, it is necessary to  evolve it by
solving the Einstein equations. Then, taking into account the
causal structure of the solution, one has to study the
asymptotic metric at future null infinity in order to assess
whether it is related, in the causal past, with that initial
condition.
This problem become intractable at   quantum level. Due to
the uncertainty principle, in non-relativistic quantum mechanics
particles do not move along well-defined trajectories and one
can only calculate the probability amplitude $\psi (t, x)$ that
a measurement at  a given time $t$ detects a particle at the
spatial point $x$. In the same way, in quantum gravity, the evolution
of an initial state does not provide a
given spacetime (that is a metric). In  absence of a spacetime, how is it
possible to introduce basic concepts  as causality, time,
scattering matrix, or black holes?

Canonical and covariant approaches provide different answers
to these issues. The first is based on the
Hamiltonian formulation of GR and is adopting a
canonical quantization procedure. The canonical commutation
relations are those that lead to the Heisenberg uncertainty principle; in
fact, the commutation of  operators on a
spatial 3-manifold at constant time is assumed, and this
3-manifold is fixed in order to preserve the notion of
causality. In the limit of asymptotically flat spacetime,  motions related to
the Hamiltonian have to be interpreted as time evolution (in
other words, as soon as  the background becomes the Minkowski spacetime,
the Hamiltonian operator assumes again its role as the generator
of  time translations). The canonical approach preserves the
geometric structure  of GR without  introducing
perturbative methods.

On the other hand, the covariant approach adopts  quantum
field theory concepts. The basic idea is that the
shortcomings mentioned above can be   circumvented by splitting
the metric $g_{\mu\nu}$ into a kinematic  part
$\eta_{\mu\nu}$ (usually flat) and  a dynamical part
$h_{\mu\nu}$. That is
\begin{equation}\label{gweak}
 g_{\mu\nu}=\eta_{\mu\nu}+h_{\mu\nu}\; .
\end{equation}
The  background geometry is given by the flat metric
tensor and it is the same of Special Relativity and standard
quantum field theory. It  allows to define  concepts of
causality, time, and  scattering. The procedure of quantization  is
  applied to the dynamical field, considered as a
little  deviation of the metric from the flat background.
Quanta are  particles with spin~two,
i.e.  {\it gravitons}, which propagate in MInkowski spacetime and are
defined by $h_{\mu\nu}$. Substituting  $g_{\mu\nu}$ into the
GR action, it follows that the gravitational Lagrangian
contains a sum whose terms contains
a different orders of approximation, the interaction of gravitons and,
eventually, terms describing   matter-graviton interaction  (if
matter is present). These terms are analyzed by  the standard 
perturbation approach of  quantum field theory.

These quantization approaches  were both developed during the 1960s
and 1970s.  In the canonical approach, Arnowitt, Deser, and
Misner developed the Hamiltonian formulation of
GR using methods proposed  by Dirac and Bergmann. In this scheme, 
the canonical variables are the
3-metric on the spatial 3-manifolds obtained by foliating
the 4-dimensional manifold. It is worth noticing that  this foliation is
arbitrary. Einstein's field  equations
give constraints between the 3-metrics and their conjugate
momenta and the evolution equation for these fields is the so-called
{\em Wheeler-DeWitt} (WDW) {\em equation}. In this picture,
GR is  the dynamical theory of
the 3-geometries, that is the {\em geometrodynamics}. The
main problems arising from this formulation are that the
quantum equations involve products of operators defined at the
same spacetime point  and, furthermore, they give rise to distributions whose physical meaning is unclear.
In any case, the main question is the absence of the Hilbert space
of states and, as consequence, the probabilistic interpretation of
the quantities is not exactly defined.

The covariant quantization  is much similar to the 
physics of particles and fields, because,  in some sense,  it has been
possible to extend QED perturbation methods  to
gravitation. This allowed the  analysis of  mutual
interaction between gravitons and of the matter-graviton
interactions. The Feynman rules for gravitons and
the demonstration that the theory might be, in principle, unitary at any order of  expansion was achieved by DeWitt.

Further progress was reached by the  Yang-Mills theories,
describing the strong, weak, and electromagnetic interactions
of particles by means of symmetries. These theories
are renormalizable because it is possible to give the particle masses through the mechanism of  spontaneous symmetry
breaking. According to this principle, it is natural to try  to
consider gravitation as a Yang-Mills theory in the covariant
perturbation approach and search for its  
renormalization. However, gravity does not fit into this scheme;
it turns out to be non-renormalizable if we consider the
graviton-graviton interactions  (two-loops diagrams) and
graviton-matter interactions (one-loop
diagrams).\footnote{Higher order
terms in the perturbative series imply an infinite number of
free parameters. At the one-loop level, it is sufficient to
renormalize only the effective constants $G_{eff}$ and
$\Lambda_{eff}$ which, at low energy, reduce to the  Newton
constant $G_N$ and the cosmological constant $\Lambda$.} In any case, the
covariant method allows to construct a  gravity theory which is
renormalizable at one-loop level in the perturbation  series
\cite{Birrell1982}.

Due to the non-renormalizability of gravity at higher orders,
the validity of the approach is restricted  to the low-energy domain,
that is, to infrared 
scales, while it fails at ultraviolet
scales. This implies that the  theory of gravity is unknown
near or at Planck scales.
On the other hand, sufficiently far from
the Planck epoch, GR and its first loop
approximation describe quite well gravitation. In this context, it makes sense to add higher order and non-minimally coupled
terms to the Hilbert-Einstein action. Furthermore, if the free
parameters  are chosen appropriately, the theory has a better
ultraviolet behaviour and  it is asymptotically free. Nevertheless,
the Hamiltonian of these theories is not bounded  from below and
they are unstable. Specifically, unitarity is violated
and probability is not conserved.

Another approach to the search for  quantum
gravity  comes from the study of the electroweak interaction. Here, gravity is treated  neglecting the other
fundamental interactions. The unification of 
electromagnetism  and   weak interaction suggests that it
could be  possible to obtain a consistent theory if
gravity is coupled  to some kind of matter. This is
the basic idea of Supergravity. In this kind of theories,
divergences due to the bosons (in this case the 2-spin gravitons) are
cancelled exactly by those due to the fermions. In this picture, it is possible  to achieve a
renormalized theory of gravity. Unfortunately, the approach works
only up to  two-loop level and  for matter-gravity couplings.
The corresponding Hamiltonian is positive-definite and the  theory is unitary. However, including higher order loops, the
infinities appear and  renormalizabilty is lost.

Perturbation methods are also adopted in string theories. In
this case, the approach is different from
the previous one since  particles are replaced by
 extended objects which are the fundamental strings. The 
physical particles, including the spin two gravitons, correspond
to excitations of the strings. The  only free parameter of the theory is the string tension  and the  interaction couplings are uniquely 
determined. As a consequence, string theory contains
all fundamental physics and it is considered a possible  {\it Theory of Everything}. String theory is  unitary and the perturbation series converges implying finite terms. This feature
follows from the characteristic  that strings are intrinsically
extended objects, so that ultraviolet divergencies, appearing at  
small scales or at large transferred impulses, are naturally
cured. This means that the natural cutoff is given by the string
length, which is of Planck size $l_P$. At  larger scales than
$l_P$, the effective string action can be written as
non-massive vibrational modes, that is, in terms of scalar
and tensor fields. This constitutes the {\it tree-level effective action}. This approach
 leads to an effective theory of gravity
non-minimally coupled with  scalar fields, which are the  so-called {\it
dilaton fields}.

In conclusion, we can summarize the previous considerations: 1)  a  unitary and renormalizable
theory of gravity  does not yet exists\footnote{It is worth to mention that recently it has been shown that an infinite derivative theory of covariant gravity, which is motivated from string theory, see \cite{Tseytlin:1995uq,Siegel:2003vt}, can be made ghost free and also singularity free \cite{Biswas:2011ar,Biswas:2005qr} (see Refs. \cite{Edholm:2016hbt,Buoninfante:2018xiw,Buoninfante:2018rlq,Buoninfante:2018mre,Buoninfante:2018xif} for some applications). }. 2) In the quantization program of  gravity, two
approaches are used: the {\em covariant approach} and the
{\em perturbation approach}. They do not lead to a
self-consistent  quantum gravity. 3) In the low-energy regime, with respect to the Planck
energy,   GR can be improved by
introducing, into the Hilbert-Einstein action,  higher
order terms of curvature invariants and non-minimal couplings
between matter and gravity. The approach leads, at least at  one-loop level, to
a consistent and renormalizable theory.

\subsection{Emergent gravity and thermodynamics of spacetime}

Recently,  several   theoretical
approaches  towards the so-called {\em emergent gravity} theories have been proposed. The
main idea is that, given the  lack of experimental
data for quantum gravity at high-energies, it is worth  approaching gravity
from  low-energies considering  some effective
theories. Gravity emerges from
fundamental constituents, as a  sort of ``atoms of spacetime'', with
 metric and  affine-connection being  collective variables
similar to hydrodynamics, where a fluid
description emerges from an aggregate of microscopic particles.
Emergent gravity  attempts the reconstruction of 
microscopic system underlying classical gravity. It is
possible to constrain the microscopic features of the
fundamental constituents by requiring that the emergent 
gravity is similar  to GR in weak field limit. This picture, however,   questions the principles constituting the   foundations of gravitational theories.

A related research line  is that of {\em analogue models}: if
gravity emerges as a collective system made of microscopic quantum 
constitutents, it could be possible to model it
with the help of physical systems where an effective metric
and a connection govern  the dynamics.  For example one can study the Hawking radiation coming from black holes adopting acoustic
analogues (the so called ``dumb holes'') \cite{Unruh1980, Visser1997,
Novello2002}, or Bose-Einstein
condensates (see \cite{Barcelo2005} for  a review of analogue
models). If  an effective metric is generated, it is 
a kinematic, in the sense that field equations are not
generated by it. However, some results are able to generate a theory of scalar gravity
\cite{Girelli2008} and progresses are possible  in this direction.
A standard feature for  emergent spacetimes is that they
exhibit Lorentz invariance at  low-energies. The Lorentz
symmetry is    broken in the ultraviolet limit where the fundamental quantum constituents of gravity cannot be avoided.

We mention these  approaches here because they question the
foundations of gravitational theory and do not state that GR
is the only theory to be reproduced at  large scale in coarse-graining: the message is that 
different theories with similar features are possible as well.

Another approach   is based on
the idea that gravity could be reproduced through
a sort of {\em spacetime thermodynamics}. This means that the Einstein field equations
should  be derived through local 
thermodynamics at equilibrium. Using thermodynamics  on
the Rindler horizons associated to the worldlines of
physical observers and assuming the  relation $S=A/4$ between entropy and horizon
area (which should be   more fundamental than
the Einstein field equations) Jacobson was able to derive the Einstein
equations  as an equation of state derived
for an ideal gas. The implication
of this result is that it does not make sense to
quantize  the field equations  to learn about
 quantum gravity. The philosophy is that
by quantizing the equation of state of an atomic hydrogen gas, we do not learn anything about the hydrogen atom and its energy levels. From this perspective,  if a
similar thermodynamics of spacetime approach  is
applied to $f(R)$ gravity, it is then necessary to  consider
near-equilibrium thermodynamics in  order to derive the 
field equations. This  demonstrates that
GR is just a  state of gravity corresponding to a given
thermodynamic equilibrium and,  when this equilibrium is
perturbed,  near-equilibrium configurations
correspond to alternative theories of gravity. According to this approach, this  justify the study of ETGs.

A result with a conceptual similar meaning is found in scalar-tensor cosmologies:  they should relax to GR during the evolution of the
universe at recent epochs.  This is another hint that
GR could  be only a particular state of equilibrium, while an entire spectrum
of theories should be considered at   higher energy excitations.

These results are very speculative and require further studies; however, they stress the  necessity to
think about gravity outside of the strict GR scheme and hint to the fact
that much more work needs to be done before claiming for a self-consistent theory of  gravity also  at lower energies.


\subsection{Kaluza-Klein theories}

The attempt to construct a unified theory of GR and  electromagnetism was first proposed by Kaluza \cite{Kaluza:1984ws} (for a review, see \cite{Duff:1994tn,Overduin:1998pn,Salam:1981xd,SatheeshKumar:2006ac}). He showed that the electromagnetism and the gravitation interactions can be described by making use of a single {\it metric} tensor if an additional spatial dimension is introduced. In a Universe with 5-dimensions, the element line reads $ds^2= G_{AB}(x, y) dx^A dx^B$, where $x^A=(x^\mu, y)$, being $y$ the additional dimension (here $A,B = 0, 1, 2, 3, 4$, and $x^\mu$, with $\mu=0, 1, 2, 3$, the usual four dimensional coordinates). In matrix form, the 5-dimensional metric tensor assumes the form
$G_{AB}=\left( \begin{array}{cc} g_{\mu\nu} & g_{\mu 4} \\ g_{4 \nu} & g_{44} \end{array} \right)$. From the metric tensor, one construct all geometric quantities such as the Riemann tensor, the Ricci tensor and scalar curvature and then the field equations. The components of the metric tensor are typically written in the following form: $G_{44}=\phi$, $G_{4\mu}=\kappa \phi A_\mu$, $G_{\mu\nu}=g_{\mu\nu}+\kappa^2 \phi A_\mu A_\nu$. The fields $g_{\mu\nu}(x, y)$, $A_\mu (x, y)$, and $\phi (x, y)$ transform as a tensor, a vector, and a scalar under diffeomorphisms (four-dimensional general coordinate transformations), respectively. The field $\phi$ is the dilaton field. Is it then natural to write down the The Einstein-Hilbert action in Kaluza-Klein five-dimensional gravity $S_{HE}=\frac{1}{2^2} \int d^5 X R_5$, where $\kappa_5$ represents the five-dimensional coupling constant while $R_5$ the five-dimensional scalar curvature. The field equations of gravity and electromagnetism can be derived from the usual variational principles.

The extra dimension $y$ is imposed to be become compact \cite{Klein:1926fj}. Hence $y$ must satisfy the boundary condition $y=y+2\pi R$. This implies that the fields ${\cal F}_A (x, y)= \{g_{\mu\nu}(x, y), A_\mu (x, y), \phi (x, y)\}$ are periodic in $y$ and may be expanded in a Fourier series as follows
 \[
  {\cal F}_A = \sum_{n=-\infty}^{+\infty} {\cal F}_{A n}\, e^{i ny/R}\,,
 \]
where $R$ is the radius of the compactified dimension. The equations of motion are  \[
\Box_5 {\cal F}_A=\left(\Box_4 +\frac{n^2}{R^2}\right){\cal F}_A=0\,, 
 \]
where $\Box_5=\Box_4-\partial^y\partial_y$ and $\Box_4=\partial^\mu\partial_\mu$ is the usual 4-dimensional D'Alembert operator. Comparing with the Klein-Gordon equation, one infers that only the massless zero modes $n = 0$ is observable at our present energy, while all the excited states (Kaluza-Klein states) have a mass and charge given by $m\sim |n|/R$ and $q\sim \kappa n/R$ [17], with $n$ the mode of excitation. In 4-dimensions, all these excited states would appear with mass or momentum $\sim {\cal O}(n/R)$. The natural radius of compactification is the Planck length $R=l_{Pl}= 1/M_{Pl}$.

Concerning the number of degree of freedom present in the Kaluza-Klein theory, owing to the fact that the metric is a $5\times 5$ symmetric tensor,
there are 15 independent components \cite{SatheeshKumar:2006ac}. The gauge fixings reduce the number of independent degrees of freedom to 5 (in 4-dimensions there are only 2 degrees of freedom for a massless graviton). Therefore the theory does contain particles other than just ordinary four dimensional graviton.
The zero-mode of five-dimensional graviton contains a four-dimensional massless graviton with 2 physical degrees of freedom, a four-dimensional massless gauge boson with 2 physical degrees of freedom, and a real scalar with 1 physical degree of freedom. The non-zero mode of five-dimensional graviton is massive and has 5 physical degrees of freedom.

Kaluza-Klein theory, although flawed and is in contradiction with experimental data, has represented, nonetheless, an important model for building up the unification the forces of nature. Many modified version of the Kaluxa-Klein theory, in fact, have been proposed in which higher and extremely small extra dimensions have been taken into account. The higher dimensional unification approaches mainly studied in literature are \cite{SatheeshKumar:2006ac} : 1) The Compactified Approach; 2) The Projective Approach. 3) The Noncompactified Approach.

The violation of the equivalence principle in Kaluza-Klein theories has been discussed in \cite{PoncedeLeon:2007bm} (see also \cite{Mukherjee:2017fqz,Chakraborty:2017qve}).


\subsection{Quantum field theory in curved spacetime}

In this Section we point out that any attempt to formulate quantum field theory
on curved spacetime necessarily leads to modifying the Hilbert-Einstein action. This means adding terms containing non-linear invariants of the curvature tensor or non-minimal
couplings between matter and the curvature originating
in the perturbative expansion. 

At  high energies, a desription of matter as a hydrodynamical  perfect fluid  is inadequate: an
accurate description asks for  quantum field theory formulated on
a  curved spacetime. Since, at  scales
comparable to  the Compton  wavelength of 
particles, matter has to be quantized, one
can adopt a semiclassical description of gravitaty where
the Einstein field equations assume  the form
 \begin{equation}
G_{\mu\nu}  \equiv R_{\mu\nu} -\frac{1}{2} \, g_{\mu\nu} R =<T_{\mu\nu}> \,,
 \label{Intro1.2.1}
\end{equation}
where the  Einstein tensor $G_{\mu\nu}$ is on the
left hand side while  the right hand  contains the
expectation value of quantum stress-energy tensor which is the source of the gravitational field. Specifically, if $|\psi>$ is a
quantum state, then $<T_{\mu\nu}> \equiv  <\psi|\hat{T}_{\mu\nu}|\psi>$,
where $\hat{T}_{\mu\nu}$ is the quantum operator associated with the classical energy-momentum tensor of the
matter field with a  regularized expectation value. 

If the background is curved, ,  quantum fluctuations of matter fields give, even in absence of classical matter and radiation, non-vanishing contributions to $ <T_{\mu\nu}> $  like it happens  in QED
\cite{Birrell1982}.  If matter fields are free, massless and conformally invariant, these corrections are 
\begin{equation}
<T_{\mu\nu}> = k_{1}\,  ^{(1)}H_{\mu\nu} + k_{3}\,^{(3)}H_{\mu\nu} \,.
 \label{Intro1.2.5}
\end{equation}
Here $k_{1, 3}$ are numerical coefficients and   
\begin{eqnarray}
^{(1)}H_{\mu\nu} & = &   2R_{;\mu\nu} -2 g_{\mu\nu} \Box R +2 R^{\sigma\tau}R_{\sigma\tau} g_{\mu\nu}
-\frac{1}{2} \,  g_{\mu\nu} R^{2}\,, \label{Intro1.2.6} \\
&&\nonumber\\
 ^{(3)}H_{\mu\nu} & = &  {R^{\sigma}}_\mu R_{\nu\sigma}
-\frac{2}{3} \, R R_{\mu\nu} -\frac{1}{2} \, g_{\mu\nu}
 R^{\sigma\tau}R_{\sigma\tau} +\frac{1}{4} \, g_{\mu\nu} R^{2}\,.
 \label{Intro1.2.7}
\end{eqnarray}
$^{(1)}H_{\mu\nu}$ is a tensor derived  by varying  the local action,
\begin{equation}
^{(1)}H_{\mu\nu} =\frac{2}{\sqrt{-g}} \frac{\delta}{\delta g^{\mu\nu}}
\left( \sqrt{-g}\, R^{2} \right) \,.    \label{Intro1.2.9}
\end{equation}
It is divergence free, that is   $^{(1)}H_{\mu;\nu}^{\nu}=0$.

Infinities coming from  $ <T_{\mu\nu}>$ are  removed by introducing an infinite number of  counterterms in the Lagrangian density of gravitation. The procedure yields a renormalizable theory. For example, one of these terms is  $C R^{2}\sqrt{-g}$, where with $C$  indicates a parameter that diverges logarithmically.  Eq. (\ref{Intro1.2.1}) cannot be generated by a finite action because the  gravitational field would be completely renormalizable,
that is, it would  eliminate a finite number of divergences to make gravitation similar to QED. On the contrary, one can only  construct a truncated quantum theory of gravity. 
The parameter used for the expansion in loop is the Planck constant $\hbar$.  It follows that the truncated theory at the one-loop level contains all terms of order $\hbar$, that is  the first quantum correction. Some points have to be stressed now: 1)  Matter fields are {\it free} and, if the  Equivalence Principle is valid at quantum level, all forms of matter couple in the same way to gravity. 2) The {\it intrinsic} non-linearity of gravity naturally arises, and then a number of loops are nrcessary
to take into account self-interactions  interactions between matter and gravitation.
In view ofremoving  divergences at one-loop order, one has to renormalize the gravitational coupling $G_{eff}$ and the cosmological constant $\Lambda_{eff}$.
One-loop contributions of $<T_{\mu\nu}>$ are the quantities introduced above, that is  $^{(1)}H_{\mu\nu}$ and $^{(3)}H_{\mu\nu}$. Furthermore, one has to consider
\begin{equation}
 ^{(2)}H_{\mu\nu} = 2{R^{\sigma}}_{\mu ; \nu \sigma}-\Box
R_{\mu\nu} - \frac{1}{2} \, g_{\mu\nu}\Box R +{R^{\sigma}}_{\mu} R_{\sigma
\nu}-\frac{1}{2} \, R^{\sigma\tau} R_{\sigma\tau}g_{\mu\nu}\,.
\end{equation}
In a conformally flat spacetime,  one has $^{(2)}H_{\mu\nu} = \frac{1}{3}^{(1)}H_{\mu\nu}$ \cite{Birrell1982}, so that only the first and third  terms of $H_{\mu\nu}$ are present in  (\ref{Intro1.2.5}).
%
%
The tensor $^{(3)}H_{\mu\nu} $ is conserved only in conformally flat spacetimes  and it cannot be obtained by varying a local action. 
%
%
The trace of the energy-momentum tensor is null  for conformally invariant classical fields while, one finds that the expectation value of the  tensor (\ref{Intro1.2.5}) has non-vanishing trace.
This result gives rise to  the so-called {\it trace anomaly} \cite{Birrell1982}.

By summing all the geometric terms in $ <T_{\rho}^{\rho}>_{ren}$, 
 deduced by
the Riemann tensor and of the same order, one derives the right hand side of
(\ref{Intro1.2.5}). In the case in which the background metric is conformally flat,
it can be expressed in terms of Eqs.  (\ref{Intro1.2.6}) and
(\ref{Intro1.2.7}). We  conclude that the trace anomaly, related to the geometric terms emerges  because the one-loop approach formulates quantum field theories on curved
spacetime.\footnote{Eqs. (\ref{Intro1.2.6}) and
(\ref{Intro1.2.7}) can include  terms containing derivatives of the metric of order
higher than fourth (fourth order being the $R^{2}$ term) if all
possible Feynman diagrams are included. For example,
corrections such as $R\Box R$ or $R^{2}\Box
R$ can be present in $^{(3)}H_{\mu\nu}$ implying equations of motion
that contain  sixth order derivatives of the metric.
Also these terms can be treated by making use of conformal
transformations \cite{Amendola1993}.}

Masses of the matter fields and their mutual interactions
can be neglected in the high curvature limit because $R\gg
m^{2}$. On the other hand, matter-graviton interactions generate  non-minimal
couplings in the effective Lagrangian. The
one-loop contributions of such terms are comparable to those given by the trace anomaly and generate, by conformal transformations, the same effects on gravity. 

The simplest effective
Lagrangian taking into account these corrections is
\begin{equation}
{\cal L}_{NMC} =- \frac{1}{2} \nabla^{\alpha} \varphi
\nabla_{\alpha} \varphi -V(\varphi)  - \frac{\xi}{2} \,
R\phi^{2} \,, \label{Intro1.2.23}
\end{equation}
where $\xi$ is a dimensionless  constant. The 
stress-energy  tensor  of the scalar field results modified accordingly  but  a conformal
transformation can be found
such that the modifications related  to curvature terms can be cast in the form of a matter-curvature
interaction. The same argument holds for the trace anomaly.
Some Grand Unification Theories lead to  polynomial couplings
of the form $ 1+\xi \phi^{2}+\zeta \phi^{4}$ that generalize the one
in (\ref{Intro1.2.23}). An exponential coupling
$\mbox{e}^{-\alpha \varphi} R$ between a
scalar field (dilaton) $\varphi$ and the Ricci scalar appears in the effective Lagrangian of strings.

Field equations derived by varying  the action ${\cal L}_{NMC}$ are
\begin{eqnarray}
\left( 1-\kappa \xi  \phi^2 \right) G_{\mu\nu} & =& \kappa
\left\{
\nabla_{\mu}\phi
\nabla_{\nu} \phi -\frac{1}{2} \, g_{\mu\nu} \,
\nabla^{\alpha}\phi \, \nabla_{\alpha} \phi -V\,
g_{\mu\nu} \right. \nonumber\\
&& \left. +  \xi \, \left[ g_{\mu\nu} \Box \left( \phi^2 \right)
-\nabla_{\mu} \nabla_{\nu}
\left( \phi^2 \right) \right] \right\} \;,  \label{pippa2} \\
&& \nonumber\\
& \Box \phi & -\frac{dV}{d\phi} -\xi R \phi =0 \; .
\label{nmKG}
\end{eqnarray}
The non-minimal coupling of the scalar field is similar to 
that derived for the 4-vector potential of curved space in   Maxwell
theory. See below Eq. (\ref{intro:deRham}).

Several   motivations can be provided for the
non-minimal  coupling in the Lagrangian ${\cal L}_{NMC}$. A
nonzero $\xi$ is  generated
by first loop corrections even if it
does not appear in the classical action
\cite{Birrell1982,Birrell1979,Nelson1982,Ford1981,Ford1987}. Renormalization of a classical theory with
$\xi=0$ shifts  this coupling constant to a value which is small
\cite{Allen1983,Ishikawa1983}. It can, however,
affect drastically an inflationary cosmological
scenario and determine its success or failure
\cite{Abbott1981,Futamase1989,Faraoni1996,Faraoni2004}.
A non-minimal  coupling   is expected at
high curvatures \cite{Ford1981,Ford1987}.
Furthermore, non-minimal coupling  solves potential problems of
primordial  nucleosynthesis  \cite{Chen2000} and, besides,
the absence  of pathologies in the  propagation of
$\varphi$-waves  requires  conformal coupling for all non-gravitational fields
\cite{Sonego1993,Grib1995a,Grib1995b,Deser1983,Faraoni1999}
\footnote{Note,
however, that the  distinction between gravitational and
non-gravitational fields  becomes representation-dependent in
ETGs, together with  the various formulations of the EP
\cite{Sotiriou2007}.}.

The conformal value $\xi=1/6$ is   fixed at infrared scales of renormalization group
\cite{Buchbinder1984,Buchbinder1985,Muta1991,Elizalde1994,Buchbinder1992,Reuter1994}.
Non-minimally coupled scalar fields have been  used
in  inflationary scenarios
\cite{Barroso1991,Fakir1991,GarciaBellido1995,Komatsu1997,Bassett1997,Futamase1997,
Salopek1988,Fakir1990,Fakir1992,Hwang1998}. The approach adopted was considering
$\xi$  as a free parameter  to  fix  problems of specific
inflationary scenarios \cite{Faraoni2001,Faraoni2004}.  Cosmological reheating  with  strong
coupling $\xi>>1$ has also been studied
\cite{Bassett1997,Tsujikawa1999,Tsujikawa2002} and  considered in relation with  wormholes
\cite{Halliwell1989,Coule1989,Coule1992}, black holes
\cite{Hiscock1990,vanderBij2000}, and boson stars
\cite{Jetzer1991,vanderBij1987,Liddle1993}. The coupling $\xi$ is not, in general, a free
parameter but depends on  the
particular scalar field 
$\varphi$ considered \cite{Voloshin1982,Hill1991,Hosotani1985,Ford1981,Ford1987, Faraoni1996,Faraoni2001,Faraoni2004}.

\subsection{Higher-order gravitational theories}

%
%

Let us take into account higher order theories and their relations to  scalar-tensor gravity
\cite{Schimming2004}. The first straightforward generalizaion of GR is 
\begin{equation} \label{Intro1}
{\cal L}=  \sqrt{-g} \, {f}(R)\,,
\end{equation}
The variation with respect to $g^{\mu\nu}$ yields the field equations
\begin{equation} \label{Intro2}
f'(R) R_{\mu\nu} -\frac{1}{2} \, f(R) g_{\mu\nu}
-\nabla_{\mu}\nabla_{\nu} f'(R) + g_{\mu\nu} \Box f'(R)=0\,,
\end{equation}
with $f' \equiv df(R)/dR$. Equation (\ref{Intro2}) is a fourth-order field equations (in metric formalism). It is convenient to introduce the new set of variables
\begin{eqnarray}
p &=& f'(R)= f' \left( g_{\mu\nu}, \partial_{\sigma} g_{\mu\nu}, \partial_{\sigma}\partial_{\rho}g_{\mu\nu} \right)\;, \\ 
\tilde{g}_{\mu\nu} &=& p \, g_{\mu\nu}\,.
\end{eqnarray}
This choice links the {\it Jordan frame}   variable $g_{\mu\nu}$ to the {\it Einstein frame} variables $ \left( p,\tilde{g}_{\mu\nu} \right)$, where $p$ is some auxiliary  scalar field.   The term ``Einstein frame'' comes from the fact that the  transformation $g \rightarrow \left( p, \tilde{g} \right)$ allows to recast  Eqs. (\ref{Intro2}) in a form similar  to the Einstein field equations of GR. In absence of matter, hence $T_{\mu\nu}^{(m)}=0$, the Einstein equations in   are
\begin{equation}  \label{Intro3}
\tilde{G}_{\mu\nu} = \frac{1}{p^{2}} \left[\frac{3}{2} \,
p_{,\mu} p_{,\nu} - \frac{3}{4} \,
\tilde{g}_{\mu\nu} \tilde{g}^{\alpha\beta}
p_{,\alpha} p_{,\beta} + \frac{1}{2} \tilde{g}_{\mu\nu}
\left(f(R)-R p \right) \right] \,.
\end{equation}
These equation can be rewritten in a more attractive way by defining
$\varphi=\sqrt{\frac{3}{2}} \ln p $, which implies
\begin{equation} \label{Intro4}
\tilde{G}_{\mu\nu} = \left[\varphi_{,\mu}\varphi_{,\nu} -
\frac{1}{2} \tilde{g}_{\mu\nu}\varphi_{,\sigma}\varphi^{,\sigma}
- \tilde{g}_{\mu\nu}V(\varphi) \right]\,,
\end{equation}
where
\begin{equation}\label{potVfR}
V(\varphi) =  \frac{Rf'(R)- f(R)}{
2 f'^{2}(R)}|_{R=R(p(\varphi))}\;.
\end{equation}
The curvature  $ R = R(p({\varphi}))$ is inferred by inverting the relation $p=f'(R)$ (provided $f''(R)\neq 0$). The field equation (\ref{Intro4}) can be
obtained from the Lagrangian (\ref{Intro1}) rewritten in terms of
$\varphi$ and the tilded  quantities
\begin{equation} \label{Intro5}
{\cal L} = \sqrt{-\tilde{g}} \left(\frac{1}{2} \tilde{R} -
\frac{1}{2} \tilde{g}^{\mu\nu}\varphi_{\mu}\varphi_{\nu} -
V(\varphi)\right) \;.
\end{equation}
which has the same form of Einstein  gravity minimally coupled to a  scalar field in presence of a self-interaction potential.
Equation Eq.~(\ref{Intro5}) clearly suggests that why the set of  variables
$\left( \tilde{g}_{\mu\nu},p\right) $ is called Einstein
frame \cite{Strominger1984,Faraoni2006a,Faraoni2006b}.    

A comment is in order. As we have seen, in the vacuumm, one can pass from the Einstein frame to  the Jordan frame. However, in the presence of matter fields, a caution is required since particles and photons have to be dealt in different ways. In the case of photons, their worldlines are geodesics both in the Jordan frame and in the Einstein frame. This is not the case for  massive particles since their geodesic in the Jordan frame are no longer transformed into geodesic in the 
Einstein frame, and vice-versa, and therefore, the two frames are not equivalent. The consequence is that the physical meaning of conformal transformations is not straightforward, although the mathematical transformations are, in principle, always possible. These considerations extend to any  higher-order or non-minimally coupled theory.

\subsection{Some aspects of the Equivalence Principle} 


As we have mentioned, in the previous Sections, the formulation of the EP is the equivalence between inertial and gravitational mass  $m_{I}=m_{G}$ (Galieleo's experiment),  which implies that all bodies fall with the same acceleration,
independently of their mass and  internal structure, in a
given gravitational field (universality of free fall or WEP).  
A more precise statement of WEP is  \cite{Will1993}

\vspace{2. mm}

``{\it If an uncharged body is placed at  an initial event in
spacetime and given an initial velocity there, then its subsequent
trajectory will be independent of its internal structure and
composition}''

\vspace{2. mm}

This formulation of WEP was enlarged by Einstein adding a new fundamental   part: according to which in a local inertial frame (the freely-falling
elevator)  not only the laws of mechanics behave in it as if
gravity  were absent, but {\em all} physical laws (except
those of gravitational
physics) have the same behaviour. The current terminology
refers to this principle as the {\em Einstein
Equivalence Principle} (EEP). A more precise statement  is \cite{Will1993}

\vspace{2. mm}

``{\it The outcome of any local non-gravitational test experiment
is independent of the velocity of the (free falling) apparatus
and the outcome of any local non-gravitational test experiment\footnote{A ``local non-gravitational experiment'' is defined as an
experiment performed in a small size freely falling laboratory,
in order to avoid the inhomogeneities of the external
gravitational field, and in which  any gravitational
self-interaction can be ignored. For example, the measurement of
the fine structure constant is a  local non-gravitational
experiment, while the Cavendish experiment is not.} is
independent of where and when in the universe it is performed} ''.

\vspace{2. mm}

From the EEP it follows that the gravitational interaction
must be described in terms of a curved spacetime, that  is  the
postulates of the so-called metric theories of gravity
have to be satisfied \cite{Will1993}:

\begin{enumerate}

\item  spacetime is endowed with a metric $g_{\mu\nu}$;

\item  the world lines of test bodies are geodesics of that
metric;

\item in local freely falling frames (called {\em local Lorentz
frames}), the non-gravitational laws of physics are those of
Special Relativity.

\end{enumerate}


These definitions characterize the most fundamental feature
of GR, hence the Equivalence Principle, as well as the physical properties
that allow to discriminate between GR and other metric theories of gravity,
In the  ETGs some additional features arise because these defintions depend on the conformal
representation of the theory adopted. More precisely, in
scalar-tensor gravity, massive test particles in the Jordan frame
follow geodesics, satisfying the WEP, but the
same particles deviate from geodesic motion in the Einstein frame
(a property referred to as  non-metricity of the theory). This
difference shows that the EP is formulated in a
representation-dependent way \cite{Sotiriou2007}. This
serious shortcoming  has not yet been addressed properly;
for the moment we proceed ignoring this problem.

\vspace{2. mm}

In what follows we shall discuss some specific features related to the Equivalence Principle:

\begin{itemize}
\item Let us assume that WEP is violated. Let us assume, for example, that the inertial masses $(m_{Ii})$ in a system  differ from the passive
ones,
\begin{equation}
m_{Pi} = m_{Ii}\left( 1+\Sigma_{A} \eta^{A}
\frac{E^{A}}{m_{Ii}c^{2}}\right)\,,
\end{equation}
where $E^{A}$ is the internal energy of the body connected to
the A-interaction and  $\eta^{A}$ is a dimensionless parameter
quantifying the violation of the WEP. It is then convenient to
introduce a new dimensionless parameter (the {\em E\"otv\"os
ratio}) considering, for example, two bodies moving with
accelerations
\begin{equation}
a_{i}=\left(1+\Sigma_{A}
\eta^{A}\frac{E^{A}}{m_{Ii}c^{2}}\right)g\;\;\;\;\;\;
(i=1,2)\;;
\end{equation}
where $g$ is now the acceleration of gravity. Then we
define the E\"{o}tvos ratio as
\begin{equation} \label{Intro8}
\eta=2 \, \frac{|a_{1}-a_{2}|}{|a_{1}+a_{2}|}=
\Sigma_{A}\eta^{A}\left(\frac{E_{1}^{A}}{m_{I1}c^{2}}-
\frac{E_{2}^{A}}{m_{I2}c^{2}}\right)\;.
\end{equation}
The measured value of $\eta$ provides information on the
WEP-violation parameters $\eta^{A}$. Experimentally, the
equivalence between  inertial and gravitational masses is
strongly confirmed  \cite{Will1993}.

\item The minimal coupling prescriptions. 
In electrodynamics the interaction is introduced  replacing the partial derivative with the covariant derivative 
$\partial_{\mu}\rightarrow {\cal D}_\mu \equiv \partial_{\mu}+ie A_{\mu}$ \cite{Gell-Mann1956} (see also \cite{Itzykson1980}). A similar scheme is used to introduce the gravitational interaction
\begin{equation} \label{Intro7}
\eta_{\mu\nu} \rightarrow g_{\mu\nu} \;, \qquad 
\partial_{\mu} \rightarrow  \nabla_{\mu} \;,
\qquad \sqrt{-\eta}\, d^{4}x \rightarrow \sqrt{-g}
\, d^{4}x\;,
\end{equation}
Here $\eta_{\mu\nu}$ is the flat Minkowski metric and $g_{\mu\nu}$ is the Riemannian one,
while $\eta$ and $g$ are their determinants \cite{Misner1974,Wald1984,Carroll2004}.

Consider the Maxwell equations in a curved spacetime 
\begin{equation} \label{Intromaxwell}
{F^{\alpha\beta}}_{;\beta}= 4\pi J^\alpha\;, \;\;\;\;\;\;\;F_{\alpha\beta ;\gamma}
+F_{\beta\gamma ;\alpha}+F_{\gamma\alpha ;\beta}=0\,,
\end{equation}
and the four-vector potential $A^{\mu}$ related to the Maxwell field by
$F_{\alpha\beta}=\nabla_{\alpha} A_{\beta}- \nabla_{\beta}A_{\alpha}$. In this framework, however, a problem arises. Using the above-mentioned rule one obtains two possible equations from the first of
eqs. (\ref{Intromaxwell}):
\begin{equation}
{A^{\beta ;\alpha}}_{;\beta}-{A^{\alpha; \beta}}_{;\beta}=4\pi J^\alpha\;,
\end{equation}
or
\begin{equation}
{A^{\beta ;\alpha}}_{;\beta}-{A^{\alpha ;\beta}}_{;\beta} + {R^\alpha}_{\beta}
A^\beta=4\pi J^\alpha\;;
\end{equation}
while  the second of eqs. (\ref{Intromaxwell}) yields, using the
Lorentz gauge $\nabla_{\mu}A^{\mu}=0$,
\begin{equation}
\left(\triangle_{dR}A\right)^\alpha=4\pi J^\alpha\,, \label{intro:deRham}
\end{equation}
where
\begin{equation}
\left(\triangle_{dR}A\right)^\alpha =-\Box A^\alpha+ {R^\alpha}_\beta A^\beta
\end{equation}
and $\triangle_{dR}$ is the de Rham vector wave operator.
Now the question is: both  Maxwell equations for the
four-potential $A^{\mu}$ are obtained using the ``comma goes to semicolon''
rule, but which is the correct one? The answer is: the
one  obtained using the de Rham operator. As consequence, we see
that  ``correspondence rules'' are not sufficient to write down
equations in curved space from known physics in flat space  when
second  derivatives are involved (that is, in most situations of
physical interest). In such cases, extra caution is
needed\footnote{As stressed, for example, in \cite{Sciama1953}, such a
prescription does not work for interactions
which do not have a ``Minkowskian''  counterpart. These interactions
are expressed in terms of the Riemann tensor or some
function of it and occur, for example, in the study of
the free fall of a particle with spin: the corresponding
equations of motion (Papapetrou equations) involve a contribution  in which the spin tensor couples to the Riemann tensor
\cite{Sciama1953}. Such a contribution can not be obtained from the
prescriptions given above. This motion is described by the
corrected geodesic equation \cite{Weinberg1972}.}

The minimal coupling prescription here discussed is 
connected with the mathematical formulation of the EEP
(actually, to  implement the EEP one needs to put in
special-relativistic form the laws under consideration and then
proceed to find the general-relativistic formulation,  switching
on gravity. In other words, we have to apply
minimal coupling prescriptions with the caveat already
discussed).

\item The last point is strictly related with the
scalar-tensor theories of gravity, do these theories satisfy the EEP?

To address this question one has to generalize the above two principles and introduce new concepts. Following Will \cite{Will1993},
one introduces the notion of ``purely dynamical metric theory'', i.e.
a theory in which {\it the behaviour of each field is
influenced to some extent  by a coupling to at least one of the
other fields in the theory}  \cite{Will1993}. In this respect, GR is a purely dynamical theory, as well as the Brans-Dicke theory since the  equations for the metric involve the scalar field, and vice-versa.

In these theories, the calculations of the metric is done in two stages: 1)  
the assignment of boundary conditions ``far'' from the local system; 2) infer the solutions of equations for the fields generated by the local
system. Owing to the coupling of the metric with fields (for given boundary conditions), the latter will  influence the metric. This implies that local gravitational experiments can depend on where the lab is located in the universe, as well as on its velocity relative to the external world. One of the consequence of such a new physical scenario is that in a Brans-Dicke theory, and more generally in Scalar Tensor Theories, the gravitational coupling ``constant'' turns out to  depend on the  asymptotic value of the scalar field.

All these considerations are strictly related to the
{\em Strong Equivalence Principle} (SEP) \cite{Will1993}:
\vspace{2. mm}

\noindent{\it (i) ``WEP is valid for
self-gravitating bodies as well for test bodies;\\
(ii) the outcome of any local test experiment is independent of
the velocity of the (freely falling) apparatus;\\
(iii) the outcome of any local test experiment is independent on
where and when in the universe it is performed''} \cite{Will1993}.
\vspace{2. mm}

The SEP differs from the EEP because it includes the self-gravitating interactions of bodies (such as planets or stars),
and because of experiments involving gravitational forces
({\em e.g.}, the Cavendish experiment). SEP reduces to the EEP when gravitational forces are ignored. In connection with the SEP, many authors have conjectured that
the only theory compatible with the Strong Equivalence Principle
is GR (that is $ SEP\longrightarrow GR-\mbox{only}$). 


\end{itemize}

\subsection{The Schiff conjecture}

The Schiff conjecture represents one of the most important topic related to the foundations of  the gravitational physics. Its original formulation asserts that   {\it every theory of gravity that satisfies the WEP and is relativistic necessarily satisfies the EEP, and is consequently a metric theory of gravity}. Hence $WEP \Rightarrow EEP$. Later, Will proposed a slight modification of Schiff conjecture:  {\it every theory of gravity that satisfies WEP and the principle of universality of gravitational red shift (UGH) necessarily satisfies EEP}. Hence in such a case $WEP + UGR \to EEP$. 

Let us discuss in some details these topics. Notice that the correctness Schiff's conjecture implies that the E\"otv\"os and the gravitational red-shift experiments would provide a direct empirical confirmation of the EEP, with the consequence that gravity can be interpreted as a geometrical (curved space-time)phenomenon. The relevance of such a fundamental aspect of the gravitational physics led to different mathematical approaches to prove the Schiff conjecture. These frameworks encompass all metric theories, as well as non metric theories of gravity.
Lightman and Lee \cite{Lightman:1973zz,Thorne:1973zz} proved Schiff's conjecture in the framework of the so called $TH\epsilon \mu$ formalism. They consider the motion of a charged particles (electromagnetic coupling) in a static spherically symmetric gravitational field $U=GM/r$ 
 \[
 S_{TH\epsilon\mu} = -\sum_a m_a \int dt\sqrt{T-H v_a^2}+\sum_a e_a \int dt v_a^\mu A_\mu (x_a^\mu)+\frac{1}{2}\int d^4 x \left(\epsilon {\bf E}^2+\frac{{\bf B}^2}{\mu}\right)\,,
 \]
where $m_a, e_a, v_a^\mu\equiv \displaystyle{\frac{dx_a^\mu}{dt}}$ represent the mass, the charge and the velocity of the particle $a$. The parameters $TH\epsilon \mu$ do depend on the gravitational field $U$, that is they essentially account for the response of the electromagnetic fields to the external potential, and may vary from theory to theory. A metric theory must satisfy the relation $\epsilon=\mu=\displaystyle{\sqrt{\frac{H}{T}}}$ for all $U$. In the case of non-metric theories, the parameters $TH\epsilon \mu$ may depend on the species of particles or on the field coupling to gravity. The metric is given by $ds^2= T(r) dt^2 - H(r)(dr^2+r^2 d\Omega)$. Lightmann and Lee showed in \cite{Lightman:1973zz} that the rate of fall of a test body made up of interacting charged particles does not depend on the structure of the body (WEP) {\it if and only if} $\epsilon=\mu=\displaystyle{\sqrt{\frac{H}{T}}}$. This implies $WEP \Rightarrow EEP$, satisfying hence the Schiff conjecture. 
Will generalized the Dirac equation in $TH\epsilon \mu$ formalism, and computed the gravitational red-shift experienced by different atomic clocks showing that the red-shift is independent on the nature of clacks (Universality of Gravitational Red-shift ($UGR$)) {\it if and only if} $\epsilon=\mu=\displaystyle{\sqrt{\frac{H}{T}}}$ \cite{Will:1985ry}. Therefore $UGR \Rightarrow EEP$, verifying in such a way another aspect of the Schiff conjecture (see also \cite{coley1982}).

W.-T. Ni was able to provide a counterexample to Schiff's conjecture by considering the coupling between a pseudoscalar field $\phi$ with the electromagnetism field ${\cal L}_{\phi F}\sim \phi \varepsilon^{\alpha\beta\gamma\delta}F_{\alpha\beta}F_{\gamma\delta}$, where $\varepsilon^{\alpha\beta\gamma\delta}$ is the completely anti-symmetric Levi-Civita symbol \cite{Ni:1977zz}. In \cite{Ohanian:1974pc,Accioly:1990hc,accioly1993} the Schiff conjecture is analyzed in the framework of gravitational non-minimally coupled theories. More specifically, the total Lagrangian density considered is given by ${\cal L}_{NMC}=\displaystyle{\frac{R}{16\pi G}+{\cal L}_M+{\cal L}_{I}(\psi^A, g_{\mu\nu})}$, where ${\cal L}_{I}(\psi^A, g_{\mu\nu})$ is the Lagrangian density of some field $\psi^A$ non-minimally coupled to gravity \cite{Accioly:1990hc,accioly1993}, while ${\cal L}_I=\chi^{\alpha\beta\gamma\delta}R_{\alpha\beta\gamma\delta}$ in \cite{Ohanian:1974pc}, where $\chi^{\alpha\beta\gamma\delta}$ depends on matter, for example $\chi^{\alpha\beta\gamma\delta} = {\bar \psi}\sigma^{\alpha\beta}\psi {\bar \psi}\sigma^{\gamma\delta}\psi, \psi^{\alpha\mu}\psi^{\beta\nu}-\psi^{\beta\mu}\psi^{\alpha\nu}$, where $\psi$ is a spin-half field and $\psi^{\alpha\beta}$ is a (nongravitational) spin-2 field. Both results show that these gravitational theories are in general, incompatible with Schiff's conjecture.

These counterexample indicate that a rigorous proof of such a conjecture is impossible. However, some powerful arguments of plausibility can be formulated.
One of them is based upon the assumption of energy conservation \cite{Haugan:1979iv}. 
Following \cite{Will2005}, consider a system in a quantum state $|A\rangle$ that decays in a state $|B\rangle$, with the emission of a photon with frequency $\nu$. The quantum system falls a height $H$ in an external gravitational field $gH=\Delta U$, so that the system in state $B$ falls with acceleration $g_B$ and the photon frequency is shifted to $\nu^\prime$. Assuming a violation of the WEP, the acceleration $g_A$ and $g_B$ of the system $A$ and $B$ are
 \[
  g_A=g\left(1+\frac{\alpha E_A}{m_A}\right)\,, \quad g_A=g\left(1+\frac{\alpha E_A}{m_A}\right)\,, \quad E_B-E_A=h\nu
  \]
that is they depend on that portion of the internal energy of the states. Here $\nu$ is frequency of the quantum emitted by the system $|A\rangle$. The conservation of energy implies that there must be a corresponding violation of
local position invariance in the frequency shift given by $\displaystyle{\frac{\nu^\prime-\nu}{\nu}=(1+\alpha)\Delta U}$, where $\nu^\prime$ is the frequency of the quantum at the bottom of the trajectory. The E\"otv\"os parameter is (for $m_A\sim m_B \sim m$)
 \[
  \eta= \frac{|g_B-g_A|}{|g_B+g_A|}\simeq \frac{\alpha(E_A-E_B)}{m}\,.
  \]
The Schiff conjecture is still nowadays an argument of a strong scientific debate and deep scrutiny.

\subsection{Mach's principle and the variation of $G$}

Following Bondi \cite{Bondi1952} there are, at least in
principle, two entirely different ways of measuring the
rotational velocity of Earth. The first is a purely terrestrial
experiment ({\em e.g.}, a Foucault pendulum), while the second is
an astronomical observation consisting of  measuring the
terrestrial rotation with respect to the fixed stars. In the
first type of experiment the  motion of the Earth is referred to
an idealized inertial frame in which Newton's laws are
verified. However, a unique general relativistic approach to define rotations has been introduced by Pirani considering the boucing photons \cite{Pirani1965,Pirani1973} (see also \cite{Pfister:2015ftd}). 
In the second kind of experiment the frame of
reference is connected to
a matter distribution surrounding the Earth and the motion of
the latter is referred to this matter distribution.
In this way we face the problem of Mach's principle,
which essentially states that the local inertial
frame is determined by some average  motion of distant
astronomical objects \cite{Bondi1952,Sciama1953}.\footnote{An
interesting discussion on this topic, also connected with different theories of space, both in
philosophy and in physics, is found in  Dicke's contribution
``The Many Faces of Mach''  in {\em Gravitation and
Relativity} \cite{Dicke1964P}. This discussion presents also the
problematic position that Einstein had on Mach's principle.}
Trying to incorporate Mach's principle into metric
gravity,  Brans and Dicke constructed a theory
alternative to  GR \cite{Brans1961}.
Taking into account the influence that the total matter has at each point  (constructing
the ``inertia''), these two authors  introduced, together with the
standard metric tensor, a new scalar field of gravitational
origin as the effective gravitational coupling. This is why the theory is referred to as a
``scalar-tensor'' theory; actually, theories in this spirit
had already been proposed years earlier by Jordan, Fierz, and
Thiery (see the book \cite{jordan1955}). An important ingredient of this
approach is that the gravitational ``constant'' is
actually a function of the total mass distribution, that is of
the scalar field, and is actually variable. In this picture,
gravity is described by the Lagrangian density
\begin{equation} \label{Intro9}
{\cal L}_{BD}=\sqrt{-g}\left[ \varphi R-
\frac{\omega}{ \varphi}
\nabla^{\mu} \varphi\nabla_{\mu} \varphi +
{\cal L}^{(m)}\right] \,,
\end{equation}
where $\omega$ is the dimensionless Brans-Dicke parameter and
${\cal L}^{(m)}$ is the matter Lagrangian including
all the non-gravitational fields. As stressed by Dicke
\cite{Dicke1961}, the Lagrangian (\ref{Intro9})  has a property
similar to  one  already discussed in the context of higher
order  gravity. Under  the conformal transformation
$g_{\mu\nu} \rightarrow \tilde{g}_{\mu\nu}=\Omega^2 g_{\mu\nu}$ with
$\Omega = \sqrt{ G_{0}\varphi} $, the Lagrangian (\ref{Intro9})
is mapped  into
\begin{equation}
{\cal L}=\sqrt{-\tilde{g} } \left(\tilde{R}+
G_{0} \tilde{\cal L}^{(m)} +
G_{0}\tilde{\cal L}^{(\Omega)}\right)\,,
\end{equation}
where
\begin{equation}
\tilde{\cal L}^{( \Omega)}=-\frac{(2\omega+3 )}{4\pi
G_{0}\Omega} (\nabla^{\alpha} \sqrt{\Omega})
(\nabla_{\alpha}  \sqrt{\Omega}) \,,
\end{equation}
and
$\tilde{\cal L}^{(m)}$ is the conformally transformed Lagrangian
density of matter. In this way the total matter
Lagrangian
$\tilde{\cal L}_{tot}=\tilde{\cal L}^{(m)}+
\tilde{\cal L}^{( \Omega )}$  has
been  introduced.  The field equations are now written in te form
of Einstein-like equations as
\begin{equation}
\tilde{R}_{\mu\nu}-\frac{1}{2} \, \tilde{g}_{\mu\nu} \, \tilde{R} =
G_{0} \tilde{\tau}_{\mu\nu} \,,
\end{equation}
where the stress-energy tensor
is now the sum of two contributions,
\begin{equation}
\tilde{\tau}_{\mu\nu} = T^{(m)}_{\mu\nu} +
\Lambda_{\mu\nu}( \Omega )\;.
\end{equation}
Dicke noted that this new (tilded, or Einstein frame) form of the
scalar-tensor  theory has certain advantages over the theory
expressed in the previous (non-tilded, or Jordan frame) form; the
Einstein frame representation, being similar to the
Einstein  standard  description is familiar and easier to handle
in some respects. But, in this new form,  Brans-Dicke theory also
exhibits unpleasant features. If we consider  the motion of a
spinless, electrically neutral, massive particle, we find that in
the conformally rescaled world its trajectory is
no longer a geodesic.  Only null rays  are left unchanged by
the conformal rescaling. This is a manifestation of the fact that the rest mass is not constant in
the conformally transformed world and the equation of motion of
massive particles is modified by the addition of an
extra force proportional to $\nabla^{\mu} \Omega$ \cite{Dicke1961}.
Photon trajectories, on the other hand, are not modified because the vanishing of the
photon mass implies the absence of a preferred physical scale and
photons stay massless under the conformal rescaling, therefore
their   trajectories  are unaffected.

This new approach to gravitation has increased the relevance of
theories with varying gravitational coupling. They are of
particular interest in cosmology since, as we discuss in  detail
in the following chapters, they have the potential to circumvent
many shortcomings of the standard cosmological model.
We list here the Lagrangians of this type
which are most relevant for this review.

\begin{itemize}

\item The low-energy limit of the bosonic string theory
\cite{Green1987,Tseytlin1989,Scherk1974} produces the Lagrangian
\begin{equation}   \label{Intro10}
{\cal L}= \sqrt{-g} \,
\mbox{e}^{-2\phi} \left( R +  4g^{\mu\nu} \phi_\mu \phi_\nu
-\Lambda\right) \;.
\end{equation}

\item The general scalar-tensor Lagrangian is
\begin{equation} \label{Intro11}
{\cal L}_{ST}=\sqrt{-g} \left[ f(\varphi) R -
\frac{\omega( \varphi)}{2}\, g^{\alpha\beta} \nabla_{\alpha} \phi
\nabla_{\beta} \phi -V(\varphi)  \right]\;,
\end{equation}
where $f(\varphi)$ and $\omega (\varphi)$ are arbitrary coupling
functions and $V( \varphi)$ is a scalar field potential. The
original Brans-Dicke Lagrangian is contained as the special case
$f(\varphi)=\varphi, \omega(\varphi)=\omega_0/\varphi$ (with
$\omega_0$ a constant), and $V(\varphi)\equiv 0$.

\item A special case of the previous general theory is that of a
scalar field non-minimally coupled to the Ricci curvature, which
has received so much attention in the literature to deserve a
separate mention,
\begin{equation}
{\cal L}_{NMC}=\sqrt{-g} \left[ \left( \frac{1}{16\pi G}-\frac{\xi}{2} \right) R -
\frac{1}{2}\, g_{\mu\nu} \nabla_{\mu} \phi \nabla_{\nu}
\phi  -V(\varphi) \right]\;,\label{Intro12}
\end{equation}
where $\xi$ is a dimensionless non-minimal coupling constant.
This explicit non-minimal coupling was originally introduced
in the context of classical radiation problems
\cite{Chernikov1968} and, later, conformal coupling with
$\xi=1/6$  was discovered to be necessary
for  the renormalizability of the $\lambda \varphi^4$ theory on a
curved spacetime \cite{Callan1970,Birrell1982}. The
corresponding stress-energy tensor (``improved energy-momentum
tensor'') and the relevant equations will be discussed later.  In
particular, the theory is conformally invariant when $\xi=1/6$
and either $V \equiv 0$ or $V=\lambda \varphi^4$
\cite{Penrose1965,Callan1970,Birrell1982,Wald1984}. 

\end{itemize}

All these theories exhibit a non-constant gravitational
coupling. The Newton constant $G_{N}$ is replaced
by the effective gravitational coupling
\begin{equation} \label{Intro13}
G_{eff}=\frac{1}{f(\varphi)} \,,
\end{equation}
in eq.~(\ref{Intro11}) which, in general, is different from
$G_{N}$ (we use $\phi$ as
the generic function describing the effective gravitational
coupling). In string theory or with non-minimally coupled
scalars, such functions are specified in
(\ref{Intro10}) and
(\ref{Intro12}). In particular, in spatially homogeneous
and isotropic cosmology, the
coupling $G_{eff}$ can only be a function of   the epoch, {\em
i.e.}, of the cosmological time.

We stress that all these scalar-tensor  theories of gravity do
not satisfy the SEP because of the above mentioned feature:
the variation of $G_{eff}$ implies that local gravitational
physics  depends  on the scalar field via $\phi$. We have
 then motivated the  introduction of  a stronger version of the
Equivalence Principle, the SEP. General theories with
such  a peculiar aspect are called {\it non-minimally coupled
theories}. This generalizes older terminology in which  the
expression ``non-minimally coupled scalar'' referred specifically
to the field described by the Lagrangian ${\cal L}_{NMC}$ of
(\ref{Intro12}), which is a special case of (\ref{Intro11}).

Let us consider, as in (\ref{Intro11}), a general scalar-tensor
theory in presence of ``standard'' matter  with total Lagrangian density $ \phi
R + {\cal L}^{( \phi)}+{\cal L}^{(m)}$, where ${\cal L}^{(m)}$
describes ordinary matter. The dynamical equations for this
matter are
contained in the covariant conservation equation $
\nabla^{\nu} T_{\mu\nu}^{(m)}=0$  for  the
matter stress-energy tensor  $T_{\mu\nu}^{(m)}$, which is
derived from the variation of the total
Lagrangian with respect to $g^{\mu\nu}$. In other
words: concerning standard matter, everything goes as in  GR
({\em i.e.}, $\eta_{\mu\nu}\rightarrow g_{\mu\nu}$,
$\partial_\mu \rightarrow \nabla_\mu$)  following the minimal coupling
prescription. What is new in these theories is the way in which the scalar and the
metric degrees of freedom appear:  now there is a direct coupling
between the scalar degree of freedom and  a
function of the tensor degree of freedom (the metric)
and its
derivatives (specifically, with the Ricci scalar of the metric
$R \left( g,\partial g,\partial^{2} g \right)$).   Then,  confining  our
analysis to the cosmological
arena, we  face two  alternatives. The first is
\begin{equation}
\lim_{t\rightarrow\infty}  G_{eff}\left( \phi(t) \right) = G_{N}\,;
\end{equation}
this is the case in which standard GR cosmology  is recovered at
the present time in the history of  the universe. The second
possibility occurs if  the
gravitational coupling is not constant today, {\em i.e.},
$ G_{eff}$ is still varying with the epoch and
$ \dot{G}_{eff}/G_{eff}|_{\mbox{now}}$ (in brief $\dot{G}/G$) is
non-vanishing.

In many theories of gravity, then, it is perfectly conceivable
that $G_{eff}$ varies with time: in some solutions $G_{eff}$ does not
even converge to the value observed today. What do we know, from the
observational point of view, about this variability?
There are three main avenues to analyze the
variability of $G_{eff}$: the first is {\it lunar laser ranging}
(LLR) monitoring the Earth-Moon distance; the
second is information from solar astronomy;  the third
consists of  data from binary  pulsars. The
LLR consists of measuring the round trip travel time and thus the
distances between transmitter and reflector, and monitoring them
over an extended period of time. The change of round
trip time contains information about the Earth-Moon system.
This round trip travel time has been measured for more than
twenty-five years in connection with the Apollo~11, 14, 15,  and
the Lunakhod~2 lunar missions. Combining these data with those
coming from the evolution of the Sun (the luminosity of main
sequence stars is quite sensitive to the value of $G$)  and the Earth-Mars radar
ranging, the current bounds  on
$\dot{G}/G$ allow at most  $0.4\times 10^{-11}$ to $1.0\times
10^{-11}$ per year \cite{Dickey1994}. The third
source of information on  $G$-variability is given by binary
pulsars systems. In order to extract data from this type of system
(the prototype is the famous binary pulsar PSR~1913+16 of Hulse
and Taylor \cite{Taylor1982}),  it has been necessary to extend the
post-Newtonian  approximation, which can be applied only to a
weakly (gravitationally) interacting $n$-body system, to
strongly (gravitationally) interacting systems. The  order
of magnitude of  $\dot{G}/G$ allowed by  these
strongly interacting systems is $2\times 10^{-11}$ yr$^{-1}$ \cite{Dickey1994}.

A general remark is necessary at this point. According to the Mach Principle, gravity can be considered as an average interaction given by the distribution of celestial bodies. This means that the same gravitational coupling can be related to the spacetime scale, then supposing a variation of $G_N$ is an issue to make more Machian the theory. From an experimental point of view, this fact reflects on the  uncertainties of the measurements of $G_N$ and it could constitute a test for any alternative theory of gravity with respect to GR.

Finally it is worth noticing that there exist also Higgs-scalar-tensor theories  (see for example, \cite{Zee:1978wi,CervantesCota:1994zf,CervantesCota:1995tz}) where inertia and gravity are strongly related. Such theories have been introduced to solve the issues raised in  the Brans-Dicke theory  where the observational results, coming from  the Mercury perihelion shift, are not matched. In view of this shortcoming, Dicke postulated the existence of a mass-quadrupole momentum giving rise to an oblateness correction of the Sun shape. Since this feature was not detected, Higgs-scalar-tensor theories were deemed necessary.


\subsection{Violation of the weak equivalence principle and quintessence}

In the previous Sections, we have pointed out that over the last years several, observations led to the conclusions that the observed Universe is dominated
by some form of (homogeneously distributed) DE. In modified gravity the DE can be described by introducing one or more than one scalar fields coupled (minimally or non-minimally) to gravity.  A candidate for DE is quintessence (the energy density associated a scalar field that evolves slowly in time) \cite{Wetterich:1987fm,Peebles:1987ek,Copeland:1997et,Ferreira:1997hj,Caldwell:1997ii}. In this scenario, fundamental coupling constants do depend on time even in late cosmology \cite{Wetterich:1987fm,Wetterich:1987fk,Wetterich:1994bg,Dvali:2001dd,Chiba:2001er}. This because, as we have seen, it is usual that in modified models of gravity the fundamental coupling constants may depend on the scalar field, that vary during the Universe evolution. Clearly, an observation of a possible time-variation of fundamental constants could be a signal in favour of quintessence, and more generally, of modified theories of gravity, since no such time dependence would be connected to DE in the case in which the latter is described by a cosmological constant.
\footnote{A low value of the electromagnetic fine structure constant $\alpha_{em}$ was reported \cite{Webb:2000mn} for absorption lines in the light from distant quasars. The data are consistent with a variation $\Delta \alpha_{em}/\alpha_{em}\simeq -0.7 \times 10^{-5}$ for a cosmological red-shift $z \approx 2$. Such a result has renewed the interest on the variation of fundamental couplings (see for example \cite{Dvali:2001dd,Chiba:2001er,Sandvik:2001rv,Olive:2001vz,Dent:2001ga,Uzan:2002vq,Damour:2002mi}).}

It is expected that, in a quintessence scenario, the gauge couplings may vary owing to the coupling between the field $\phi(x)$ and the kinetic term for the gauge fields in a GUT [20]. For example, for the electromagnetic field one has $L_F = \frac{1}{4} Z_F (\phi(x)) F^2$, where $F^2=F_{\mu\nu} F^{\mu\nu}$. Such a coupling preserves all symmetries and makes the renormalized gauge coupling $g \sim Z_F^{-1/2}$ dependent on time through the evolution of the field $\phi(x)$ \cite{Wetterich:2002ic}. As argued in \cite{Wetterich:2002ic}, the coupling of the field $\phi(x)$ with matter induces a new gravity-like force that does depend on the composition of the test bodies. In this respect a violation of the equivalence principle arises \cite{Peccei:1987mm}.

Along these lines, very recently it have been proposed new and general models in which a light scalar field (playing the role of scalar Dark Matter) is introduced in the gravity action (similar to Eq. (72)). In the most and simplest general case, in fact, the light scalar field couples non-universally to the standard matter fields, leading as a consequence to a violation of the Einstein equivalence principle (EEP). As discussed in the previous Sections, the scalar fields are predicted in high dimensional theories,  in particular in string theory with the dilaton and the moduli fields \cite{Green1987,Damour1994,Damour2002}. It is worth to mention that these models based on light scalar field provide galactic and cosmological predictions for low masses, ranging from $10^{-24}$eV to $10^{-22}$eV (see for example, Refs. \cite{Hu:2000ke,Beyer:2014uja,Marsh:2015xka,Hui:2016ltb}. Here we recall the total action in which a microscopic modeling for the coupling between the scalar field and standard matter has been conveniently introduced
\cite{Damour:2010rp,Damour:2010rm}
\begin{equation}\label{DamourDonoghue}
  S=\int d^4x [{\cal L}_{NMC}+{\cal L}_{SM}+{\cal L}_{int}]\,,
\end{equation}
where ${\cal L}_{NMC}$ is the Lagrangian density (72), ${\cal L}_{SM}$ the Lagrangian density of the Standard Model, and finally ${\cal L}_{int}$ is the Lagrangian density of the interaction, which can be of two form \cite{Stadnik:2015kia,Damour:2010rp,Damour:2010rm}
\begin{equation}\label{LintDamourDonoghue}
  {\cal L}_{int}=\phi^a \left[\frac{d_e^{(a)}}{4\mu_0}F^2-\frac{d_g^{(a)}\beta_3}{2g_3}(F^A)^2-\sum_{i=e,u,d}\left(d_{m_i}^{(a)}+\gamma_{m_i}d_g^{(a)}\right)m_i {\bar \psi}_i \psi_i\right]
\end{equation}
Here $a=1$ and $a=2$ correspond to the linear and quadratic \cite{Stadnik:2015kia,Stadnik:2014tta,Kalaydzhyan:2017jtv} coupling between scalar and matter field, respectively,
while $F_{\mu\nu}$ and $F_{\mu\nu}^A$ are the electromagnetic and the gluon strength tensors, $\mu_0$ the magnetic permeability, $g_3$ the QCD gauge coupling, $\beta_3$ the $\beta$ function for the running of $g_3$, $m_i$ the mass of the fermions (electron and light quarks $u, d$), $\gamma_{m_i}$ the anomalous dimension giving the energy running of the masses of the QCD coupled fermions, and finally $d^{a}$ are the constants characterizing the interaction between the light scalar field and the different matter sectors. The main consequence of the model based on (\ref{DamourDonoghue}) is that the constants of nature turn out to be linearly or quadratically depending on the scalar field \cite{Damour:2010rp,Damour:2010rm}. For the electromagnetic fine structure constant $\alpha_{EM}$, the masses $m_i$ of the fermions, and the QCD energy scale $\Lambda_3$, one obtains
 \begin{eqnarray*}
   \alpha_{EM}(\phi) &=& \alpha_{EM}\left[1+\frac{d_e^{(a)}\, \phi^a}{a}\right] \\
   m_i (\phi) &=& m_i\left[1+\frac{d_{m_i}^{(a)}\, \phi^a}{a}\right]\qquad i=e, u, d \\
   \Lambda_3(\phi) &=& \Lambda_3\left[1+\frac{d_g^{(a)}\, \phi^a}{a}\right]
 \end{eqnarray*}
with $a=1,2$ for the linear and the quadratic coupling. The dependence of the particle masses on the scalar field suggests to study tests of the universality of free fall. Following \cite{Hees:2018fpg}, one gets that the differential acceleration between two bodies $A$ and $B$ located at the same position in a gravitational field generated by a body $C$, is
 \begin{equation}\label{DeltaaWolf}
   \Delta {\bf a}\equiv {\bf a}_A-{\bf a}_B=-[\alpha_A(\phi)-\alpha_B(\phi)][\nabla \phi+{\bf v} {\dot \phi}]\,,
 \end{equation}
where $\alpha_{A,B}=\displaystyle{\frac{\partial \ln m_{A,B}(\phi)}{\partial \phi}}$ and ${\bf v}$ the particle velocity. Using the expressions for the scalar field derived in the case of a spherically symmetric extended body with radius $R$ and constant matter density with mass $M$, one infers the explicit expression for the E\"otv\"os parameter $\eta$ (Eq. (64)) \cite{Hees:2018fpg}
\begin{equation}\label{etaWolf}
  \eta=2\frac{|{\bf a}_A-{\bf a}_B|}{|{\bf a}_A+{\bf a}_B|}=\left\{  \begin{tabular}{ll} $\Delta {\tilde a}^{(1)} s_C^{(1)}e^{-r/\lambda_\phi}\left(1+\frac{r}{\lambda_\phi}\right)$ & $\quad$ (\text{linear coupling}) \\
     & \\
  $\Delta {\tilde \alpha}^{(2)} s_C^{(2)} \frac{\phi_0}{2} \left(1-s_C^{(2)}\frac{G M_C}{r}\right) $ & $\quad$ (\text{quadratic coupling}) \end{tabular}\right.
\end{equation}
Here $\Delta {\tilde \alpha}^{(a)}={\tilde \alpha}_A^{(a)}-{\tilde \alpha}_A^{(a)}$, with $a=1,2$ and ${\tilde \alpha}^{(a)}$ is a combination of the coefficients $d_{e, m, g}^{(a)}$ and the dilatonic charges associated to the bodies $A$ and $B$, $s_C^{(1)}=3{\tilde \alpha}_C^{(1)}\displaystyle{\frac{x\cosh x-\sinh x}{x^3}}$, with $x=\displaystyle{\frac{R}{\lambda_\phi}}$ ($\lambda_\phi=m_\phi^{-1}$ is the Compton wavelength of the scalar field), and $s_C^{(2)}={\tilde \alpha}^{(2)}_C J_{\pm}(y)$, with $y=\sqrt{3|{\tilde \alpha}_C^{(2)}|G M_A/R_A}$ and $J_\pm = \pm 3\displaystyle{\frac{y-\tanh y}{y^3}}$, and finally $\phi_0$ is the amplitude of the scalar field.

An interesting aspect of these results is that in the neighborhood region of a central body and in the limit of strong coupling, for the quadratic coupling Eq. (\ref{etaWolf}) assumes the form $\eta\simeq \displaystyle{\Delta {\tilde \alpha}^{(2)} s_C^{(2)}\phi_0^2 \frac{h}{R_C+h}}$, where $h$ is the altitude with respect to the radius $R_C$. On the other hand, for small coupling and far from the gravitational source, one gets $\eta\simeq s_C^{(2)}\Delta {\tilde \alpha}^{(2)}\phi_0^2$, that is the E\"otv\"os parameter is independent on the location of the two masses. As argued in \cite{Hees:2018fpg}, this particular forms of the E\"otv\"os parameter could be potentially tested in dedicated experiments.

Finally we comment the possibility to violate the Einstein equivalence principle by measuring the frequency ratio between two clocks located at the same position and working on different atomic transition. Defining $Y=X_A/X_B$, where $X_{A,B}$ are the specific transitions for each clocks, one finds \cite{Hees:2018fpg}
\begin{equation}\label{YY0}
  \frac{Y(t, {\bf x})}{Y_0 }  = \left\{  
  \begin{tabular}{ll} $K +\Delta \kappa^{(1)}
  \left[\phi_0\cos(\omega t-{\bf k}\cdot {\bf x}+\delta)- s_A^{(1)}\frac{GM_A}{r}\,e^{-r/\lambda_\phi}\right]$ & \quad (\text{linear coupling}) \\
     & \\
  $K +\Delta \kappa^{(2)}\frac{\phi_0^2}{2} \left[\left(1-s_A^{(2)}\frac{GM_A}{r}\right)^2 +\cos(2\omega t+2\delta) \left(1-s_A^{(2)}\frac{G M_C}{r}\right)^2\right]$ & \quad (\text{quadratic coupling}) \end{tabular}\right.
\end{equation}
where $K$ is an unobservable constant and $k^{(a)}$, $a=1,2$ depend on the constants $d_{e, m, g}^{(a)}$.


\subsection{Equivalence principle in screening mechanisms}

As extensively discussed in the previous Sections, the introduction of the extended theories of gravity have been motivated
by the necessity to explain the observed cosmic acceleration, hence to provide a "geometric" interpretation of the DE.
In these models, gravity is modified on large distances.
However, although modifications to GR must be relevant on large scales, they are strongly constrained in Solar System (in what follows we shall refer to \cite{Sakstein:2017pqi}). In fact, any deviation
is subdominant in Solar System tests by a factor $\lesssim 10^{-5}$, and the latter is further reduced in some specific theories (in \cite{Touboul:2017grn,Berge:2017ovy} is discussed the case of theory that predicts strong violations of the
weak equivalence principle for which deviations are constrained by a factor $\lesssim 10^{-15})$.
As an example of extended theories of gravity, consider once again the Brans-Dicke gravity (the scalar field $\phi$ couples to
gravity and is parameterized by the parameter $\omega_{BD}$). In
the non-relativistic limit, one finds the equation of motion for $\phi$
\begin{equation}\label{BDeq}
  \nabla^2 \phi = - \frac{8\pi G \rho}{2+3\omega_{BD}}
\end{equation}
from which one derives the PPN parameter $|\gamma-1|=(2+\omega_{BG})^{-1}$.
The Cassini bound $|\gamma-1| < 2.1 1\times 10^{-5}$ \cite{Bertotti:2003rm} implies $\omega_{BD} > 4\times 10^4$. From (\ref{BDeq})
it follows that the effective coupling to matter is
$\alpha_{eff} \sim 1/\omega_{BD} \lesssim 10^{-4}$. As a consequence, any Brans-Dicke like
modifications of GR must be subdominant on all scales by a factor $\sim 10^4$, hence such theories
are cosmologically irrelevant. A similar conclusions follows if one assumes that the scalar field is massive, so that the field equation
(\ref{BDeq}) gets modified a $(\nabla^2 +m^2)\phi = - 8\pi G \alpha \rho$,
yielding, for a a static, spherically symmetric body, a Yukawa-like potential $V(r)=\displaystyle{\frac{GM}{r}\left(1+2\alpha^2 e^{-m r}\right)}$
(experiments constrained Yukawa-like potentials on distances ranging from the Earth-Moon scale \cite{Bertotti:2003rm,Adelberger:2017low}  to micron scales \cite{Kapner:2006si,Adelberger:2003zx}, so that $m > (\mu \text{m})^{-1}$ is required to evade Solar system tests).

These two examples show that solar system tests constraint these models with the consequence that they do not have any cosmological relevance because
the force must either be too weak, or too short ranged. Such difficulties are avoided by
screening mechanisms by nonlinear modifications of the
Poisson equation. The modifications are such that deviations from GR in the Solar system are dynamically suppressed,
without requiring a fine-tuning of the mass or the coupling to matter. Screening mechanisms studied in literature are:
\begin{itemize}
  \item Chameleon screening \cite{Khoury:2003aq,Khoury:2003rn} (the mass of the field changes dynamically mediating short ranged forces in the Solar System but may have effects on cosmological scales).
  \item Symmetron screening \cite{Khoury:2003aq,Khoury:2003rn} (the coupling to matter varies dynamically so that it is uncoupled in the Solar System and may induces deviations from
GR on cosmological scales).
  \item Vainshtein's mechanism \cite{Barreira:2012kk} (nonlinear kinetic
terms alter the field profile sourced by massive
bodies. In such a case fifth forces are highly suppressed in the
Solar System, while on cosmological scales, theories that
exhibit this mechanism can self-accelerate without a
cosmological constant, which makes them interesting
alternatives to $\Lambda$CDM cosmologies).
\end{itemize}

An interesting aspect of screening mechanisms, is that they may violate the equivalence principle \cite{Hui:2009kc} (see also \cite{Kraiselburd:2015vyf}). For example, in chameleon theories one can define a scalar charge for an object \cite{Hui:2009kc}
\[
Q_i = M_i \left(1-\frac{M_i(r_s)}{M_i}\right)
 \]
so that the force on an object due to an externally applied
chameleon field is $F_{Ch}=\alpha Q_i \nabla \phi_{ext}$ (this is analogous to
the gravitational charge $M$ so that $F_{grav}= M \nabla \phi_N^{ext}$
where $\phi_N^{ext}$ is an external Newtonian potential). Two
objects of different masses and internal compositions will
have different scalar charges and will therefore fall at
different rates in an externally applied chameleon field,
signifying a breakdown of the weak equivalence principle
(WEP). The chameleon force between two bodies, A and B,
is \cite{Mota:2006fz}
\[
F_{AB}=\frac{GM_AM_B}{r^2}\left(1+2\alpha Q_A Q_B e^{-m_{eff} r}\right)
\]
and as a result of this the PPN parameter $\gamma$ is
$\gamma=\displaystyle{\frac{2}{1+2\alpha Q_A Q_B e^{-m_{eff} r}}-1}$
(see also \cite{Burrage:2017qrf,Zhang:2016njn,Hees:2011mu}). Here A refers to the body responsible for the
deflection/time delay of light while the body B is a separate body
used to measure the mass of the body A (for example, for light
bending by the Sun one would take A as the Sun and B as
the Earth).


\subsection{Long-range forces and spin-gravity coupling terms}
\label{th:Long-range forces and spin-gravity}

In this Section we discuss the possibility that the spin of particles can be present in gravitational potentials. There are essentially some reasons for searching long-range forces that are depending on spin of particles: 1) The role of spin in gravitation (see for example \cite{Leitner:1964tt,HariDass:1976xsb,Hayashi:1979qx}). 2) The interaction associated with the exchange of a light or massless pseudoscalar boson or similar interactions \cite{Weinberg:1977ma,Wilczek:1977pj,Dine:1981rt,Shifman:1979if,Kim:1979if,Moody:1984ba}. 

In fact, new particles predicted in theories that extend the standard model may induce modifications to spin-spin interaction between fermions \cite{Dobrescu:2006au}. As an example, we recall the pseudoscalar fields, such as the axion \cite{Moody:1984ba}, and the axial-vector fields, such as paraphotons \cite{Dobrescu:2004wz} and extra Z bosons \cite{Dobrescu:2006au,Appelquist:2002mw}, the first associated with theories with spontaneously broken symmetries \cite{Weinberg:1977ma,Wilczek:1977pj,Dine:1981rt}, the latter in new gauge theories (these new particles, predicted also in string theories \cite{Arvanitaki:2009fg}, are typically introduced to explain the DE \cite{Friedland:2002qs,Flambaum:2009mz} and the DM \cite{Bertone:2004pz}). 3) A number of Kaluza-Klein theories \cite{Bars:1986gt,Barr:1986uv} and supersymmetric theories \cite{Lazarides:1985bj}, in the low-energy limit, predict couplings in which the spins of particles are involved.

As an example we report the Yukawa-like potential between fermions in the case in which they exchange a (new) vector or axial vector\footnote{
It is worth to recall that gravitational interactions between two objects that do not conserve the discrete symmetries were proposed in \cite{HariDass:1976xsb}
 \[
U(r)= \frac{GM}{r}\left[\alpha_1 \frac{{\bf s}^{(1)} \cdot {\bf {\hat r}}}{r}+ \alpha_2 \frac{{\bf s}^{(1)} \cdot {\bf v}}{r}+\alpha_3 \mu
{\bf {\hat r}} \cdot {\bf {v}} \right]\,,
 \]
where $\alpha_{1,2,3}$ are generic coefficients, $M$ is the total mass, $\mu$ the reduced mass, ${\bf r}$ the relative displacement, ${\bf v}$ the relative velocity, and ${\bf s}^{(1)}$ is the intrinsic spin of one of the objects (see also \cite{Lambiase:2004sm,Papini:2001un}).} $A$ \cite{Moody:1984ba,Dobrescu:2006au}
 \begin{equation}\label{VspinA}
   V_A (r)= \xi_A \, {\bf s}_1 \cdot {\bf s}_2\, \frac{e^{-r/\lambda}}{r}\,,
 \end{equation}
where $\xi_A = \displaystyle{\frac{g_A^{(e)}g_A^{(e)}}{4\pi}}$ is the dimensionless axial-vector coupling constant between the electrons, ${\bf s}_1$ and ${\bf s}_2$ represent the spins of the electrons, and $r$ is the inter-particle separation, while the dipole-dipole potential between electrons corresponding to the exchange of a new axion-like pseudoscalar particle $P$ is \cite{Moody:1984ba,Dobrescu:2006au}
 \begin{equation}\label{VspinspinP}
   V_P(r)=\xi_P \, \frac{e^{-r/\lambda}}{4m_e^2}\, 
   \left[{\bf s}_1\cdot {\bf s}_2\left(\frac{4\pi}{3}\delta^{3}(r)+\frac{1}{\lambda r^2}+\frac{1}{r^3}\right)-({\bf s}_1 \cdot {\bf {\hat r}})({\bf s}_2\cdot {\bf {\hat r}})\left(\frac{1}{\lambda^2 r}+\frac{3}{\lambda r^2}+\frac{3}{r^3}\right)\right]\,,
 \end{equation}
where $\xi_P=\displaystyle{\frac{g_P^{(e)}g_P^{(e)}}{4\pi}}$ is the dimensionless pseudoscalar coupling constant between the electrons, $m_e$ the mass of the electron mass, and ${\bf {\hat r}}={\bf r}/r$.

The general analysis of long-range forces between macroscopic objects (polarized spin medium) mediated by light particles that include spin and velocity terms have been performed in \cite{Moody:1984ba,Dobrescu:2006au}. Experiments aimed to tests such new terms will be discussed in next Sections.


\subsection{The equivalence principle in Poincare Gauge Theory and Torsion}

As we have seen the equivalence principle sates that the effect of gravity on matter is locally equivalent to the effect of a non-inertial reference frame in special relativity.
%
The dynamical content of the equivalence principle can be understood by considering an inertial frame in\footnote{In this Section, we follow the notation in \cite{Blagojevic:2002du}. A space $(L_4; g)$ with the most general metric compatible linear connection $\Gamma$ is called Riemann-Cartan space $U_4$. If the torsion vanishes, a $U_4$ becomes a Riemannian space $V_4$ of GR; if, alternatively, the curvature vanishes, a $U_4$ becomes Weitzenbock's teleparallel space $T_4$. The condition $R^\alpha_{\,\, \beta\gamma\chi} = 0$ transforms a $V_4$ into a Minkowski space $M_4$, and $T^\alpha_{\,\, \beta\gamma} = 0$ transforms a $T_4$ into an $M_4$.} $M_4$, in which matter field $\phi$ is described by the Lagrangian $L_M(\phi; \partial_i \phi)$. Passing to a noninertial frame, $L_M$ transforms into $\sqrt{-g} L_M(\phi; \nabla_i \phi)$, with $\nabla_i = e_i^\mu (\partial_\mu+\omega_\mu)$ the covariant derivative (this is the minimal substitution discussed in the previous Section). The gravitational field (equivalent to the non-inertial reference frame) appears in the quantities $\sqrt{-g}$ and $\nabla_i$, and can be eliminated on the whole spacetime by reducing to the global inertial frame, while for real gravitational fields one has that they can be eliminated only locally. For introducing a real gravitational field, hence, Einstein replaced $M_4$ with a Riemann space $V_4$. However, also a Riemann-Cartan space $U_4$ could have been chosen \cite{Blagojevic:2002du}.

Another formulation of the Equivalence Principle asserts that the effect of gravity on matter can be locally eliminated by a suitable choice of reference frame, and matter behaves following the laws of Special Relativity \cite{Blagojevic:2002du}, i.e. {\it at any point P in spacetime an orthonormal reference frame ${\bf e}_i$ can be chosen such that $\omega^{ij}_{\quad \mu}=0$ and $e_i^\mu =\delta_i^\mu$ at P}. The important consequence of this statement is that it holds not only in GR (i.e. $V_4$), but also in Poincare Gauge Theory (i.e. $U_4$) \cite{Heyde1975,Hartley:1995dg}. The Equivalence Principle is not violated in manifolds with torsion, fitting in natural way into a $U_4$ geometry of spacetime. It holds in $V_4$, as well as in $T_4$. Notes however that in more general geometries, characterized by a symmetry of the tangent space higher than the Poincare group, the usual form of the Equivalence Principle can be violated, and local physics differs from Special Relativity \cite{Blagojevic:2002du,Hehl:1994ue}.


\subsection{The violation of the equivalence principle for charged particles}

Let us discuss now the tests of Universality of Free Fall (UFF) for charged particles. The interest for these studies follows from the fact that, in some frameworks, a violation of the UFF is related with charge non-conservation \cite{Landau:2001qb}.
Considering a connection of UFF and Universality of the Gravitational Red-shift (UGR) \cite{Dvali:2001dd}, the most favourable model for a violation of the UGR is a time dependent fine-structure constant caused by a time-varying electron charge. Therefore, tests of the UFF for charged matter can be interpreted as UGR tests, too.

To test the validity of the EP is analogue to test the minimal coupling procedure, hence to search for an anomalous coupling of the gravitational field (as an extension of the standard minimal coupling procedure, previously discussed).
%
%
In the non-relativistic regime, the Hamiltonian of a charged particle in a gravitational field is given by
\begin{equation}\label{HamiltonianQWEP}
  H=-\frac{\hbar^2}{2m}\left(\nabla+\frac{iq}{\hbar c}{\bf A}\right)^2 + m U + \kappa q U + \lambda q^2 U\,,
\end{equation}
where $\kappa$ and $\lambda$ are free parameters with dimensions $[\kappa]=[mass/charge]$ and $[\lambda]=[mass/charge^2]$ respectively. In the Hamiltonian (\ref{HamiltonianQWEP}), one can define an effective mass $m_{eff} = (m+\kappa q) U$ that can be interpreted as the charge dependence of the gravitational mass.
Since the charge of a particle is related to space–time symmetries through the CPT theorem, the problem of violation of  EP for charged particles assumes a particular interest for anomalous charge couplings.

The stability of an non-pointlike electron requires an effective dependence on the square of the electron charge \cite{rohrlich2000}. Furthermore, the generalized Maxwell equations, in general, violate the UFF in a way in which appears once more  the square of the charge \cite{Ni:1977zz}. According to these considerations,  it makes sense to take into account
a general model  having charge–dependent inertial and gravitational masses \cite{Dittus2004}. One can choose the parameters in such a way that neutral systems, made up of bound charged particles, exactly fulfill the UFF while isolated charged particles may violate it. Thus, one can introduce an E\"otv\"os coefficient that depends on the charge of particles, i.e.  $\eta = \eta_0 +\kappa_1\frac{q_1}{m_1} - \kappa_2\frac{q_2}{m_2}$. Here only the linear charge dependence is considered and   $\eta_0$ indicates the ordinary E\"otv\"os parameter for the masses. These considerations then suggest a comparison between the free fall of a charged and a neutral particle described by $\eta=\eta_0+\kappa \frac{q}{m}$. By shielding all electromagnetic fields, neutral and charged particles, without internal structure, must fall following the same path. Let us note that, in this framework, experiments in space seem to be favoured in order to reduce the disturbances induced by the stray fields \cite{Dittus2004}.


\subsection{Equivalence principle violation via quantum field theory}

In this Section we discuss the EP violation in a QFT and GR framework~\cite{Donoghue1984a,Donoghue1984b} (for modified gravity, see for example Refs.~\cite{Hui2009,ArmendarizPicon2011} and Ref.\cite{Scardigli2016} for the generalized uncertainty principle). The system consists of an electron with mass $m_{0}$ (the renormalized mass of the particle when the temperature is zero) in thermal equilibrium with a photon heat bath. The aim of the
analysis is the evaluation of electron's gravitational and inertial mass
in the low-temperature limit (namely, $T\ll m_{0}$). The presence of a non-zero temperature is crucial since $m_{g}=m_{i}$ for $T=0$.

The gravitational and inertial masses  are derived by adopting a Foldy\textendash Wouthuysen
transformation~\cite{Foldy1949} on the Dirac equation which allows to derive a Schr\"odinger equation (non-relativistic limit of particles with spin half)
in which the expression for the mass is easily recognizable.

In order to operationally define the inertial mass, one applies an electric field to charged particle and study the consequent acceleration \cite{Donoghue1984b,Donoghue1984a}. One has therefore to evaluate the finite temperature (radiative) corrections to the electromagnetic vertex. After the renormalization procedure and taking into account the finite temperature contributions, one obtains \cite{Donoghue1984b,Donoghue1984a}
\begin{equation}
\left({\slashed p}-m_{0}-\frac{\alpha}{4\pi^{2}} {\slashed I}\right)\psi=e\Gamma_{\mu}A^{\mu}\psi.\label{eq:92}
\end{equation}
Here we ahve used the notation ${\slashed a}\equiv \gamma^\mu a_\mu$, $\alpha$ is the fine-structure constant, $\gamma^{\mu}$
are the Dirac matrices, $A^{\mu}$ is the electromagnetic four-potential, and the quantity $I_\mu$ is defined as
\begin{equation}
I_{\mu}=2\int d^{3}k\frac{n_{B}\left(k\right)}{k_{0}}\frac{k_{\mu}}{\omega_{p}k_{0}-\mathbf{p}\cdot\mathbf{k}},\label{eq:89}
\end{equation}
with $k_{\mu}=\left(k_{0},\mathbf{k}\right)$ and where $\omega_{p}$ and $\mathbf{p}$ are connected by
$\omega_{p}=\sqrt{m_{0}^{2}+|\mathbf{p}|^{2}}$. 
In Eq.~(\ref{eq:89}) $n_{B}(k)$ represents the Bose-Einstein distribution:
\begin{equation}
n_{B}(k)=\frac{1}{e^{\beta k}-1},\label{eq:91}
\end{equation}
where $\beta={1}/{k_{B}T}$, with $k_{B}$ being the Boltzmann constant.
Finally, $\Gamma_{\mu}$ accounts for the finite temperature corrections to the
electromagnetic vertex
\begin{equation}
\Gamma_{\mu}=\gamma_{\mu}\left(1-\frac{\alpha}{4\pi^{2}}\frac{I_{0}}{E}\right)+\frac{\alpha}{4\pi^{2}}I_{\mu}.\label{eq:94}
\end{equation}
Applying the Foldy\textendash Wouthuysen transformation,
Eq.~(\ref{eq:92}) reduces to a Schr\"odinger-like equation
\begin{eqnarray}
i\frac{\partial\psi_{s}}{\partial t} &=& \left[m_{0}+\frac{\alpha\pi T^{2}}{3m_{0}}+\frac{|\mathbf{p}|^{2}}{2\left(m_{0}+\frac{\alpha\pi T^{2}}{3m_{0}}\right)}+e\phi+\frac{\mathbf{p}\cdot\mathbf{A}+\mathbf{A}\cdot\mathbf{p}}{2\left(m_{0}+\frac{\alpha\pi T^{2}}{3m_{0}}\right)}+\ldots\right]\psi_{s}\label{eq:95} \\
 & & \nonumber \\
 &=& H\psi_s \nonumber
\end{eqnarray}
To identify the inertial mass one calculates the acceleration
 \[
 {\bf a}=-[H, [H, {\bf r}]]=\frac{e{\bf E}}{m_0 + \frac{\alpha \pi T^2}{3m}}
 \]
from which one identifies the inertial mass
\begin{equation}
m_{i}=m_{0}+\frac{\alpha\pi T^{2}}{3m_{0}}.\label{eq:96}
\end{equation}
This relation shows that the difference between the
inertial mass of an electron at finite temperature and $m_{0}$ is
due exclusively to the thermal radiative correction of Eq.~(\ref{eq:96}).
The fact that the inertial mass $m_i$ increases with $T$ is expected since it represents the increased inertia needed to travel through the background heat bath.

An analogous procedure can be also performed for the gravitational mass $m_g$.
Calculations of Refs. \cite{Donoghue1984b,Donoghue1984a} rely on the weak field approximation, i.e., to first order in the gravitational field (see Eq. (\ref{gweak})),  and consider the radiative corrections calculated in flat space. The Dirac equation that takes into account the gravitational interaction reads \cite{Donoghue1984a,Donoghue1984b}
\begin{equation}
\left({\slashed p}-m_{0}-\frac{\alpha}{4\pi^{2}} {\slashed I}\right)\psi=\frac{1}{2}h_{\mu\nu}\tau^{\mu\nu}\psi,\label{eq:97}
\end{equation}
where $\tau^{\mu\nu}$ is the renormalized stress-energy
tensor while  $h_{\mu\nu}~=~2\,\phi_g\,\mathrm{diag}\left(1,1,1,1\right)$, with $\phi_g$ the gravitational potential. Once again, a Foldy\textendash Wouthuysen transformation yields the Schr\"odinger-like equation
\begin{eqnarray}
i\frac{\partial\psi_{s}}{\partial t} &=& \left[m_{0}+\frac{\alpha\pi T^{2}}{3m_{0}}+\frac{|\mathbf{p}|^{2}}{2\left(m_{0}+\frac{\alpha\pi T^{2}}{3m_{0}}\right)}+\left(m_{0}-\frac{\alpha\pi T^{2}}{3m_{0}}\right)\phi_{g}\right]\psi_{s},\label{eq:100} \\
 & & \nonumber \\
 &=&  H_g \psi_s \nonumber
\end{eqnarray}
The calculation of the acceleration gives
 \[
 {\bf a}=-[H_g, [H_g, {\bf r}]]=\frac{m_0-\alpha\pi T^2/3m_0}{m_0+\alpha\pi T^2/3m_0}
 \]
 from which the identification of the gravitational mass gives
\begin{equation}
m_{g}=\left(m_{0}-\frac{\alpha\pi T^{2}}{3m_{0}}\right).\label{eq:101}
\end{equation}
Clearly, there is no difference between $m_{g}$ and $m_{i}$ at zero temperature, so that only radiative corrections render the violation
of the equivalence principle feasible. In principle, this result would  yield a violation of the equivalence principle in an E\"otv\"os-type experiment, although at accessible temperatures the effect is small. In fact, from Eqs.~(\ref{eq:96}) and (\ref{eq:101}) one gets
\begin{equation}
\frac{m_{g}}{m_{i}}=1-\frac{2\alpha\pi T^{2}}{3m_{0}^{2}},\label{eq:103}
\end{equation}
in the first-order approximation in $T^{2}$. At temperature of the order $T\sim 300$K the corrections is $\sim 10^{-17}$. We point once more that these results hold in the approximation $T\ll m_e$. Equation (\ref{eq:103}) is a direct consequence of the fact that Lorentz invariance of the finite temperature vacuum is broken, which means that it is possible to define an absolute motion
through the vacuum (i.e. the one at rest with the heat bath). 
The case of gravitational coupling of leptons in a medium  has been studies in \cite{Mitra:2001ar,Nieves:1998xz}.

\subsection{Equivalence principle violation via modified geodesic equation}

Let us now discuss a different method, proposed in \cite{Gasperini1987}, that reproduces the previous results, in particular Eq.~(\ref{eq:103}).

The starting point is the analysis of a charged test particle of renormalized mass at zero temperature $m_{0}$ in
thermal equilibrium with a photon heat bath in the low-temperature
limit $T\ll m_{0}$. The dispersion relation reads ~\cite{Donoghue1984a}
\begin{equation}
E=\sqrt{m_{0}^{2}+|\mathbf{p}|^{2}+\frac{2}{3}\alpha\pi T^{2}},\label{eq:104}
\end{equation}
which can be easily identified with the first-order correction in
$T^{2}$ that descends from the finite temperature analysis. The stress-energy tensor $T^{\mu\nu}$
related to the test particle, whose world line can be contained in
a narrow ``world tube'' in which $T^{\mu\nu}$ is non-vanishing. The conservation equation
for the stress-energy tensor can be integrated over a three-dimensional
hyper-surface $\Sigma$ and defined as:
\begin{equation}
\int_{\Sigma}d^{3}x'\sqrt{-g}T^{\mu\nu}\left(x'\right)=\frac{p^{\mu}p^{\nu}}{E},\label{eq:105}
\end{equation}
where $p^{\mu}$ is the four-momentum and $E=p^{0}$ the energy, given by
$E=\int_{\Sigma}d^{3}x'\sqrt{-g}T^{00}\left(x'\right).$
%
%
These equations hold in the limit where the world tube radius goes to zero~\cite{Papapetrou1951}.

As shown in  ~\cite{Donoghue1984a}, the source of gravity, at finite temperature and in weak-field approximation, turns out to be (in the
rest frame of the heat bath)
\begin{equation}
\Xi^{\mu\nu}=T^{\mu\nu}-\frac{2}{3}\alpha\pi\frac{T^{2}}{E^{2}}\delta_{\;\;0}^{\mu}\delta_{\;\;0}^{\nu}T^{00},\label{eq:107}
\end{equation}
where $\Xi^{\mu\nu}$ contains not only the information on the Einstein tensor $G^{\mu\nu}$, but also thermal corrections to it\footnote{Eq. (\ref{eq:107}) is explicitly derived after the choice of the privileged reference frame at rest with the heat bath. The latter give rises to
a Lorentz invariance violation of the finite temperature vacuum. In fact, in the tangent space (flat space), one cannot consider a  Minkowski vacuum anymore owing to the fact that it is replaced by a thermal bath. As a consequence, Lorentz group is no longer the symmetry group of the local tangent space to the Riemannian manifold, even though general covariance still holds there. According to this, one can proceed keeping in mind that the situation under investigation is slightly different from the usual GR scheme \cite{Gasperini1987}.}

The generalization of (\ref{eq:107}) to a curved space-time is \cite{Gasperini1987}
\begin{equation}
\Xi^{\mu\nu}=T^{\mu\nu}-\frac{2}{3}\alpha\pi\frac{T^{2}}{E^{2}}e_{\;\;\hat{0}}^{\mu}e_{\;\;\hat{0}}^{\nu}T^{\hat{0}\hat{0}},\label{eq:108}
\end{equation}
where $e_{\;\;\hat{0}}^{\mu}$ denotes the vierbein field and the hatted indexes are the ones related to the flat tangent space. The Einstein field equations are hence given by $G^{\mu\nu}=\Xi^{\mu\nu}$.
The Bianchi identity $\nabla_{\nu}G^{\mu\nu}=0$ implies
\begin{equation}
\nabla_{\nu}T^{\mu\nu}=\nabla_{\nu}\left(\frac{2}{3}
\alpha\pi\frac{T^{2}}{E^{2}}e_{\;\;\hat{0}}^{\mu}e_{\;\;\hat{0}}^{\nu}T^{\hat{0}\hat{0}}\right),
\label{eq:110}
\end{equation}
so that,  using $\overset{.}{x}^\mu\equiv dx^\mu/ds$ and $E=m\overset{.}{x}^{\hat{0}}=m\overset{.}{x}^{\rho}e_{\rho}^{\;\;\hat{0}}$
one gets \cite{Gasperini1987}
\begin{equation}\label{eq:116}
\overset{..}{x}^{\mu}+\Gamma_{\alpha\nu}^{\;\;\;\;\mu}\overset{.}{x}^{\alpha}\overset{.}{x}^{\nu}=\frac{2}{3}\alpha\pi T^{2}\left[\frac{\overset{.}{x}^{\nu}\partial_{\nu}e_{\;\;\hat{0}}^{\mu}}{mE}-\frac{e_{\;\;\hat{0}}^{\mu}
\left(\overset{..}{x}^{\nu}e_{\nu}^{\;\;\hat{0}}+\overset{.}{x}^{\nu}\overset{.}{x}^{\beta}\partial_{\beta}e_{\nu}^{\;\;\hat{0}}\right)}{E^{2}}+
\frac{\Gamma_{\alpha\nu}^{\;\;\;\;\mu}e_{\;\;\hat{0}}^{\alpha}e_{\;\;\hat{0}}^{\nu}}{m^{2}}\right].
\end{equation}
Eq.~(\ref{eq:116}) represents a generalization of the geodesic equation to the case in which the temperature is non-vanishing.

\subsection{Application to the Schwarzschild metric}

We now analyze Eq.~(\ref{eq:116}) for the Schwarzschild geometry. The metric tensor is given by
\begin{equation}
g_{\mu\nu}=\mathrm{diag}\left(e^\nu,-e^\lambda,-r^2,-r^2\mathrm{sin}^2\theta\right),
\label{eq:117}
\end{equation}
where
\[
\qquad e^{\nu}=e^{-\lambda}=1-2\phi=1-\frac{2M}{r}.
\]

For our purpose, we shall consider only radial motion ($\overset{.}{\vartheta}=\overset{.}{\varphi}=0$).
The non-vanishing vierbeins for the Schwarzschild metric are $e_{\;\;\hat{0}}^{0}=e^{-\frac{\nu}{2}}$, $e_{\;\;\hat{1}}^{1}=e^{-\frac{\lambda}{2}}$.

The geodesic equation for $\mu=0$ is (here ${}^\prime\equiv \partial/\partial r$)
\begin{equation}
\overset{..}{t}+\nu'\overset{.}{r}\overset{.}{t}=-\frac{2}{3}\alpha\pi T^{2}\left[\frac{\overset{.}{r}\nu'}{2mE}+\frac{\overset{..}{t}+\frac{\overset{.}{r}\overset{.}{t}\nu'}{2}}{E^{2}}e^{\frac{\nu}{2}}\right]e^{-\frac{\nu}{2}},\label{eq:122}
\end{equation}
and since $\frac{E}{m}=\overset{.}{x}^{\hat{0}}=\overset{.}{x}^{\alpha}e_{\alpha}^{\;\;\hat{0}}=\overset{.}{t}\,e^{{\nu}/{2}}$, Eq.~(\ref{eq:122}) can be cast in the form
\begin{equation}\label{eq:125}
\left(1+\frac{2\alpha\pi T^{2}}{3E^{2}}\right)\left(\overset{..}{t}+\overset{.}{\nu}\overset{.}{t}\right)=0.
\end{equation}
The radial contribution can be computed involving Eq.~(\ref{eq:116}) for $\mu=1$
\begin{equation}\label{eq:127}
\overset{..}{r}+\frac{\nu'}{2}\left(\overset{.}{t}^{2}e^{\nu-\lambda}-\overset{.}{r}^{2}-\frac{2\alpha\pi T^{2}}{3m^{2}}e^{-\lambda}\right)=0.
\end{equation}
An integration  of Eq. (\ref{eq:127}) gives \cite{Gasperini1987}
\begin{equation}
e^{\lambda}\overset{.}{r}^{2}-e^{\nu}\overset{.}{t}^{2}-\frac{2\alpha\pi T^{2}}{3m^{2}}\nu=\mathrm{const}.\label{eq:136}
\end{equation}
The constant is determined from the condition of normalization of the 4-velocity $\overset{.}{x}^{\mu}\overset{.}{x}_{\mu}=-1$ , i.e.
\begin{equation}\label{eq:138}
e^{\lambda}\overset{.}{r}^{2}-e^{\nu}\overset{.}{t}^{2}=-1\,.
\end{equation}
In the limit of vanishing gravitational field (namely, $\nu,\lambda\rightarrow 0$ as $r\rightarrow\infty$),
Eq.~(\ref{eq:138}) reduces to $\overset{.}{r}_{\infty}^{2}-\overset{.}{t}_{\infty}^{2}=-1$, which 
compared with Eq.~(\ref{eq:136}), implies
\begin{equation}
e^{\lambda}\overset{.}{r}^{2}-e^{\nu}\overset{.}{t}^{2}-\frac{2\alpha\pi T^{2}}{3m^{2}}\nu=-1.\label{eq:140}
\end{equation}
In the weak-field approximation and owing to Eq.~(\ref{eq:140}), it is immediate to find that Eq.~(\ref{eq:127}) results modified as:
\begin{equation}\label{eq:145}
\overset{..}{r}=-\frac{M}{r^{2}}\left(1-\frac{2\alpha\pi T^{2}}{3m^{2}}\right)\,.
\end{equation}
To first-order approximation in $T^{2}$, as in the previous QFT treatment, one obtains
\begin{equation}\nonumber
\frac{m_{g}}{m_{i}}=1-\frac{2\alpha\pi T^{2}}{3m_{0}^{2}},\label{eq:146}
\end{equation}
which is exactly Eq.~(\ref{eq:103}).

\subsection{Application to the Brans-Dicke metric}

In the case of the Brans-Dicke action, the action reads
\begin{equation}
S_{BD}=\int d^4x\sqrt{-g}\left(\varphi R-\omega\frac{1}{\varphi}g^{\mu\nu}\partial_{\mu}\varphi\partial_{\nu}\varphi+\mathfrak{L}_\mathrm{matter}\left(\psi\right)\right).\label{eq:50}
\end{equation}
where
\begin{equation}
\varphi=\frac{1}{16\pi G_{eff}},\label{eq:51}
\end{equation}
and such a result is traduced in the introduction of a new ``effective''
gravitational constant that has to be identified with the scalar field.
Here one assumes that $\varphi$ is spatially uniform, and it must vary slowly
with cosmic time (this is consistent with experimental data)

Field equations derived from Eq.~(\ref{eq:50}) are
\begin{equation}
2\varphi G_{\mu\nu}=T_{\mu\nu}+T_{\mu\nu}^{\varphi}-2\left(g_{\mu\nu}\nabla_{\mu}\nabla_{\nu}\right)\varphi,\label{eq:54}
\end{equation}
and
\begin{equation}
\Box \varphi=\zeta^{2}T,\label{eq:55}
\end{equation}
where $\zeta^{-2}=6+4\,\omega$
and $T=g^{\mu\nu}T_{\mu\nu}$. The symbol $\Box$ denotes
the usual D'Alembert operator.
In Eq.~(\ref{eq:54}), $T_{\mu\nu}$ and $T_{\mu\nu}^{\varphi}$ are extracted
by varying $\mathfrak{L}_{matter}$ and the kinetic term of $S_{BD}$, respectively.
As expected, field equations for the metric tensor becomes the ones derived by GR in the limit $\varphi=\mathrm{const}={1}/{16\pi G}$.

The field equations admit a static and isotropic solution so that the line element is:
\begin{equation}
ds^{2}=e^{v}dt^{2}-e^{u}\left[dr^{2}+r^{2}\left(d\vartheta^{2}+sin^{2}\vartheta d\Phi^{2}\right)\right],\label{eq:eff57}
\end{equation}
with
\begin{equation}
e^{v}=e^{2\alpha_{0}}\left(\frac{1-\frac{B}{r}}{1+\frac{B}{r}}\right)^{\frac{2}{\lambda}}, \qquad e^{u}=e^{2\beta_{0}}\left(1+\frac{B}{r}\right)^{4}\left(\frac{1-\frac{B}{r}}{1+\frac{B}{r}}\right)^{\frac{2\left(\lambda-C-1\right)}{\lambda}},\label{eq:eff58}
\end{equation}
with $\alpha_{0}$, $\beta_{0}$, $B$, $C$ and $\lambda$ being constants
that can be connected to the free parameter of the theory $\omega$.
Since it is a scalar-tensor theory, a solution for $\varphi$ must also
be found; in the considered case, the outcome turns out to be
\begin{equation}
\varphi=\varphi_{0}\left(\frac{1-\frac{B}{r}}{1+\frac{B}{r}}\right)^{-\frac{C}{\lambda}},\label{eq:eff61}
\end{equation}
where $\varphi_{0}$ is another constant.

Repeating the previous analysis leading to (\ref{eq:145}), for BD theory one gets
\begin{equation}
\overset{..}{r}=-\frac{v'}{2}\left\{ 1+\left(e^{-v}-1\right)\left(\frac{\lambda B}{r}-C\right)-\frac{2\alpha\pi T^{2}}{3m^{2}}\left[1+v-\left(\frac{\lambda B}{r}-C\right)v\right]\right\} e^{-u}.\label{eq:eff279}
\end{equation}

From Eq.~(\ref{eq:eff279}), one observes that there is
not only the radiative correction to the ratio ${m_{g}}/{m_{i}}$,
but also another contribution which exclusively depends on $\omega$ and that
correctly vanishes in the limit $\omega\rightarrow\infty$, that is when GR is recovered.
The evaluation of the second quantity of Eq.~(\ref{eq:eff279})
allows to put a lower bound to the parameter
of the Brans-Dicke theory. In fact, in the weak field regime, imposing  $|({m_{g}-m_{i}})/{m_{i}}|<10^{-14}$~\cite{Baessler1999} and using \cite{Barros1997}
\begin{equation}
\alpha_{0}=\beta_{0}=0;\;\;C=-\frac{1}{2+\omega};\;\;B=\frac{GM\lambda}{2};\;\;\lambda=\sqrt{\frac{2\omega+3}{2\omega+4}}.\label{eq:299}
\end{equation}
one infers
\begin{equation}
\omega>\frac{2GM}{r}\cdot10^{14},\label{eq:305}
\end{equation}
which is the final expression for the lower bound of the
Brans-Dicke parameter  in the weak-field approximation.
For the Earth  $M_{\oplus}=5.97\cdot10^{24}\;Kg;\;\;\;R_{\oplus}=6.37\cdot10^{6}\;m.$
so that \cite{Blasone2018b}
\begin{equation}
\omega>1.40\cdot10^{5},\label{eq:308}
\end{equation}
that is similar to a bound recently obtained~\cite{Will2005}, which gives $\omega>3\cdot10^{5}$.
For the sake of completeness, it is useful to look at a table that
contains a prediction of the most reliable bounds for $\omega$~\cite{Arun2013}.

\begin{table}[ht]
\begin{center}
\caption{This table includes expected bounds on the parameter $\omega$ from different experiments (see \cite{Arun2013} and references therein).}
\begin{tabular}{|c|c|c|}
\hline
Detector & System & Expected bound on $\omega$\tabularnewline
\hline
\hline
aLIGO & $\left(1.4+5\right)M_{\odot}$ & $\sim100$\tabularnewline
\hline
Einstein Telescope & $\left(1.4+5\right)M_{\odot}$ & $\sim10^{5}$ \tabularnewline
\hline
Einstein Telescope & $\left(1.4+2\right)M_{\odot}$ & $\sim10^{4}$ \tabularnewline
\hline
eLISA & $\left(1.4+400\right)M_{\odot}$ & $\sim10^{4}$ \tabularnewline
\hline
LISA & $\left(1.4+400\right)M_{\odot}$ & $\sim10^{5}$ \tabularnewline
\hline
DECIGO & $\left(1.4+10\right)M_{\odot}$ & $\sim10^{6}$ \tabularnewline
\hline
Cassini & Solar System &   $\sim10^{4}$\tabularnewline
\hline
\end{tabular}
\end{center}
\end{table}

\subsection{Standard Model Extensions  and the Weak Equivalence Principle}

As  stated before, the EEP asserts that in any local Lorentz frame about any point in spacetime, the laws of physics are described by the special relativity (including the standard model of particle physics) \cite{Misner1974}.

As widely believed, general relativity and the standard model can be considered as the low energy limit of some fundamental theory of physics at high energy
scales, that, in turn, might give rise to violations of EEP at some scale \cite{Bell1996,Kostelecky1988,Colladay1998}, although its exact form is not well defined. In this framework, the standard model extension (SME) \cite{Colladay1998} represent a flexible and widely applied \cite{Kostelecky2008} context for describing violations of EEP. The SME is an effective field theory that extend the standard model action by adding new terms that break local Lorentz invariance and other tenets of EEP \cite{Kostelecky2010}. In this model, the energy conservation, gauge invariance, and general covariance are preserved. As in other models \cite{Bell1996}, EEP
violation in the SME can manifest in different ways (for example, it may be strongly suppressed in normal matter relative to antimatter \cite{Kostelecky2010}).

In the framework of SME, in Ref. \cite{Hohensee2013} the authors show that EEP violation in antimatter can be constrained by means of tests in which  bound systems of normal matter are used. More specifically, an anomaly that violates the WEP for free particles generates anomalous gravitational redshifts in the energy of systems in which they are bound. For a nuclear shell model one can estimate the sensitivity of a variety of atomic nuclei to EEP violation for matter and antimatter.

Focusing on conventional matter (made up of protons,
neutrons, and electrons), the spin-independent violations of EEP in the SME acting
on a test particle of mass $m^w$ are described by the action \cite{Kostelecky2010} (see also \cite{Kostelecky2008,Hohensee2011})
\begin{equation}\label{ActionSME}
S= -\int \left[ \frac{m^w  \sqrt{[g_{\mu\nu}-2({\tilde c}^w)_{\mu\nu}]dx^\mu dx^\nu}}{1+\frac{5}{3}({\tilde c}^w)_{00}}+
({\tilde a}^w)_{\mu} dx^\mu\right] \,,
\end{equation}
where the superscript $w= p, n, e$ (for proton, neutron,
or electron) indicates the type of particle in question, $g_{\mu\nu}$ is
the metric tensor, $dx^\mu$ is the interval between two points in
spacetime. The $({\tilde c}^w)_{\mu\nu}$ tensor describes a fixed background field that modifies the effective
metric that the particle experiences, and thus, its
inertial mass relative to its gravitational mass. The four vector
$({\tilde a}^w_{eff})_{\mu}=((1-\alpha U)({\tilde a}^w_{eff})_{0},({\tilde a}^w_{eff})_{j})$, where $U$ is the
Newtonian potential, represents the particles coupling to a field with a nonmetric interaction $\alpha$ with gravity. As
$({\tilde a}^w_{eff})_{\mu}$  is CPT odd \cite{Colladay1998}, this term enters with opposite sign in the action for an antiparticle ${\tilde w}$. Both
$({\tilde c}^w)_{\mu\nu}$ and $({\tilde a}^w)_{\mu}$  vanish
if general relativity is valid. For convenience, Eq. (\ref{ActionSME})
includes an unobservable scaling of the particle mass by
$1+\frac{5}{3}({\tilde c}^w)_{00}$.
Consider the isotropic subset of the model \cite{Kostelecky2010}, i.e.
$({\tilde c}_{w})_{\mu\nu}$ is diagonal and traceless, and the spatial terms in the vector $({\tilde a}^w_{eff})_{\mu}$ vanish.
In the nonrelativistic, Newtonian limit, the
single particle Hamiltonian produced by the action (\ref{ActionSME}) is
given by
\begin{equation}\label{HSME}
  H=\frac{1}{2}m^w v^2-m_g^w U
\end{equation}
where the effective gravitational mass $m_g^w$ is given by
 \begin{equation}\label{gravmassSME}
   m_g^w=m^w\left[1-\frac{2}{3}({\tilde c}^w)_{00} +\frac{2\alpha}{m^w}({\tilde a}^w_{eff})_0\right]\,.
 \end{equation}
Experimentally observable EEP violations are proportional to the particles gravitational to inertial mass ratio
\begin{equation}\label{inertialgramratio}
  \frac{m_g^w}{m^w} = 1-\frac{2}{3}({\tilde c}^w)_{00} +\frac{2\alpha}{m^w}({\tilde a}^w_{eff})_0\equiv
  1+\beta^w
\end{equation}
and are described by the parameter $\beta^w$ \cite{Hohensee2011}. From Eq. (\ref{inertialgramratio}), it follows that $({\tilde c}^w)_{00}$ and $({\tilde a}^w_{eff})_0$ are responsible for violations of the WEP, an aspect of EEP \cite{Will2005}, since they produce
particle-dependent rescalings of the effective gravitational potential. In addition, EEP violation is not apparent in the
nonrelativistic motion of a free particle if $\alpha ({\tilde a}^w_{eff})_0=\frac{m^w}{3} ({\tilde c}^w)_{00}$, although it remains manifest in the motion of the antiparticle ${\tilde w}$, for which $\beta^{\tilde w}=-\frac{2\alpha}{m^w}({\tilde a}^w_{eff})_0-\frac{2}{3}({\tilde c}^w)_{00}$, a limit discussed in \cite{Kostelecky2010}. The antimatter anomaly $\beta^{\tilde w}$ does contribute to tests involving
nongravitationally bound systems of matter, owing to the
anomalous gravitational red-shift produced by $({\tilde c}^w)_{00}$ in the
energies of bound systems (for details, see \cite{Hohensee2011}).

\subsection{Strong Equivalence principle in modified theories of gravity}

As we have discussed in the previous Sections, in modified or alternative theories of gravity, General Relativity is generalized including extra degrees of freedoms, such as scalar, vector or tensor fields, higher orders terms in the scalar invariants, and so on \cite{Will1993}.
Typically, in these models the new degree of freedoms couple non-minimally with, referring to Section 2, scalar curvature. More explicitly, this is the case of the Brans-Dicke theory, the prototype of scalar tensor-theories, in which the scalar field $\phi$ couples minimally to scalar curvature $R$, so that the action reads (\ref{eq:50}). The effects of the non-minimal coupling is, in some regime, to generate new (gravitational) interactions among masses, modifying in a different way the values of the perturbations of the metric (weak-field approximation) $h_{00}=-2GM_G/r$, related to the gravitational mass $M_G$, and
$h_{ij}$, related to the inertial mass $M_I$ \cite{Ohanian1971,Ohanian1973}, i.e. $M_I=\frac{1}{16\pi G}\int d^4 x \sqrt{-g} (h^i_{i, j}-h^i_{j, i})dS^j$. In General Relativity, since $h_{ij}=-2GM_G/r$ ($h_{00}=h_{ij}$, one gets $M_I=M_G$, while in Brans-Dicke theory (and hence in more general theories of gravity), since $h_{00}\neq h_{ij}$ (weak-field limit of (\ref{eq:eff58})), one gets $M_G=M_I+f(\omega, E_\phi)$, where $f(\omega, E_\phi)$ depends on the parameter $\omega$ and the self-energy of the scalar field $E_\phi$ \cite{Ohanian1971,Ohanian1973}.

\section{Experimental tests of the weak equivalence principle
\label{sec:TestsWEP}}
The WEP has been experimentally verified to remarkable accuracy. This is made possible by the fact that the universality of free fall (UFF) can be tested in null experiments, as the physical quantity of interest is the
relative acceleration between two freely falling proof masses. If the gravitational mass $m_g$ of a body differs from its inertial
mass $m_I$, the acceleration $\mathbf{a}$ of the body in a gravitational
field $\mathbf{g}$ is given by $\mathbf{a}= (m_g/m_I)\mathbf{g}$.
Experiments determine upper limits to the differential acceleration $|a_1-a_2|$ between two freely falling test
masses of different composition. Possible violations of WEP are then quantified by the  E\"{o}tv\"{o}s parameter 
defined in Eq. (\ref{Intro8})
\begin{equation}
\eta=2\left |\frac{a_1-a_2}{a_1+a_2}\right |.
\end{equation}
Tests with increasing accuracy correspond to decreasing upper limits on $\eta$. 
As long as UFF is valid, the differential acceleration and thus $\eta$ must be null within experimental uncertainties. As for any null experiment, no specific model is required to obtain the physical quantity of
interest to be compared with the measured signal.

Various kinds of null experiments are possible to test WEP, differing in the magnitude of the potential signal and in the impact of noise sources and systematic effects. In the following of this paper we describe past, ongoing and future  WEP test experiments by grouping them into three main classes. Section \ref{sec:macroscopic} describes experiments in which the test masses are macroscopic bodies. In section \ref{sec:LLR} we present UFF tests by the observation of celestial bodies and their movement with
respect to each other. In section \ref{sec:atoms} we discuss experiments with microscopic test masses, i.e. atoms, molecules, and elementary particles. WEP tests can be also classified according to different criteria. Sections \ref{sec:macroscopic} and \ref{sec:atoms} include ground laboratory tests as well as experiments in space. Macroscopic proof masses in ground experiments can be either suspended or left in free fall.  
The differences between experimental classes are discussed in the following sections.

It is worth mentioning that other experiments, that strictly speaking cannot be considered as tests of WEP, deeply rely on it for their validity. Relevant examples are the measurement of the Newtonian gravitational constant $G$ performed with freely falling samples \cite{Schwarz1998,Rosi2014} or the comparison of different gravimeters for metrological purposes \cite{Francis2015}.


\subsection{Lab experiments with macroscopic masses}
\label{sec:macroscopic}

Laboratory WEP tests based on macroscopic masses 
are either performed with freely falling masses or with suspended masses. The latter class of experiments compares the acceleration experienced by two masses of different composition as they fall in the gravity field of the Earth. In this case, the signal to be detected, namely a non-zero differential acceleration resulting from a WEP violation, is maximum as it is proportional, via the E\"{o}tv\"{o}s parameter, to the full gravitational acceleration of the Earth. Unfortunately, the typical free fall time on Earth cannot be longer than a few seconds to keep the height of the instrument within a reasonable size. This imposes a major limitation to the measurement sensitivity. In addition, free-fall experiments are very much dependent from the initial conditions (position and velocity) of the test masses as they are released and therefore to external perturbations acting on the instrument.

Experiments with suspended masses are done, with a few exceptions, using a torsion balance, with test masses of different composition suspended at the opposite ends of the beam. When the beam is oriented along the East-West direction, the differential acceleration responsible for a WEP violation is proportional to the centrifugal acceleration, which provides a driving signal for a WEP violation about three orders of magnitude smaller than in a free-fall experiments. Despite the lower signal, torsion balances are today providing the best laboratory tests of the Weak Equivalence Principle due to the long measurement time at equilibrium and to the excellent control of systematic effects that they can offer by spinning the instrument around its axis. 

Experiments with suspended and freely falling macroscopic masses are described in the next sections.

\subsubsection{Tests with suspended masses \label{sec:balances}}

As a cornerstone of mechanical theories, WEP has been experimentally investigated since the dawn of modern age. First experimental tests of the UFF date back to the early 1600's, when Galileo Galilei compared the oscillation periods of two simple pendulums with different composition \cite{Galilei1638}. Considering that the
two masses are in free fall along the tangent to the trajectory
of their respective oscillation, Galilei managed to test the UFF with an accuracy at the $10^{-3}$ level \cite{Fuligni1993}. Newton repeated the experiment to test the equivalence of inertial and gravitational mass with similar precision \cite{Newton1687}, and two centuries later Bessel improved it to an accuracy of $2\times10^{-5}$ \cite{Bessel1832} with a more precise determination of the pendulum length, and by comparing many different materials including gold, silver, lead,  quartz, marble, clay, loadstone, water. Pendulum WEP tests were also performed on radioactive materials in the early XX century:  Thomson \cite{Thomson1909} reached a $5\times10^{-4}$ precision for radon, and  Southerns \cite{Southerns1910}, achieved a $5\times10^{-6}$ precision for uranium oxide. Further evolution of this method led to the remarkable precision of $3\times10^{-6}$ in the experiment of H. H. Potter in 1923 \cite{Potter1923}.
Simple pendulum experiments are intrinsecally limited by the large impact of dissipative damping forces from suspension and from air, as well as by geometrical asymmetries between the two pendulums to be compared. Moreover, with increasing precision the anharmonic terms of the pendulum dynamics
become relevant, and the period depends on the amplitude
which then has to be controlled with high precision. 

A breakthrough occurred in the late 19th century, due to the intuition of E\"{o}tv\"{o}s to employ a Cavendish torsion balance (more precisely, Boy's modification) to compare inertial and gravitational mass in a null experiment. E\"{o}tv\"{o}s'
first series \cite{Eotvos1890}, published in 1890, reached a precision of  $5\times10^{-8}$. Two decades later, with D. Pekar and E. Fekete, E\"{o}tv\"{o}s improved it to  $3\times10^{-9}$ \cite{Eotvos1909}. 
In E\"{o}tv\"{o}s' experiments, the inertial acceleration is given by the centrifugal
force due to the Earth's rotation, while the gravitational acceleration is
the component of $g$ necessary to compensate it. Two masses of different composition are suspended at opposite ends of the torsion balance beam; the centrifugal forces on
the two weights due to the Earth's rotation are balanced against a component
of the Earth's gravitational field. A  WEP violation would produce a rotation of the torsion
balance: if the ratio of passive gravitational mass to
inertial mass should differ from one test mass to the other, there would be a torque
tending to twist the torsion balance. 
When the beam is oriented along the East-West direction, the differential acceleration responsible for a WEP violation is proportional to the centrifugal acceleration $a_c=\Omega^2_{\oplus}R_{\oplus}\cos{\theta}\sin{\theta}$, where $R_{\oplus}$ and $=\Omega_{\oplus}$ are the radius of our planet and the angular velocity of its rotation motion, and $\theta$ is the latitude at the instrument location. For $\theta=\pi/4$, the centrifugal acceleration amounts to $16.8$\,mm/s$^2$ providing a driving signal for a WEP violation about $600$ times smaller than in a free-fall experiments. 
One major limitation is given by the
fact that a potential violation would produce a static (DC) signal.
The effect of a non-zero signal can be detected by exchanging
the position of the two masses, so that the sense of the twist from WEP violation would be reversed.

Improved versions of E\"{o}tv\"{o}s' experiment were designed to produce an AC signal from potential WEP violations, by spinning the instrument around its axis. The spin motion introduces a modulation of the WEP violating signal without intervening on the balance configuration \cite{Roll1964,Braginskii1971}. This is achieved mainly in two different ways: by locking the torsion balance on the gravitational field of Sun at
equilibrium with the inertia of the Earth that rotates
around it, so that the violation signal is modulated by the Earth's
spin with a 24\,h period; or by actively rotating the torsion balance around the suspension wire to up-convert possible violation signals from DC to the rotation frequency. Main advantage of the former method is the natural modulation of the potential signal without potential systematics and technical noises from active mechanical rotation of the apparatus. The main advantage of the latter method is the higher modulation frequency of the signal, allowing to remove many low-frequency noise sources and systematics.  In particular, mechanical losses due to internal damping are lower at higher frequencies, and up-conversion brings the signal to a region of reduced thermal noise. Combinations of different up-conversions have also been designed, e.g. with rotating torsion balances in the field of the Sun.  

Earth rotation offers a natural platform for spinning a torsion balance with daily period against the Sun. 
If the beam is aligned with the north-south
direction, a WEP-violating differential acceleration would produce a maximum torque when the Sun is at the astronomical horizon. The horizontal component of the gravitational acceleration toward
the Sun is at most  $6$\,mm/s$^2$. Thus the signal for UFF tests in the gravitational field of the Sun is smaller than for tests in the Earth's field by about a factor $3/8$.
Dicke's torsion balance experiment provided the first UFF test in
the field of the Sun \cite{Roll1964}, reaching  $10^{-11}$, followed by Braginsky and
Panov down to  $10^{-12}$.
The latter experiment provided the best
estimate of the E\"{o}tv\"{o}s parameter for nearly 30 years. A variant of the torsion balance is obtained by replacing the suspension wire with a so-called fluid fiber, introduced by Keiser and Faller at the end of the 1970s \cite{Keiser1979}. In this kind of setup, test
masses made of hollow metal bodies float on fluids, and their position is controlled by an
electrostatic system.  A potential WEP violation, inducing a differential acceleration between the solid and the liquid, is measured on the control
signal needed to keep the test masses in constant position. WEP tests with fluid fibers were performed in 1979
with an accuracy of $10^{-10}$ \cite{Keiser1979} and in 1982 with an accuracy of
$4\times10^{-11}$ \cite{Keiser1982}. The potential accuracy of such method was estimated at levels between $10^{-13}$ and $10^{-14}$ \cite{Faller1983}.
A similar experiment was performed by Thieberger in 1986
by observing the horizontal drift of a hollow copper sphere floating
freely in water \cite{Thieberger1987}. The driving horizontal gravitational force was generated by placing the setup near a steep cliff; a differential acceleration between copper and water would result in a drift velocity of the sphere. Indeed Thieberger measured a net differential acceleration,
indicating a potential WEP violation with $\eta<1.3\times 10^{-11}$ arising from a fifth force. 
A comparative experiment of the same kind was performed with a more symmetric setup by Bizzeti et al. in 1989 \cite{Bizzeti1989}; no WEP violation was found, up to an accuracy of $2.4\times10^{-12}$.

A disadvantage with the Sun as source is a weaker driving signal as compared to that in the field of the Earth. Spinning a torsion balance by means of a uniformly rotating turntable allows the Earth to be used as the attractor \cite{Wagner2012}. Moreover, driving force modulation can be kept at higher frequencies, reducing the thermal noise \cite{Saulson1990} and disentangling the WEP violating signal from other effects, e.g. temperature variations,  that naturally occur at the diurnal frequency,
including  thermal effects, microseismicity, local mass motions. Such effects originate from the Sun  through  radiative heating of the Earth's surface and atmosphere,  with a typical thermal time delay, rather than by gravitational interaction. Though Earth tidal forces have no daily periodicity and can be neglected, gravity gradients originating from solar tides occur mostly at twice the diurnal frequency, resulting in a spurious WEP violation of the order $10^{-12}$ for a balance arm of 15\,cm.

Torsion balance experiments provided the best limits on potential WEP violations for ground tests
so far (see \cite{Wagner2012}, Table 3). Such experiments have confirmed UFF both in the field of the
Sun, up to about
$10^{-12}$, and in the field of the Earth, up to about $10^{-13}$, as well as in the field of local source masses, up to about $5\times10^{-12}$.

The most precise torsion balance to date was realised by the so
called E\"{o}t-Wash research group. A first
experiment in 1989 provided a WEP test in the gravitational field of the Earth with  $1\times10^{-11}$ accuracy \cite{Heckel1989}. In the same year, a test
with $5.3\times10^{-12}$ accuracy was done in the field of a local mass distribution by placing the torsion balance near a river lock, resulting in a 12 min periodic modulation of $\sim2\times10^8$\,kg of water with known distribution as an attractor for copper and lead test masses \cite{Bennet1989}. 
The precision was improved to $1.9\times10^{-12}$ in 1994 with test masses made of Beryllium and of an Aluminum/Copper alloy in the  field of the Earth, and later to
$1.2\times10^{-12}$ with Si and Al+Cu test masses in the field of
the Sun. 
In 1999 the E\"{o}t-Wash group measured the differential acceleration
of a Cu test mass toward a Pb attractor to be $a_{\hbox{Cu}}-a_{\hbox{Pb}}<(1.0\pm2.8)\times 10^{-15}$\,m/s$^2$.
Comparing to the corresponding
gravitational acceleration of $9.2\times 10^{-7}$\,m/s$^2$ this leads
to $\eta<10^{-8}$ \cite{Smith1999}. 
An experiment performed in 2001 in the field of the Sun
\cite{Adelberger2001} improved the result of Braginsky and Panov
to  $\eta<10^{-13}$. In 2008 the same group obtained $\eta< (0.3\pm1.8)\times10^{-13}$ with Be and Ti test masses in the field of the Earth \cite{Schlamminger2008}. 
The latter result represents the most accurate WEP test on ground. 

Another recent torsion balance experiment provided a WEP test at the $10^{-13}$ level on chiral masses \cite{Zhu2018}, using a pair of lef-handed and right-handed quartz crystals. 

Current experiments  with torsion balances are mainly limited by systematic effects arising from gravity gradients coupling to geometrical asymmetries in the the torsion pendulum, and thus producing differential directions for the forces on the test bodies. Several environmental parameters can produce effects that mimic a WEP-violating
signal. Tilts of the rotation axis with respect to local vertical, couple the pendulum
to gravity gradients; the same applies to temperature fluctuations, thermal gradients, and magnetic fields. The main bias terms can be subtracted to some extent using the method described in \cite{Su1994, Heckel2008}. For each driving term, the corresponding parameter is modulated with large amplitude to calibrate 
its effect on the WEP-violating signal; calibration factors and measured parameters are combined in post-processing of the actual WEP data to correct for the contribution of bias driving terms. Gravity gradients can be measured with a gradiometer and compensated with a suitable configuration of local source masses. 
Rotating the compensation system by 180$^{\circ}$ about its vertical principal axis doubles the effect of
ambient gradient. This allows to determine the corresponding systematic error on the torsion balance WEP test, which is measured from the ratio of the torsion balance and gradiometer signals in the two compensator positions. 
Additional sources of systematic errors originate from fibre twisting due to residual tilts of the setup or wobbles of the rotary axis in combination with asymmetries in the upper suspension point. Such effects can be corrected by carefully measuring the residual tilt, e.g with a dual-axis tilt sensor placed above the upper attachment of the fibre and beneath the pendulum, and controlling the rotation axis to be along the vertical direction.
Temperature gradients and magnetic effects are usually  mitigated by multi-stage passive shielding.
Changes in the balance spinning frequency $\omega_s$ are another source of systematic errors. A spurious signal would be proportional to the component of the rate variation at the Fourier frequency $\omega_s$. As the corresponding torque scales as $\omega_s^2$, the effect can be measured by operating the torsion pendulum at different spinning frequencies, and can be mitigated by choosing the lowest spinning rate compatible with the technical noise floor.

A modern variant of Galileo's simple pendulum WEP test in the field of the Sun has been proposed recently, \cite{Iafolla2016}. The experiment is based on a differential accelerometer with zero baseline, measuring the relative acceleration of two test masses of different materials suspended on a pendulum. Ensuring a precise centering of the test masses the system should provide a high degree of attenuation of the local
seismic noise. With a cryogenic differential accelerometer under vacuum, the experiment should provide a WEP test with $10^{-14}$ precision.

\subsubsection{Freely falling masses}
\label{sec:droptests}

Unlike in torsion balances, mass drop WEP tests are done by leaving two test masses in free fall at the same time, and measuring their relative displacement as a signature of differential acceleration. This method was never used  in high-precision experiments until the late 1980s. The legendary UFF test by Galileo was indeed never done by dropping masses from the Pisa leaning tower, but rather using pendulums

Free fall experiments in evacuated tubes, which are nowadays popular in science teaching and outreach, date back to the
17th century, when Boyle  performed free fall
tests with feathers or pieces of paper \cite{Boyle1964}. 
Similar experiments with coins and feathers are reported in 1717, when Desaguliers demonstrated the UFF  to King George I and to the Royal Society led by Newton \cite{Desaguliers1734}.
 In 1971 Astronaut Scott performed a UFF test on the Moon during the Apollo 15
mission, by dropping an
Aluminum hammer and a falcon feather from a height
of about 1.6\,m and observing them hit the ground
simultaneously \cite{Allen1972}.

A revival of UFF tests with freely falling test masses occurred during the 1980's, in the attempt to improve E\"{o}tv\"{o}s' WEP tests \cite{Fischbach1986}. After Dicke's and Braginsky's experiments it was clear that substantial progress required a rotation of the torsion balance; however the control of systematic effects in rotating the setup in the laboratory, that was necessary for tests in the field of the Earth, was considered extremely challenging.  On the contrary, the recent progress in laser interferometry ranging made mass drop tests more attractive. Moreover, free fall tests have the advantage of a much higher driving acceleration ($g\simeq9.8$\,m/s$^2$ in mass drop tests versus $g<1.69\cdot10^{ -2}$\,m/s$^2$ on the torsion balance). 
A first precision mass drop WEP test was performed by Worden in 1982  at the $10^{-4}$ uncertainty level, with a test mass constrained to 1D motion  by means of a magnetic bearing \cite{Worden1982}.  
This result was improved in 1984 by Sakuma with an accuracy
of $1\times10^{-8}$ \cite{Sakuma1984}. In 1986, by observing the rotation
of a freely falling disc made of two halves of different materials, Cavasinni et al. confirmed the validity of WEP with $1\times10^{-10}$ accuracy \cite{Cavasinni1986}. 
A high-precision mass drop experiment with two separate test masses
was performed for the first time in 1987 by Niebauer et al. \cite{Niebauer1987}. Measuring the position of freely falling Uranium and Copper test masses with an interferometer, they proved the WEP up to
$5\times10^{-10}$. The same experiment
with different materials was repeated by Kuroda and Mio in 1989, reaching an accuracy of $1\times10^{-10}$ \cite{Kuroda1989, Kuroda1990}.
The disc experiment of \cite{Cavasinni1986} was repeated in 1992 by Carusotto et al. with Aluminum
and Copper, reaching an accuracy of $7.2\times10^{-10}$ \cite{Carusotto1992,Carusotto1996}.

So far, the precision of drop tests was limited to a few parts in $10^{10}$, in spite of the 600-fold larger driving signal strength compared to torsion balances. The improvements over E\"{o}tv\"{o}s' result was limited to one order of magnitude, in contrast to the much more sophisticated technologies employed. More recently this has been overwhelmed by the E\"{o}t-Wash  rotating balance. As discussed in\cite{Niebauer1987,Carusotto1992,Carusotto1996} the main limitations of mass drop experiments are from errors in initial conditions at release coupling to the gravity gradient of the Earth to provide a differential acceleration error that mimics a violation signal. In experiments with separate test masses, a laser interferometer tacks the differential trajectory 
\begin{equation}
  \delta x(t)=\delta x_0+\delta v_0  t(1+\gamma_vt^2/6)+\gamma_ht^2/2
  \label{eq:drop}
\end{equation}
where $\delta X_0$, $\delta v_0$, $\gamma_v$, $\gamma_h$ represent the initial differential displacement, initial differential velocity, vertical gravity gradient, and horizontal gravity gradient.
The effect of vertical gradient is partly removed by fitting the measured trajectory with Eq. \ref{eq:drop}. Further mitigation of the systematic error from vertical velocity differences is obtained by alternating the order in which the objects are dropped. Errors arising from the horizontal gravity  gradient are mitigated by alternating the position of the two masses.
For experiments with disk test masses, a major systematic error arises from the disk precession around its angular momentum. The effect can be partly corrected by measuring the two components of the angular momentum in the disk plane just after the release.

A way to improve drop tests is by increasing the free fall time, since the effect of a violation increases quadratically with time. Experiments on balloons \cite{Iafolla1998} and sounding rockets \cite{Reasenberg2012} have been proposed, which would allow a free fall time of several tens of seconds. The effect of gravity gradients, which would be a major source of systematic errors, can be separated from potential WEP violation signals by spinning the system around a horizontal axis: in such way the signal from gravity gradient appears at twice the rotation frequency while a violation signal would be at the rotation frequency.

In principle there is still more potential in mass drop experiments, especially with a longer free fall time and sensors with a higher resolution.  Similar tests are possible on ground in facilities such as the drop tower of the ZARM center at the University of Bremen, where a free fall time of 4.74\,s is achievable (9.3\,s using the catapult). A mass drop WEP test with $10^{-7}$ accuracy was performed in 2001 with highly sensitive
SQUID sensors \cite{Vodel2002}.  In free fall experiments the control of starting
conditions for positions and velocities of the test masses is crucial. An Electrostatic Positioning System (EPS) was developed to this purpose \cite{Sondag2016}. For optimal
conditions an accuracy of $10^{13}$ is expected with this setup.

Another free fall experiment is the project Principle Of
the Equivalence Measurement (POEM) at the Harvard-
Smithsonian Center for Astrophysics \cite{Reasenberg2001, Reasenberg2007}. Two test masses in 0.5\,m distance
in a co-moving vacuum chamber were bounced on a kind
of trampoline 0.9\,m up and down several times. To test
the WEP, the shifting between the test masses is measured.
In principle, with a time average over several bounces a sensitivity of
$5\times10^{-13}$ can be reached and an improvement on ground
up to $1\times10^{-14}$ is possible.

Free fall experiments with microscopic test masses enable in principle a much better control over systematic effects. They are discussed in section \ref{sec:atoms}.

\subsubsection{Experiments with macroscopic masses in space}

Testing the Weak Equivalence Principle in a ground based laboratory has some obvious limitations that can be overcome by going to space.

As discussed in the previous sections, experiments with freely falling masses exploit the full gravitational acceleration of the Earth to maximize the strength of a WEP violation; unfortunately, their sensitivity is hampered by the short measurement time achievable in a ground-based laboratory and their accuracy is limited by the poor control of systematic effects depending on the initial conditions (position and velocity) of the test masses. On the contrary, experiments based on masses suspended on a torsion balance allow for a long integration time and provide a much better control of systematic effects, but their sensitivity is limited by the driving gravitational signal being three orders of magnitude lower than the Earth gravity acceleration.

The next evolutionary step of these instruments is clearly space. The laboratory inside a freely falling spacecraft is indeed the ideal environment to push WEP tests to their ultimate limits.

Under weightlessness conditions, the classical free fall experiment can still benefit from a WEP violating signal proportional to the local acceleration of gravity (at the spacecraft height). More importantly, the experiment can now be executed in a compact apparatus, with the test masses still in the spacecraft reference frame, where their relative motion can be observed over a long and unperturbed measurement time. The reduced volume of the instrument, compared to the much larger drop towers or free-fall capsules used on Earth, allows to better control the experiment against external perturbations. 

A torsion balance in space would as well benefit from the full gravitational acceleration as a driving signal, gaining almost three orders of magnitude on the effect to be measured with respect to an Earth-based experiment having the same sensitivity to differential accelerations.

Concentric test masses are the common denominator of all the instruments proposed for a space test of the Weak Equivalence Principle. This is the case for the space missions that will be described in this section, MICROSCOPE, STEP and GG, which is the natural evolution of a torsion balance for space. Such a design is possible in space because of the extremely small coupling forces needed to control the masses position under weightlessness conditions. Small coupling constants directly translate into higher instrument sensitivity to differential accelerations and therefore to WEP violations. 

More importantly, a spinning spacecraft can be used to introduce a modulation of the differential acceleration resulting from a WEP violation, both for a free-fall and a torsion balance experiment, to distinguish the WEP violating signal from other effects appearing at different frequencies. In this case, as the platform is rotating with the instrument itself, the mass distribution in the immediate vicinity of the test bodies does not introduce any signal modulation as soon as there are no moving parts or changes in the mass distribution of the spacecraft.

Still, gravity gradients remain one of the predominant sources of systematic error imposing an ad-hoc design of the experimental setup and the test masses. Test masses of different shape couple differently to gravity gradients due their different multipole moments. This effect produces a differential acceleration competing with a violation of the Weak Equivalence Principle. Test masses shall therefore be designed to approach the shape of a gravitational monopole or to have matching gravitational multipole moments \cite{Connes1997}. As a consequence, manufacturing processes shall ensure precise control on the shape of the masses and the material itself shall be selected to be highly homogeneous and easily machinable \cite{Touboul_2019}. On the other hand, the gravity environment generated by the spacecraft surrounding the test masses, and thus primarily interacting with them, can also be controlled. As already demonstrated by the LISA Pathfinder mission, a protocol-based measurement of the mass and the distance of all satellite parts can ensure a balance of the gravitational accelerations at the sub~nms$^{-2}$ level \cite{Armano_2016a}. Such techniques have proven to be very effective in reducing the systematic errors introduced by gravity gradients.

Differential accelerometers based on a nested test mass design are also affected by the radiometer effect. The infrared radiation of the Earth is absorbed by the satellite and consequently by the instrument housing, thus producing a temperature gradient that depends on the satellite's orientation with respect to the Earth-to-satellite direction. Due to the residual gas around the test masses, this temperature gradient is responsible for a differential acceleration that is directly proportional to the pressure of the residual gas and to the temperature gradient and that cannot be distinguished from a WEP violating signal. The thermal design of the spacecraft and the instrument head as well as the design of the vacuum system enclosing the test masses is therefore important to minimize this effect. In STEP, where cryogenic temperatures are reached in a He dewar, the residual gas pressure and the temperature gradients can be better controlled. In \cite{Nobili2001} the radiometer effect is calculated for MICROSCOPE, STEP and GG and discussed with respect to the specific design of the three instruments.   

In space, the test masses of the differential acceleration sensor can accumulate charges due to the interaction with high energy charged particles travelling through the solar system. In the presence of stray charges, the source mass interacts with the caging mechanism and readout system via Coulomb forces introducing noise and bias on its position and on the measurement signal readout. Different methods can be used to discharge the masses and counteract this effect. MICROSCOPE stray charges are managed via a thin (0.7~$\mu$m) gold wire connected to a sole plate and driven by a control voltage \cite{Touboul_2019}. A different discharging system has been demonstrated in space by the LISA Pathfinder mission. In this case, an ultra-violet lamp illuminating both the test masses and the surrounding environment is used to generate a current of photoelectrons \cite{Sumner2009} that can be tuned to null the charge  of the test mass itself. In this way, the corresponding noise and bias can be reduced to negligible levels. 

Finally, external perturbations can be accurately controlled in space. Stabilization loops can be implemented to reduce temperature fluctuation at the instrument head below 100~$\mu$K \cite{Touboul2017}. The Newtonian noise, generated by fluctuations of terrestrial gravity and representing one of the most important limitations of ground-based tests of the Weak Equivalence Principle, is totally absent on a spacecraft. In space, other perturbations, such as air drag or solar radiation pressure, can introduce noise and bias affecting the WEP test. However, several drag free systems have already demonstrated their ability to reduce residual accelerations below $3\times10^{-11}$ms$^{-2}$Hz$^{-1/2}$, as in the MICROSCOPE mission \cite{Touboul2017}, or even better, down in the $10^{-15}$ms$^{-2}$Hz$^{-1/2}$ regime, as in LISA Pathfinder~\cite{Armano2016,Armano2018}.

The MICROSCOPE (MICROSatellite \`a train\'ee Compens\'ee pour l'Observation du Principe d'Equiv\-a\-lence) mission provides the most accurate test of the Weak Equivalence Principle \cite{Touboul2017,Touboul_2019}.
The satellite was launched from Kourou on 25 April 2016 on a Soyuz rocket and injected into a dawn-dusk Sun-synchronous orbit with an altitude of 710~km. The spacecraft embarks two differential accelerometers, each of them based on two hollow cylinders. They are aligned along the symmetry axis, precisely centered, and kept in their equilibrium position by capacitive electrodes. The two differential accelerometers only vary for the test masses composition: Pt:Rh alloy for both cylinders of the reference sensor unit (SUREF); Pt:Rd and Ti:Al:V for the inner and outer test mass of the unit sensitive to WEP violations (SUEP). The test masses have been precisely machined to a relative difference between the momenta of inertia smaller than $10^{-3}$ and a density homogeneity better than 0.1\%, thus reducing  differential accelerations due to gravity gradients to negligible levels \cite{Touboul_2019}. The same set of electrodes provides measurement and control of both the position and the attitude of the test masses. They are machined on a silica substrate to ensure high position stability. The voltage applied at the electrodes, which is proportional to the force exerted on the test masses to keep them centered, represents the main data output of the instrument from which the differential acceleration between the test masses is extracted. Once the test masses are correctly aligned, if General Relativity holds, zero differential acceleration at both the SUREF and SUEP sensor heads shall be read. The magnetic environment is controlled by a magnetic shield surrounding the complete payload and modelled by a finite element calculation. The tight housing allows the sensors to operate in the $10^{-5}$~Pa regime, where the radiometer effect is strongly reduced. Radiometer effect and radiation pressure disturbances are kept below the damping introduced by the 7~$\mu$m wire connecting the test masses and the cage to control electrical charging effects~\cite{Touboul2012}. Cold gas thrusters actuated by the accelerometers' measurements reduce the effect of air drag and, more generally, of non-gravitational forces acting on the spacecraft. The drag-free control system relies on the linear and angular accelerations measured at one of the test masses. Residual accelerations below $3\times10^{-11}$ms$^{-2}$Hz$^{-1/2}$ could be measured in closed-loop configuration. This result is about a factor 10 better than originally specified. Star tracker measurements are also used to determine the spacecraft attitude. To increase the modulation frequency of the gravitational signal provided by the Earth, the satellite is rotated ($\sim$1~mHz) around the orthogonal direction to the sensitive axis of the differential accelerometers. The SUREP and SUEP power spectral density of differential acceleration measurements along the axial direction (sensitive axis of the instrument) are $5.6\times10^{-11}$ms$^{-2}$Hz$^{-1/2}$ and $1.8\times10^{-11}$ms$^{-2}$Hz$^{-1/2}$, respectively, at the modulation frequency of a few mHz expected for the WEP violation. This floor level is limited by the damping noise of the thin gold wire connected to the test masses to control charging effects. Systematic errors are currently dominated by thermal effects, which could be evaluated to $<67\times10^{-11}$ms$^{-2}$. The instrument sensitivity was determined by applying temperature variations both at the electronics and at its baseplate. Measurements revealed that the sensor unit temperature coefficient is 2 orders of magnitude higher than expected. This issue is still under investigation. On the positive side, the temperature stability of the instrument baseplate and the electronics was measured to be better than 20~$\mu$K over 120 orbits, about 2 orders of magnitude smaller than initial estimates, thus mitigating temperature-related effects. The contribution of self-gravity and stray magnetic fields to the measurement error was estimated from finite element models and found negligible. 
After analyzing the data corresponding to 120 satellite orbits, an E\"{o}tv\"{o}s parameter of $\left[ -1\pm9\,(\text{stat})\pm9\,(\text{syst}) \right]\times10^{-15}$ could be estimated for titanium and platinum \cite{Touboul2017}, improving by one order of magnitude previous results obtained from torsion balance \cite{Wagner2012} and lunar laser ranging \cite{Williams2012} experiments. This measurement also establishes new constraints to modifications of the Newton's law of gravity by a Yukawa-like coupling and improves existing constraints on WEP violations by a light scalar field \cite{Berge2018}. The MICROSCOPE mission has been decommissioned on 18 October 2018, after accumulating about 1900 orbits of science data on the SUEP sensor, 900 on the SUREF sensor, and 300 orbits for calibration. This also includes 750 orbits of measurements for characterizing the on-board temperature sensors and further reduce the systematic effects due to temperature variations. After the first results reported in 2017, the complete data set delivered by the MICROSCOPE mission is still under scrutiny to improve both the statistical and systematic error on the WEP test, hopefully going below the $1\times10^{-15}$ accuracy level.

The STEP (Satellite Test of the Equivalence Principle) mission concept is similar to MICROSCOPE, but it relies on a completely different technology, which is expected to push the accuracy of WEP tests down to 1 part in $10^{18}$ \cite{Sumner2007,Overduin2012}. The STEP mission was selected for a phase A study in 1990. An engineering model of the accelerometer to test the technology was built in 2004. The payload is composed of 4 differential accelerometers (DAs) operating simultaneously with the following combination of test masses: Be and Pt:Ir for DA1; Be and Nb for DA2; Nb and Pt:Ir for DA3; Be and Pt:Ir for DA4. DA1 to DA3 measurements allow to check that the sum of the differential acceleration measured at the 3 sensor heads between the three materials - Be, Pt:Ir, and Nb - is zero thus providing control on measurements systematics. DA1 and DA4 differ for the shape of the test masses and for their coupling to Earth and spacecraft gravity. As for MICROSCOPE, the differential accelerometers are composed of two hollow masses with cylindrical symmetry, precisely centred and aligned along their axis. STEP DAs are arranged in a helium dewar and operated at 2~K. The cryogenic environment is providing very good thermal and mechanical stability for the DAs operation, ultra-high vacuum and reduced thermal noise from gas damping, excellent shielding from external magnetic fields, reduced radiation pressure effects due to temperature gradients. More importantly, it allows to use SQUID technology for high-sensitivity position readout and for generating the weak reaction force that centers the test masses along the axial direction. The gas generated by boiling helium is used by thrusters to stabilize the spacecraft against non gravitational accelerations. In addition, when in drag-free, common mode accelerations can be measured with respect to the spacecraft reference frame to $10^{-15}-10^{-12}$~ms$^{-2}$, well below the MOND (MOdified Newtonian Dynamics) acceleration scale $a_0$. Based on this performance, STEP has recently been proposed for a test of MOND theories and of the Strong Equivalence Principle \cite{Pereira2016}. As discussed in section \ref{sec:droptests}, alternative WEP tests in microgravity are also possible on ground laboratories such as the Bremen drop tower 
\cite{Vodel2002}.

Galileo Galilei (GG) is an alternative proposal designed to test the Einstein WEP to better than 1 part in $10^{17}$ \cite{Nobili2018a, Nobili2018b}. Differently from MICROSCOPE and STEP, GG can be considered as the space version of a beam balance. The test masses are two hollow cylinders of different composition that are weakly coupled by means of mechanical suspensions. Once properly set into equilibrium by piezo actuators, the beam of the balance is aligned along the symmetry axis of the cylinders, thus defining the plane orthogonal to this direction as the sensitive plane of the instrument. This configuration provides a rejection of common mode acceleration noise as high as $10^5$, thus drastically relaxing the level of drag control required at the spacecraft. Rapid rotation of the instrument is important to reduce the thermal noise in the detection of WEP violations and to efficiently decouple it from systematic effects appearing at different frequencies. In GG, the spin axis of the spacecraft coincides with the symmetry axis of the instrument. Therefore, after initial spin up, the spacecraft co-rotates with the cylindrical test masses around the principal axis of the system and it is passively stabilized to very fast rotation rates ($\sim$1~Hz) by angular momentum conservation. Due to the high mass of the GG test cylinders and the large gap between them, thermal noise due to gas damping and to the radiometer effect become two orders of magnitude smaller than in MICROSCOPE. Finally, the displacement of the GG test masses is read by a laser interferometer gauge, which provides very low noise and fast integration times. An accuracy budget of the GG instrument and an evaluation of the systematic effects is provided in \cite{Nobili2018a}. A laboratory demonstrator of the space instrument has been built and it is presently under test. On the ground the instrument has reached a sensitivity to differential acceleration measurements of $\sim 7\times10^{-11}$~ms$^{-2}$ (at $1.7\times10^{-4}$~Hz upconverted by rotation to 0.2~Hz), currently limited by Newtonian noise, mainly tilt, acting on the ball bearings \cite{Nobili2012}. An optimized design based on low noise air bearings, low coupling joints, and a laser interferometer readout system is under study to push the instrument performance down to the $10^{-16}-10^{-15}$ regime.          
 
\subsection{Tests based on the measurement of the Earth-to-Moon and Earth-to-satellite distance \label{sec:LLR}}

Lunar Laser Ranging (LLR) experiments are performed since 1969, when the first array of corner cube reflectors was positioned on the Moon by Apollo~11. A review of LLR tests of gravity con be found in \cite{Merkowitz2010}. To date, 5 arrays of retro-reflectors are operational on the Moon surface and routinely used for ranging experiments: Apollo 11, 14 and 15, Lunokhod 1 and 2. Among them, Apollo 15 is the one with the largest lidar cross section and therefore the most widely used for LLR (about 75\% of normal point data). 
In a LLR measurement, a short laser pulse is fired by a ground-based Satellite Laser Ranging (SLR) station towards one of the Moon corner cube reflector arrays. The back reflection is collected by the SLR station and the interval between the fire time and the reception time is recorded. Round-trip travel time measurements are then fitted to a model of the solar system ephemeris including tidal effects, relativistic effects, propagation in the atmosphere, plate motion, etc. The SLR stations mostly contributing to LLR data are the Observatoire de la C\^ote d’Azur (OCA) in France, the McDonald Laser Ranging System (MLRS) and the Apache Point Observatory Lunar Laser-ranging Operation (APOLLO), both in the US. Thanks to the 3.5~m diameter telescope and to the array of high-efficiency avalanche detectors, the APOLLO station has today reached millimeter ranging precision and accuracy to the Moon \cite{Murphy2012, Adelberger2017, Viswanathan2018}. To this level, effects like regolith motion, thermal expansion of the retro-reflectors array, oceans and atmosphere loading effects start to become relevant. 

Earth and Moon are two celestial bodies freely falling in the gravitational field of the Sun (primary body). If the Universality of Free Fall principle is violated, their accelerations towards the Sun are different, thus introducing a polarization of the lunar orbit~\cite{Nordtvedt1968}. This effect manifests itself with the appearance of a modulation of the Earth-Moon distance (LLR measurements) along the Earth-Sun direction at the synodic period (29.53~day). For a relative differential acceleration between the Earth and the Moon of $\Delta a/a$, the perturbation $\delta r$ to the Earth-Moon distance expressed in meters is given by \cite{Damour1996} 
\begin{equation}
\delta r=-2.9427\times10^{10}\frac{\Delta a}{a} \cos D\,[\textrm{m}] \,,
\end{equation}
where $D$ is the synodic angle. After the first LLR test of the Equivalence Principle in 1976 \cite{Williams1976,Shapiro1976}, the accuracy of relative differential acceleration measurements $\Delta a/a$ between Earth and Moon has progressively improved to $1.4\times10^{-13}$ \cite{Williams2004a,Williams2004b,Williams2012}, to recently reach $7\times10^{-14}$ \cite{Viswanathan2018} and $5\times10^{-14}$ \cite{Hofmann2018}. Such results could be obtained after modelling the effects of the gravitational interaction of the Sun and the planets on the Moon, now treated as an extended body. High order terms of the Earth-Moon gravitational interaction and the effect of solid Earth tides on the Moon orbit were also improved.

Celestial bodies have non negligible gravitational self-energy. This was already clear in 1968 \cite{Nordtvedt1968}, when Nordtvedt proposed to use the Earth-Moon system to test the Strong Equivalence Principle. In this case, the relative differential acceleration responsible for a violation of the Universality of Free Fall principle can be expressed as
\begin{equation}
\frac{\Delta a}{a} = \eta_{CD} + \eta_{SEP}\left(\frac{U_E}{M_Ec^2}-\frac{U_M}{M_Mc^2}\right)\,,
\end{equation}
where $\eta_{CD}$ is the composition-dependent violation parameter, $\eta_{SEP}$ is the Nordtvedt parameter measuring SEP violations, $U$ and $M$ represent the gravitational self-energy of the test body and its mass. Therefore, to exclude any cancellation effect between a composition-dependent WEP violation and an equal and opposite SEP violation, an independent test of the Weak Equivalence Principle based on test masses having similar composition to the Earth and Moon interior, but with negligible gravitational self-energy is required. The experiment, performed in 1999 with the torsion balance apparatus of the Washington group, confirmed the validity of WEP for two test bodies reproducing the Earth and the Moon composition to $1.4\times10^{-13}$ \cite{Baessler1999, Adelberger2001}. Combined with with LLR measurements, this test can be used to constrain SEP violations. The best estimate of the Nordtvedt parameter based on laser ranging measurements of the Earth-Moon system has reached an uncertainty of $1.1\times10^{-4}$ \cite{Hofmann2018}.

The Universality of Free Fall can also be tested by tracking satellites orbiting around the Earth, e.g. LAGEOS, LAGEOS II and LARES \cite{Lucchesi2015}. As also discussed in \cite{Nobili2008}, the sensitivity of a test based on the Earth-LAGEOS system in the gravitational field of the Sun is a factor 300 worse than for the Earth-Moon system. Indeed, LAGEOS and LARES satellites are much closer to the Earth compared to the Moon. As a consequence, the effects of the Sun gravitational potential on the Moon orbit are significantly stronger than for a satellite orbiting the Earth at low altitude. Even if not competitive with LLR experiments, SLR tests still remain of interest to evaluate the impact of different systematic errors in the final result. As an example, non-gravitational perturbations, which play a major role in the determination of the LAGEOS and LARES orbits, are completely negligible for the Moon. On the contrary, some gravitational perturbations (tidal effects) are important for the Moon and negligible for an Earth-orbiting satellite. Finally, SLR measurements could also be combined with LLR measurements in a grand-fit procedure to better estimate common parameters thus improving LLR and interplanetary ranging \cite{Nordtvedt2001}.

Similar tests can be performed by ranging other gravitating bodies in the solar system. The MESSENGER mission with its Doppler tracking measurements collected over 7 years allows the precise determination of Mercury's ephemeris. This wealth of data has recently been used to test the Strong Equivalence Principle with reduced uncertainty \cite{Genova2018}. Spacecraft and planet orbits are numerically integrated to provide a global solution from which parameters relevant for General Relativity, planetary physics, and heliophysics can be extracted. This analysis is today constraining the Nordtvedt parameter to $7\times10^{-5}$.

\subsection{Tests with microscopic particles: atoms, molecules, neutrons, antimatter \label{sec:atoms}}

This section is  devoted to a review of the  tests of the WEP with microscopic particles, mainly atoms. 

Based on recent advances in cold  atom optics, atomic sensors, namely atom interferometers \cite{Cronin2009,Tino2014} and atomic clocks \cite{Poli2013,Ludlow2015}, established themselves as new powerful tools for precision measurements and fundamental tests in physics  \cite{Safronova2018}.

Atom interferometry  enabled the realization of precision tests of the WEP that were previously performed only with macroscopic classical masses.
As will be clear from the data reported in this review, the sensitivity of atomic experiments did not reach yet the one of the classical experiments but predictions are that similar or even higher levels of sensitivity will be obtained both in Earth laboratories and in experiments in space.

An important advantage of using atoms is, in a properly designed apparatus, the control of possible systematics thanks to the well known and reproducible properties of the atoms, the possibility of realizing an atomic probe of extremely small size and precisely controlling its position, the potential immunity from stray field effects, and the availability of different states and different isotopes that in some cases allows the rejection of common-mode spurious effects and/or a cross-check of the results. Perhaps still more important is that new kinds of tests are possible that exploit the specific quantum features of atomic probes:  qualitatively new experiments can be performed with test masses having well-defined properties in terms of, e.g., proton and neutron number, spin, internal quantum state, bosonic or fermionic nature.

In the final part of this section, tests with neutrons, with charged particles, and with anti-matter are also described  because of their fundamental interest but the  precision achieved so far in these cases is still much lower compared to the other tests.

It can be expected that in the future the development of matter-wave interferometry with molecules will enable also the comparison of the free fall for such systems with  different conformations, different internal states, different chiralities; this will not be discussed in the present review because sensitivities are still too low to be significant in this context but preliminary results and a discussion of future prospects can be found in \cite{Rodewald2018}.

It should be noticed that all experiments so far were performed with systems consisting of particles of the first elementary particle family and that direct tests for particles of the second and the third families are missing  until now (see, e.g., \cite{Antognini2018} and references therein).

\subsubsection{Precision measurements of gravity with atom interferometry} 

The  idea of an atom interferometer can be easily understood from the analogy with an optical interferometer: using suitable atom optics made of  material structures or, more often, laser light, an atomic wave packet is split, reflected and recombined: at the output, interference can be observed. In a more general view, it can be considered as a quantum interference effect arising from the different paths connecting the initial and final states of a system. Any effect affecting in a different way the different paths will produce a change in the interference pattern at the output; by detecting this change, the effect can be measured.

In most experiments, the best performances have been achieved using atom interferometry schemes in which  the wavepackets of freely falling samples of cold atoms are split and recombined using laser pulses in a Raman \cite{Kasevich1991,Kasevich1992}
or Bragg \cite{Giltner1995,Muller2008b}  configuration.
The effect of gravity leads to a phase change $\Delta \phi = k g T^2$ where $k$ is the effective wave-vector of the light used to split and recombine the wave packet, $g$ is gravity acceleration, and $T$ is the time of  free-fall of the atom between the laser pulses. This corresponds indeed to the free-fall distance measured in terms of the laser wavelength.

Other schemes were developed to measure $g$ based on Bloch oscillations (~\cite{Ferrari2006} and references therein). In this case, the atoms are not falling freely under the effect of gravity but the combined effect of gravity and the periodical potential produced by the laser standing wave leads to oscillations in momentum space with a frequency $\nu_{BO}=mg\lambda/2h$, where $m$ is the atomic mass, $\lambda$ is the wavelength of the laser producing the lattice and $h$ is Planck's constant. By measuring the frequency of the Bloch oscillation $\nu_{BO}$, $g$ is determined.
This method can also be interpreted as the measurement of the gravitational potential difference between adjacent lattice wells which are separated by $\lambda/2$. A few wells are filled with ultracold atoms so that the gravimeter has a sub-millimiter size down to a few micrometers. For this reason, it was also proposed as a method to test the $1/r^2$ Newtonian law for gravity at micrometric distances \cite{Ferrari2006,Sorrentino2009}.

Atom interferometers enabled precise measurements of several physical effects; in particular,  in gravitational physics, 
the measurement of gravity acceleration~\cite{Kasevich1992, Peters1999, Muller2008, LeGouet2008,Hu2013,Sorrentino2012,Freier2016,Hardman2016,Xu2019}, 
gravity gradient~\cite{Snadden1998, Mcguirk2002,Sorrentino2012,Sorrentino2014, Duan2014, Pereira2015, Wang2016}
and curvature~\cite{Rosi2015, Asenbaum2017}, 
determination of the gravitational constant $G$
\cite{Mcguirk2002, Tino2002,Fattori2003, Bertoldi2006, Fixler2007, Lamporesi2008, Rosi2014, Prevedelli2014}, 
investigation of gravity at microscopic distances \cite{Ferrari2006,Sorrentino2009,Tackmann2011},
search for dark energy,  chameleon and theories of modified gravity \cite{Hamilton2015,Jaffe2017,Sabulsky2019}, 
applications to geodesy, geophysics, engineering prospecting, inertial navigation~\cite{Bresson2006, DeAngelis2009,Wu2019}. 

In~\cite{Muller2010a}, experiments measuring $g$ with different atom interferometry methods ~\cite{Peters1999,Ferrari2006,Clade2005} were reinterpreted as measurements of the Einstein's gravitational redshift, thus claiming an improvement in precision by 4 orders of magnitude with respect to the Gravity Probe A test reported in~\cite{Vessot1980}. This paper started a controversy on such an interpretation and on the nature of the phase shift measurement in an atom interferometer \cite{Muller2010a, Wolf2010, Muller2010b, Hohensee2011, Wolf2011, Sinha2011, Hohensee2012, Wolf2012, Unnikrishnan2012, Schleich2013b}.

Atom interferometers and optical atomic clocks were  proposed for the detection of gravitational waves on ground and in space \cite{Chiao2004,Borde2004,Roura2006,Tino2007,Dimopoulos2008,Tino2011,Yu2011,Hogan2011,Graham2013,Kolkowitz2016} and the first prototypes are presently under construction \cite{Canuel2018, MAGIS100, Zhan2019}.

Experiments in space based on cold atom sensors were proposed since long \cite{Tino2007space,Tino2013,Altschul2015,Tino2019}, the required technology development is in progress \cite{Aguilera2014}, and proof-of-principle experiments were recently performed \cite{Becker2018}.


After the early observation of free fall of atoms using long  beams of potassium and cesium atoms \cite{Estermann1947}, several experiments have  compared the free fall of different atoms to test the WEP, as described in detail in the following: $^{85}$Rb vs $^{87}$Rb~\cite{Fray2004,Bonnin2013, Zhou2015}, $^{39}$K vs $^{87}$Rb~\cite{Schlippert2014}, the bosonic $^{88}$Sr vs the fermionic $^{87}$Sr~\cite{Tarallo2014}, atoms in different spin orientations~\cite{Tarallo2014,Duan2016}. The relative accuracy of these measurements, reaching so far $10^{-8}$ - $10^{-9}$, is expected to improve by several orders of magnitude in the near future thanks to the rapid progress of atom-optical elements based on multi-photon momentum transfer~\cite{Muller2009, Chiow2011} and of large-scale facilities providing a few seconds of free fall during the interferometer sequence~\cite{Dimopoulos2007,Zhou2011,Kovachy2015}. Experiments testing the free fall of anti-hydrogen are in progress~\cite{Kellerbauer2008,Charman2013, Aghion2014}. WEP tests with a precision $\sim 10^{-15}$ using  atom interferometers in space were proposed~\cite{Tino2007space,Tino2013,Altschul2015}.

Concerns have  been raised on the potential of atom interferometry for high precision tests of the WEP~\cite{Nobili2016}. A scheme to overcome these possible limitations was proposed in~\cite{Roura2017} and experimentally demonstrated in \cite{DAmico2017,Overstreet2018}.

\subsubsection{Atoms vs macroscopic objects}

In different experiments, gravity acceleration for atoms has been compared with the one for macroscopic masses.

Already in the first demonstration of a high-precision atom interferometry gravimeter with Cs atoms, a classical gravimeter based on a Michelson optical interferometer with a vertical arm containing a freely falling corner-cube was used for comparison. The atom gravimeter was realized with a Raman interferometry scheme. An uncertainty of $\Delta g/g = 3 \times 10^{-9}$ was achieved with a free-fall time $2T=320$~ms. The comparison between the two gravimeters was interpreted as the demonstration that the macroscopic glass mirror falls with the same acceleration, to within 7 parts in $10^{9}$, as the quantum-mechanical Cs atom \cite{Peters1999}.

In \cite{Merlet2010,Francis2015}, comparisons between a mobile Raman atom gravimeter with $^{87}$Rb and classical absolute gravimeters were performed with comparable uncertainties. 

A conceptually different scheme was used in \cite{Poli2011}. The experiment  was based on Bloch oscillations of Sr atoms in a vertical optical lattice.   In order to increase the sensitivity, in this work a method to measure the frequency of higher harmonics of the Bloch frequency was adopted. The value of the acceleration measured with this atomic sensor was compared with the one obtained in the same lab with a classical FG5 gravimeter. The two values agreed within 140 ppb.

\subsubsection{Different isotopes}

Experiments were performed testing WEP for different isotopes of an atomic species. Compared to the experiments discussed in the following in which different atomic species are compared, these are somehow simpler; the similar masses and nearby transition frequencies make the apparatus and the control of systematics less complex.

In \cite{Fray2004}, an atom interferometer based on the diffraction of atoms from standing optical waves acting as effective absorption gratings was used to compare the two stable isotopes of rubidium, $^{85}$Rb and $^{87}$Rb, with a relative accuracy of $1.7 \times 10^{-7}$. In this work, a test for a possible difference of the free fall acceleration as a function of relative orientation of nuclear and electron spin was also performed with a differential accuracy of $1.2 \times 10^{-7}$ by comparing interference patterns for $^{85}$Rb atoms in two different hyperfine ground states (see sect. \ref{diffQS}). A comparable precision for the differential free fall measurement of $^{85}$Rb and $^{87}$Rb was later obtained in \cite{Bonnin2013} using Raman atom interferometry.

About one order of magnitude improvement in precision was obtained in \cite{Zhou2015}. A four-wave double-diffraction Raman transition scheme was used for the simultaneous dual-species atom interferometer to compare $^{85}$Rb and $^{87}$Rb. The value obtained for the E\"otv\"os parameter is $\eta = (2.8 \pm 3.0 \times 10^{-8})$.

Ongoing experiments in large baseline interferometers aim to a final precision in the $10^{-15}$ range and beyond \cite{Dimopoulos2007,Zhou2011,Kovachy2015}. Limiting factors due to the gravity gradients were discussed \cite{Nobili2016} and possible solutions were proposed in \cite{Roura2017} and  demonstrated in \cite{DAmico2017,Overstreet2018}. In \cite{Overstreet2018} the gravity gradient compensation in a long duration and large momentum transfer dual-species interferometer with $^{85}$Rb and $^{87}$Rb allowed to reach a relative precision of $\Delta g/g \approx 6 \times 10^{-11}/$shot or $3 \times 10^{-10}/\sqrt{Hz}$ that makes such a WEP test realistically feasible at the $10^{-14}$ level.

In \cite{Tarallo2014}, the WEP was tested for the bosonic and fermionic isotopes of strontium atoms, namely, $^{88}$Sr and $^{87}$Sr. As in \cite{Poli2011}, gravity acceleration for the two isotopes was determined by measuring the frequency of the Bloch oscillations for the atoms in a vertical optical lattice. By detecting the coherent delocalization of matter waves induced by an amplitude modulation of the lattice potential at a frequency corresponding to a multiple of the Bloch frequency, the limit obtained in this work for the E\"otv\"os parameter is $\eta = (0.2 \pm 1.6 \times 10^{-7})$. As discussed in the following, the results of this experiment are also relevant as a WEP test for bosons vs fermions and for the search of spin-gravity coupling.

\subsubsection{Different atomic species}

Recently, experiments were performed to test WEP with different atomic species. This requires the development of more complex experimental setups and a more difficult control of  systematics. The theoretical framework to interpret the experimental results can be found in Refs. \cite{Damour2012,Hohensee2013b,Hartwig2015}.

The possibility of a test using rubidium and potassium atoms was discussed in \cite{Varoquaux2009}. The first results were reported in \cite{Schlippert2014}; in this work,  $^{87}$Rb and $^{39}$K were compared using two Raman interferometers. The result was an E\"otv\"os ratio $\eta = (0.3 \pm 5.4 \times 10^{-7})$ mainly limited by the quadratic Zeeman effect and the wave front curvature of the Raman beams. The choice of atomic species in this paper was compared with others in terms of sensitivity to possible violations of the EP predicted by a dilaton model \cite{Damour2012} and by standard-model extensions \cite{Hohensee2013b}.  

The ongoing activity for a test with rubidium and ytterbium atoms in a 10-m baseline atom interferometer was discussed in \cite{Hartwig2015} with the goal to reach an E\"otv\"os ratio in the $10^{-12}-10^{-13}$ range.

\subsubsection{Atoms in different quantum states}\label{diffQS}

While it can be argued that some of the experiments with atoms described above are not qualitatively different from the ones performed with macroscopic classical systems as far as the physics which is tested is concerned, the experiments described in this section take full advantage of the quantum nature of the atoms as probes of the gravitational interaction.

\begin{itemize}

\item  Atoms in different energy eigenstates and in superposition states\\
The mass-energy relation $E=mc^2$ in special relativity  implies that the internal energy of a system affects its mass. It is then of interest to verify the validity of the equivalence of the inertial and gravitational mass for systems in different internal quantum states. This was theoretically discussed in \cite{Zych2015,Zych2018} and possible experimental tests with atoms in different internal states were proposed. In particular, the importance of tests involving atoms in superpositions of the internal energy eigenstates was highlighted because this corresponds to a genuine quantum test. Another possible experimental test of the quantum formulation of the equivalence principle was proposed in \cite{Orlando2016}. 

A first experimental test of the equivalence principle in this quantum formulation was reported in \cite{Rosi2017}. A Bragg atom interferometer was used to compare the free fall of $^{87}$Rb atoms prepared in two hyperfine states $|1\rangle= |F=1, m_F=0\rangle$ and $|2\rangle=|F=2, m_F=0\rangle$, and in their coherent superposition $|s\rangle=(|1\rangle + e^{i\gamma}|2\rangle)/\sqrt{2}$. In order to increase the measurement sensitivity, the atom interferometer was operated at the 3rd Bragg diffraction order, corresponding to $6 \hbar k$ total momentum transfer between the atoms and the radiation field. \\
The comparison of the free-fall acceleration for atoms in the $|1\rangle$ and $|s\rangle$ states led to the first experimental upper bound of $5 \times 10^{-8}$ for the parameter corresponding to a  violation of the WEP for a quantum superposition state. \\
Based on models in which WEP violations increase with the energy difference between the internal levels\cite{Damour2012}, in this paper the prospect to use states with an energy separation larger than the hyperfine splitting was also proposed considering optically separated levels in strontium for which the relevant atom interferometry schemes were already demonstrated \cite{Mazzoni2015,Hu2017,Hu2020}.\\
The comparison of gravity acceleration for atoms in the $|1\rangle$ and $|2\rangle$ hyperfine states led to an E\"otv\"os ratio $\eta_{1-2} = (1.0 \pm 1.4 \times 10^{-9})$ that corresponds to an improvement by about two orders of magnitude with respect to the previous limit set in \cite{Fray2004}.
A further improvement by a factor of 5 in the precision of this test was recently reported in \cite{Zhang2018}, approaching the $10^{-10}$ level.
%
%
\item Atoms in entangled states\\
In \cite{Geiger2018} a quantum test of the WEP with entangled atoms was proposed. In the proposed experiment, a  measurement of the differential gravity acceleration between the two atomic species would be performed by entangling two atom interferometers operating on the two species. The example of $^{85}$Rb and $^{87}$Rb was analyzed in detail showing that an accuracy better than $10^{-7}$ on the E\"otv\"os parameter can be achieved.\\
Although no theoretical model is available predicting a WEP violation in the presence of entanglement, this is clearly a case of a purely quantum system to be further investigated.\\
 The free fall of particles in quantum states without a classical analogue and in particular for Schr\"odinger cat states in the configuration space was studied theoretically in \cite{Viola1997}.

\item Atoms in different spin states\\
As discussed above (see in particular Sect. \ref{th:Long-range forces and spin-gravity}), 
spin-gravity coupling and torsion of space-time were extensively investigated theoretically \cite{Peres1978,Ni2010,Capozziello2011}. \\
Different experiments were performed using macroscopic test masses \cite{Ni2010,Heckel2008,Zhu2018}, atomic magnetometers \cite{Venema1992,Kimball2013}, and by measuring hyperfine resonances in trapped ions \cite{Wineland1991}. The differential free-fall experiments with atoms in different hyperfine states are also relevant in this frame \cite{Fray2004,Rosi2017,Zhang2018}. \\
Recently, experiments were performed using atom interferometry to search for the coupling of the atomic spin with gravity.

In \cite{Tarallo2014}, the experimental comparison of the gravitational interaction for the bosonic isotope of strontium $^{88}$Sr, which has zero total spin in its ground state, with that of the fermionic isotope $^{87}$Sr, which has a half-integer nuclear spin $I=9/2$, was performed based on the measurement of the frequency of Bloch oscillations for the atoms in a vertical optical lattice under the effect of Earth's gravity.\\
A modified gravitational potential including a possible violation of WEP and the presence of a spin-dependent gravitational mass was considered in the form
$V_{g,A}(z)=(1+\beta_A+kS_z)m_Agz$, 
where $m_A$ is the rest mass of the atom, $\beta_A$ is the anomalous acceleration generated by a nonzero difference between gravitational and inertial mass due to a coupling with a field with nonmetric interaction with gravity, $k$ is a model-dependent spin-gravity coupling strength, and $S_z$ is the projection of the atomic spin along gravity direction. \\
As already described above, the Bloch frequency corresponds to the
site-to-site energy difference induced by the gravitational interaction; by measuring the frequency of Bloch oscillations for $^{88}$Sr and $^{87}$Sr an E\"otv\"os parameter $(0.2 \pm 1.6) \times 10^{-7}$ was obtained.
Since the frequency of Bloch oscillations depends on the mass of the particle, in the analysis of the data the  $m^{88}/m^{87}$ mass ratio was taken into account which is known with a relative precision $\sim 10^{-10}$.
The analysis of the Bloch resonance spectrum for $^{87}$Sr provided an upper limit for the spin-gravity coupling strength $k=(0.5 \pm 1.1) \times 10^{-7}$. This result also sets a bound for an anomalous acceleration and a spin-gravity coupling for the neutron either as a difference in the gravitational mass depending on the spin direction or as a coupling to a finite-range interaction \cite{Venema1992,Ni2010}.

In \cite{Duan2016}, a Mach-Zehnder-type Raman atom interferometer was used to compare the gravity acceleration of freely-falling $^{87}$Rb atoms in different Zeeman sublevels $m_F=+1$ and $m_F=-1$, corresponding to opposite spin orientations. The experiment required a special care to control the high sensitivity of these states to magnetic field inhomogeneity. The E\"otv\"os parameter obtained in this experiment was $(0.2 \pm 1.6 )\times 10^{-7}$. The data were also interpreted as providing an upper limit of $5.4 \times 10^{-6}$ m$^{-2}$ for a possible gradient field of the spacetime torsion.

In \cite{Rodewald2018}, based on recent advances of matter-wave interferometry with large molecules, the prospect of a test of WEP for molecules with opposite chiralities was proposed.

It should be noted that a complete analysis connecting theoretically the models tested in the different experiments performed so far in this frame is still missing. 

\item Atoms in a Bose-Einstein condensate\\
Possible differences in the gravitational interaction for bosons and fermions were investigated theoretically \cite{Barrow2004} and tested experimentally \cite{Tarallo2014}.

Violations of the WEP for atoms in a quantum state such as a Bose-Einstein condensate were discussed in (\cite{Goklu2008,Herrmann2012} and references therein). Since in quantum physics particles are described by an extended wave packet, the validity of the WEP which refers to point-like particles can be questioned. A model based on spacetime fluctuations allows to predict a possible difference in the observed free fall for different particles because the different spatial extensions of the wavefunction of particles of different masses would lead to an averaging of the metric fluctuations over different spatial volumes.
Also, the metric fluctuations would produce decoherence.\\
Such elusive effects, if ever observable, would require atom interferometers with extremely high sensitivity, that is, a very long evolution time. For this and other scientific goals, the technology needed to perform experiments in microgravity is being developed \cite{Debs2011,Kuhn2014,Altschul2015,Sondag2016,Becker2018} as described in detail in the following.
   
\end{itemize}

\subsubsection{Experiments with atoms in microgravity \label{sec:atomspace}}
The ultimate performance of atomic sensors for WEP tests can be reached in a space-based laboratory. In space atoms can rely on a very quiet environment where Newtonian noise is absent and microvibrations and non-gravitational accelerations can be reduced to very low levels. Very long and unperturbed free fall conditions can be obtained, allowing atomic wavepackets to evolve, sense the space-time metric, and record its signature in their phase. At the same time, very long and unperturbed interaction times between the atomic ensemble and the interrogation fields can be achieved. This is translating into a significant increase of the instrument sensitivity and a better control of the systematic errors.

As an example, the phase accumulated in a Mach-Zehnder interferometer, $\Delta \phi = k g T^2$, is directly proportional to the square of the free evolution time $T$ between the three laser pulses of the interferometry sequence. The typical duration of an atom interferometry sequence on the ground is $2T\approx1$~s, which corresponds to a free fall distance of about 10~m in the gravity field of the Earth. In space, both the atoms and the instrument platform are in free fall and interrogation times of $2T=10$~s or longer can be achieved, improving the instrument sensitivity by a factor 100 or more with respect to a similar instrument operated on the ground.

Achieving a free evolution time of $2T\approx10$~s on the ground would require an atomic fountain apparatus with several hundred meters of free-fall length, showing another important aspect of atom-based sensors designed for space compared to their laboratory counterpart, i.e. the compactness. In space, atoms interrogation can take place in a small vacuum chamber with a typical size of a few liters. This volume can be better controlled against external perturbations, such as temperature, magnetic fields, etc. As an example, the development of large size mu-metal shields to accurately control the external magnetic field along the free evolution trajectories of a long atomic fountain (10~m or longer) remains a non negligible technology challenge.  

Finally, a technique to counteract the effect of gravity gradients has recently been developed \cite{Roura2017} and experimentally demonstrated \cite{DAmico2017}, reducing to a negligible level one important source of instability and systematic error in precision measurements by atom interferometry. 

STE-QUEST (Space-Time Explorer and QUantum Equivalence Space Test) is a mission designed to test different aspects of the Einstein Equivalence principle in space \cite{Altschul2015}. The STE-QUEST scientific objectives include an atom interferometry test of the Weak Equivalence Principle, an absolute measurement of the Einstein's gravitational redshift, and tests of Standard Model Extension (SME). Here, we will only focus on the WEP test. The on-board instrument dedicated to this measurement is a differential atom interferometer. Originally designed to compare the free fall of the 85 and 87 rubidium isotopes, the instrument has recently been re-adapted to operate on potassium and rubidium that, due to the larger difference in neutrons and protons, are expected to provide higher sensitivity in the detection of a WEP violation \cite{Damour2010}. The two atomic ensembles would be cooled down to very cold temperatures (100~pK regime) and simultaneously interrogated in the atom interferometry sequence by using the double-diffraction technique \cite{Leveque2009}. The simultaneous interrogation provides rejection ratio of common mode acceleration noise (e.g. air drag and mechanical vibrations), which can vary from $10^{-9}$ for $^{85}$Rb-$^{87}$Rb simultaneous interferometers \cite{Aguilera2014} to $<10^{-3}$ for the $^{87}$Rb-$^{39}$K couple. The requirements on the control of non-gravitational accelerations acting on the spacecraft are therefore very modest for a $^{85}$Rb-$^{87}$Rb differential interferometer and significantly more stringent for the $^{87}$Rb-$^{39}$K one, but still well within the available technology as demonstrated in the MICROSCOPE \cite{Touboul2017} and LISA Pathfinder missions \cite{Armano2018}. A design description of the STE-QUEST differential atom interferometer can be found in \cite{Schuldt2015}. The expected error budget is presented in \cite{Aguilera2014}. The instrument will be able to measure differential accelerations down to $8\times10^{-15}$ms$^{-2}$ corresponding to a WEP test at the $1\times10^{-15}$ level. 

A similar instrument has also been proposed for a WEP test on the International Space Station (ISS) \cite{Williams2016}. The ISS is a harsh environment for what concerns non gravitational accelerations, rotations, tilt noise, and mechanical vibrations. The instrument is therefore designed to ensure optimal control on systematic errors and high rejection of common mode effects. The differential accelerometer compares the free fall acceleration of $^{85}$Rb and $^{87}$Rb atomic samples in a symmetric configuration with two separate source regions. Bragg lasers tuned to the wavelength for which the two rubidium isotopes have the same polarizability are used to simultaneously interrogate the atomic samples in the interferometric sequence. This approach ensures a very high suppression of laser noise and common mode acceleration noise. The instrument will be accommodated on a rotating platform to control gravity gradient effects. 

The SAGE (Space Atomic Gravity Explorer) mission proposal \cite{Tino2019} has the scientific objective to investigate gravitational waves, dark matter, and other fundamental aspects of gravity such as the WEP as well as the connection between gravitational physics and quantum physics using optical atomic clocks and atom interferometers based on ultracold strontium atoms.

Several experiments and test activities are currently in progress to demonstrate the maturity of atom-based sensors for space operation and to evaluate the ultimate stability and accuracy that can be reached in differential acceleration measurements for WEP tests. 

In 23 January 2017, the MAIUS-1 experiment was launched in a sounding rocket to a height of 243~km. During the lift-off phase and the 360~min of free-fall conditions, 110 experiments involving atoms cooling and manipulation were performed. They include laser cooling and trapping of atoms, observation of the BEC phase transition, BEC transport on the atom chip, and study of BEC collective oscillations under weightlessness conditions \cite{Becker2018}. This experiment demonstrates the building blocks of future atom interferometry experiments in space. 

The Cold Atom Lab (CAL) is a multiuser facility launched to the ISS in 21 May 2018. The CAL instrument is designed to produce ultracold atomic samples of $^{39}$Rb, and $^{41}$K \cite{Elliott2018} down to quantum degeneracy. In the microgravity environment of the ISS, it is possible to decompress the atomic traps to very low levels thus achieving ultra-low densities and picokelvin temperatures. The experiment will test different atomic sources for atom interferometry in weightlessness conditions.

Significant progress has also been achieved by making use of microgravity facilities available on the ground, in particular the Bremen drop tower and the zero-gravity parabolic airplane. 

Mach-Zehnder interferometry experiments on a Bose-Einstein condensate have been performed in the Bremen drop tower \cite{Muntinga2013}. The drop tower capsule was operated both in drop and catapult mode, providing a free fall duration of 4.7~s and 9.4~s, respectively. The atom interferometer could then demonstrate a shot-noise limited resolution of $6.2\times10^{-11}$~ms$^{-2}$ in the drop mode and $5.5\times10^{-12}$~ms$^{-2}$ in the catapult mode. With this performance, a sensitivity of a few parts in $10^{13}$ for a WEP test should be possible in less than 100 drops (also see \cite{Herrmann2012}). Unfortunately, the study of systematic effects of a WEP test would result very unpractical in the drop tower facility.    

A WEP test on $^{87}$Rb and $^{39}$K has been performed in the microgravity conditions of an airplane in parabolic flight. The E\"{o}tv\"{o}s ratio was measured to $3.0\times10^{-4}$, limited by the noisy acceleration environment ($10^{-2}g$~Hz$^{-1/2}$). This result, certainly not competitive with respect to other WEP tests, remains important as it demonstrates the possibility of using correlated interference fringes to perform a WEP test with an accuracy two orders of magnitude below the level of ambient vibration noise. The experiment could therefore confirm the expected rejection to common mode vibration for a $^{87}$Rb-$^{39}$K differential interferometer \cite{Varoquaux2009, Barrett2015}.

\subsubsection{Tests with neutrons}

As for the atoms, the first low-precision measurements of gravity acceleration for neutrons were performed by measuring the drop of collimated beams of thermal neutrons  \cite{McReynolds1951,Dabbs1965}. They were also interpreted as tests of the universality of free fall. In \cite{Dabbs1965}, a test of a possible dependence of neutron acceleration on the two vertical neutron-spin projections $\pm \frac{1}{2}$ was performed finding no difference within the experimental sensitivity.

After the first observation of gravitationally induced interference in a neutron interferometer \cite{Colella1975}, tests of WEP were performed using neutron interferometers reaching a precision of $10^{-3}$ \cite{Staudenmann1980,Werner1988}, an accelerated interferometer \cite{Bonse1983}, and by a slow neutron gravity refractometer with a quoted relative uncertainty of $3 \times 10^{-4}$ \cite{Koester1976,Sears1982}.

More recently, gravity acceleration for neutrons was measured with a  cold neutron interferometer \cite{vanderZouw2000} and with a spin-echo spectrometer \cite{deHaan2014} with a relative precision of $10^{-3}$.

In \cite{Sarenac2018} prospects to achieve a relative precision $\Delta g/g \sim 10^{-5}$ using a three phase-grating moir{\'e} large area neutron interferometer were discussed proposing also a measurement of the value of the gravitational constant $G$ with neutrons.\\

\subsubsection{Tests with antimatter and with charged particles}

In principle, gravitation for antimatter may obey different laws than for ordinary matter. Measuring and comparing the gravitational
properties of matter and antimatter may probe different aspects of SME \cite{Kostelecky2010} and quantum vacuum \cite{Hajdukovic2010}. Theoretical considerations based
on energy conservation in the gravitational field and on
arguments from QFT constrain the validity of the WEP for antimatter
up to an accuracy of $10^{-14}$ \cite{Goldman1987, Nieto1991}. Nevertheless, these arguments are indirect and
need some experimental validation by comparing the effect of gravity on antiparticles on and their corresponding ordinary matter particles.

Experiments to test gravity on elementary particles and antiparticles were proposed and developed since the 1960's.
Electrically charged antiparticles (e.g. positrons and antiprotons) can be either observed in beams under free fall conditions, or trapped within a combination
of magnetic and electric fields, as in Penning traps \cite{vandyck1976}. 
In 1967 Witteborn and Fairbank observed the time-of-flight distribution of electrons and positrons in free fall inside a drift tube. 
Two decades later, an experiment was performed at the CERN Low Energy Antiproton Ring to measure the gravitational acceleration of antiprotons \cite{Holzscheiter1995}. $\bar p$ were collected and cooled in a Penning trap,  then released in a  vertical drift tube.
However both freely falling and trap-based
systems of charged antiparticles are affected by errors from residual stray electric and magnetic fields \cite{Darling1992} which make any 
gravitational measurement extremely difficult. The static electric field  $mg/q$ required to compensate gravity acceleration is only 56\,pV/m for positrons, and about 100\,nV/m for antiprotons. Even in case of perfect shielding from stray fields, the gravitational sag of electrons in a drift tube produces charge density anisotropies resulting in a electric forces of the same order of gravity \cite{Schiff1970}.
WEP tests
with electrons under weightless conditions have been proposed to get rid of gravity-induced electric
fields \cite{Dittus2004}.  
On the other hand, current experiments to test WEP on anti-matter are focusing on neutral systems such as positronium \cite{Cassidy2014}, muonium \cite{Kirch2014,Antognini2018}, antihydrogen \cite{Walz2004,Kellerbauer2008,Sacquin2014,Hamilton2014,Perez2015}, and on particle-antiparticle pairs \cite{Apostolakis1999}, for which the effect of stray fields is strongly suppressed.

The first evidence of antihydrogen was achieved at the European Organization for
Nuclear Research (CERN) in 1995 \cite{Baur1996} and confirmed two years later at Fermilab \cite{Blanford1998}. The recombination of electron-positron pairs produced from the collision of relativistic antiprotons with Xenon targets formed $\bar H$ atoms at relativistic energies, unsuitable for precision measurements. 
A breakthrough occurred when $\bar H$ at thermal energies (few hundreds K) was first produced in 2002 by the ATHENA experiment  \cite{Amoretti2002}, via three-body reaction by mixing
trapped antiprotons ($\bar p$) with positrons ($e^+$) at low energies. This result, shortly followed by a similar achievement from the ATRAP experiment \cite{Gabrielse2002}
opened the possibility to test WEP on
neutral antimatter. 

However the gravitational force is so weak that a WEP experiment requires further cooling of antimatter down to cryogenic temperatures. 
Several second-generation experiments with antimatter
have been then developed at CERN, where the only intense source of low energy antiprotons is available worldwide, i.e. the Antiproton Decelerator (AD) currently under upgrade to Extra Low ENergy Antiproton ring (ELENA). 
Such experiments must face several challenges. Neutral antimatter particles are produced from their charged constituents. This requires  complex experiments combining advanced methods from high-energy physics for particle beams optics and detectors, with advanced methods for ion trapping and atom optics. Moreover neutral antiparticles are produced at much lower rates and at much higher temperatures than in the typical quantum sensors described in section  \ref{sec:atoms}. 
The various experiments at AD developed different methods to produce $\bar H$ at rates and temperatures suited for precision measurements. 

The ALPHA experiment, designed to perform high resolution spectroscopy on $\bar H$, generates anti-hydrogen by three-body recombination between trapped, evaporatively cooled antiprotons and trapped positrons. The low-energy tail of the $\bar H$ distribution is captured in a magnetic trap at a rate of about 10 atoms on cycles of 4 minutes \cite{Ahmadi2017}. ALPHA performed a preliminary measurement of the Earth’s gravitational effect on magnetically trapped $\bar H$. The resulting gravitational acceleration of $\bar H$ was constrained to within 100 times
the $g$ value for matter \cite{Amole2013}.

The GBAR experiment aims to generate ultracold antihydrogen through the anti-ion $\bar H^+$ \cite{Walz2004, Sacquin2014}. The $\bar H^+$ ion is produced via two cascaded charge exchange processes
from the interaction of $\bar p$ with a positronium target, then of the generated $\bar H$ with the same target.  $\bar H^+$ ion can be sympathetically cooled with laser cooled Be$^+$ ions down to $\mu$K temperatures. The excess positron can then be laser detached in
order to recover the neutral $\bar H$ with very low temperature. A
high-intensity positron source has been developed for $\bar H^+$ production. 

The projects Antimatter
Experiment: Gravity, Interferometry, Spectroscopy
(AEgIS) \cite{Kellerbauer2008}, which is operating since 2012, has been designed to measure the gravitational acceleration  with matter-wave interferometry on a pulsed  $\bar H$ beam at sub-kelvin temperatures. 
In AEgIS, antihydrogen atoms are produced via a charge exchange
reaction between Rydberg-excited positronium atoms and cold antiprotons
within an electromagnetic trap. The resulting Rydberg antihydrogen atoms will
 be horizontally accelerated by an electric field gradient (Stark effect), then they will
pass through a moir\'e deflectometer. The vertical deflection caused by the Earth's gravitational
field will provide a Weak Equivalence Principle test for antimatter.
Detection will be undertaken via a position sensitive detector. Around $10^3$ antihydrogen
atoms are needed for the gravitational measurement to be completed.
The generation of antihydrogen via charge-exchange process was already demonstrated in the ATRAP
experiment \cite{Storry2004}, where Ps were excited toward a Rydberg state by collisions with laser-excited
cesium atoms. AEgIS rather plans to directly laser
excite positronium.

An alternative to testing UFF on bound antimatter systems is the search for mass differences on particle-antiparticle pairs. In particular, neutral kaon is the only system where particle-antiparticle differences are detected; this is explained as arising from CP-violating terms in the $K^0-\bar K^0$  mass matrix. 
In \cite{Apostolakis1999} upper limits on possible $K^0-\bar K^0$  mass difference were determined from the analysis of data on tagged $K^0$ and $\bar K^0$ decays into $\pi^+\pi^-$ from the CPLEAR experiment over three years. The results are in agreement with the Equivalence Principle to a level of 6.5, 4.3 and $1.8\times10^{-9}$ respectively, for scalar, vector and tensor potentials originating from the Sun. Such determination of mass difference for kaon is ten orders of magnitude more precise than for $p-\bar p$ \cite{Gabrielse1990}.

High precision gravity measurements will require the application of interferometric methods. While the most precise quantum sensors are based on light pulse matter-wave interferometry, see section \ref{sec:atoms}, such method is not readily applied on antimatter systems. This is mostly due to the comparably high temperatures currently achievable and to the extremely short wavelength of resonance optical transitions in $\bar H$ and in $ps$. 
Inertial sensing with Talbot-Lau interferometry  \cite{Sala2016} allows to work with low-intensity, weakly coherent beams. This method has been recently demonstrated on a  beam of low energy positrons \cite{Sala2019}.










\section{Conclusions and outlook}


General Relativity and metric theories of gravity are  based on the validity of the Equivalence Principle, according to which the gravitational acceleration is (locally) indistinguishable from acceleration caused by mechanical (apparent) forces. The consequence of the Equivalence Principle is that gravitational mass is equal to inertial mass, $m_g=m_I$. This identity was already pointed out by  Galileo and Newton, but Einstein recognizes it as a fundamental aspect involving also accelerations and forces and then  elevating it to a principle.  Equivalence Principle allowed  Einstein to construct a theory capable of  explaining  gravity  and acceleration under the same physical standard.  Based on this assumption, he stated the following   fundamental postulated: in a free falling frame, all non-gravitational laws of physics (hence not only the mechanical ones) behave as if gravity was absent. More generally, Equivalence Principle asserts that objects with different (internal) composition are subject to the same acceleration when moving in a gravitational field. This new principle of nature led Einstein to the revolutionary interpretation of gravitation: gravity can be described as a curvature effect of  space-time. as a consequence, the Einstein Equivalence Principle plays a crucial role in all metric theories of gravity, as well as in the Standard Model of particle physics which is not in conflict with GR.

As we have seen in this review, the Equivalence Principle essentially encodes the local Lorentz invariance (clock rates are independent of the clocks velocities), the local position invariance (the universality of red-shift) and the universality of free fall (all free falling point particles follow the same trajectories independently of their internal structure and composition).  The first to two principle, i.e. the Lorentz and position invariance, are hence related to the local properties of physics, so that they can be tested by using atomic clocks and measurements of spectroscopy, while the third  one, the universality of free falling point-like particles, can be tested by tracking trajectories, hence freely falling test masses as discussed in the experimental part of this Review paper.

The Equivalence Principle can be formulated in two different forms: the weak and the strong form.

The weak form of the Equivalence Principle states that the gravitational properties of the interaction of particle physics of the Standard Model, hence the strong and electro-weak interactions, obey the EP.
As we have pointed out, the equality $m_g=m_i$ implies that, in an external gravitational field, different (and neutral) test particles undergo to the same free fall acceleration, and in a free falling inertial frame only tidal forces may appear (apart the latter, free falling bodies behave as if the gravity is absent). However, it is worth  noticing that in many extensions of the Standard Model, new interactions (quantum exchange forces) are introduced, and, in general, they may violate the weak equivalence principle owing to the coupling with generalized charges,
rather than mass/energy as happens in gravity.

The second form, the strong Equivalence Principle, is such that it extends the weak one including the gravitational energy. In GR,  the strong Equivalence Principle is fulfilled thanks to the gravitational stress-energy tensor, while it can be  violated in some extension of GR (as, for example, scalar tensor theories discussed above where a  scalar field is present in the gravitational interaction and can be non-minimally coupled with geometry).

As a final remark, an important issue has to be discussed. It is  related to the parameterized post-Newtonian formalism \cite{Will1993,Turyshev2008} and the Equivalence Principle. In a nutshell, the PPN approximation is a method for obtaining the
motion of the system in terms of higher powers of the small parameters (gravitational potential and velocity square) with respect the ones given by Newtonian mechanics. The relevant aspect is that this formalism allows to describe the motion of $M_q$ celestial bodies that is common to many theories of gravity. The acceleration of a body can be written in the form  ${\ddot {\bf r}}_p={\ddot {\bf r}}^{GR}_p+\delta {\ddot {\bf r}}^{PPN}_p$, where ${\ddot {\bf r}}^{GR}_p$ is the usual acceleration derived in GR, while ${\ddot {\bf r}}^{PPN}_p$ is the PPN corrections \cite{Will1993,Turyshev2008}
 \begin{eqnarray*}
 {\ddot {\bf r}}^{PPN}_p &=& \sum_{q\neq p}\frac{GM_q({\bf r}_q-{\bf r}_p)}{|{\bf r}_q-{\bf r}_p|^3}\left\{
 \left(\left[\frac{M_g}{M_I}\right]_q-1\right)
 - \frac{2(\beta+\gamma-2)}{c^2}\sum_{r\neq p}\frac{GM_r}{|{\bf r}_r-{\bf r}_p|}
 -\frac{2(\beta-1)}{c^2}\sum_{s\neq q}\frac{GM_s}{|{\bf r}_s-{\bf r}_q|} \right. \\
 & + & \left. \frac{\gamma-1}{c^2}|{\dot {\bf r}}_q-{\dot {\bf r}}_p|^2 +\frac{\dot G }{G} (t-t_0) \right\}
 +\frac{2(\gamma-1)}{c^2}\sum_{q\neq p}\frac{GM_q}{|{\bf r}_p-{\bf r}_q|}\left[{\ddot {\bf r}}_q+
 \frac{[({\bf r}_p-{\bf r}_q)\cdot ({\dot {\bf r}}_p-{\dot {\bf r}}_q)]({\dot {\bf r}}_p-{\dot {\bf r}}_q)}{|{\bf r}_p-{\bf r}_q|^2} \right]\nonumber
 \end{eqnarray*}
Some comments are in order: ${\it 1)}$ The correction ${\ddot {\bf r}}^{PPN}_p$ vanishes in the case of GR, that is when the PPN parameter $\gamma$ and $\beta$ assume the values: $\gamma=1=\beta$. ${\it 2)}$ The expression for ${\ddot {\bf r}}^{PPN}_p$ does contain the variation of the gravitational constant $G$, the ${\dot G}/G$-term, typical of  scalar tensor theories of gravity; ${\it 3)}$ In ${\ddot {\bf r}}^{PPN}_p$ a term related to the strong equivalence principle appears, i.e. $\left[\frac{M_g}{M_I}\right]-1$, which is typically expressed in the form $\left[\frac{M_g}{M_I}\right] = 1 + \eta\left(\frac{E}{mc^2}\right)$, where the parameter $\eta$, depending on PPN parameters as $\eta=4\beta-\gamma-3$, encodes deviations from GR, and it is therefore related to the violation of the strong Equivalence Principle ($\eta=0$ in GR), while $m$ and $E$ are the mass and gravitational self-energy of the body, respectively. Taking, for example, a uniform sphere of radius $R$ one gets
 \begin{equation}
 \left[\frac{E}{mc^2}\right]=-\frac{G}{2mc^2}\int d^3 x\, d^3 x^\prime \frac{\rho(x)\rho(x')}{|{\bf x}-{\bf x}'|}=-\frac{3}{5}\frac{GM}{c^2 R}\,.
 \end{equation}
This relation shows that, for Solar System it is $\left[\frac{E}{mc^2}\right]_\odot\sim 10^{-6}$, while for bodies of lab test one gets $\left[\frac{E}{mc^2}\right]_{lab}\sim 10^{-25}$. This suggests that planet-size bodies are required for testing the strong Equivalence Principle with a certain confidence level.

Finally, we discussed  the possibility to violate the Equivalence Principle considering systems at finite temperature.
Although the Equivalence Principle asserts (in the weak form) that the gravitational acceleration is identical for all
bodies, i.e.  $m_I=m_g$, the latter equality can be violated in  quantum field theory considering a  finite temperature framework.  In fact,  as shown in a series of seminal papers \cite{Donoghue1984a}, a {\it fraction of the mass of a particle arises through the finite-temperature component of the radiative corrections}. This result is a consequence of the Lorentz non-invariance of the finite temperature vacuum. According to this result, theories at finite temperature could be the straightforward way to test Equivalence Principle at fundamental quantum level.

In conclusion, various issues in modern physics, both from  gravitational and particle physics sectors, predict violations of the Equivalence Principle. Given the importance of this question, the  experimental challenge is to look for frameworks where possible  violations could manifest. Besides, one has to  improve the limits of experimental tests. As we have shown in this review, matching experiments from  laboratory and satellites is  an important complement  to probing fundamental physics at very high precision level, and, in turn, possible results  could open new and unexpected scenarios.


\section{Acknowledgements}
G.M.T. acknowledges useful discussions with Enrico Iacopini and Marco Tarallo.
S.C. and G.L. thank Maurizio Gasperini for useful discussions on the presented topics and {\it Istituto Nazionale di Fisica Nucleare} (INFN)-{\it iniziativa specifica} MOONLIGHT2.
This work was supported by the Italian Ministry of Education, University and Research (MIUR) under the PRIN 2015 project ``Interferometro Atomico Avanzato per Esperimenti su Gravit\`{a} e Fisica Quantistica e Applicazioni alla Geofisica''.






\end{document}